\documentclass[aps,prc,twocolumn,groupedaddress,showpacs]{revtex4-1}

\usepackage{hyperref}
\usepackage{dcolumn}
\usepackage{xspace}
\usepackage{amssymb}
\usepackage{amsmath}
\usepackage{graphicx}% Include figure files
\usepackage{bm}% bold math
\usepackage{color}

\definecolor {Blue}                {rgb}{.9,.9,.99}

\def\y{{\bm y}}
\def\tri{{{}^3{\rm H}}}
\def\hel{{{}^3{\rm He}}}
\def\het{{{}^3{\rm He}}}
\def\heq{{{}^4{\rm He}}}

\def\jac{x}

\def\phan{\phantom{9}}
\newcommand{\jacb}{{\mbox{\boldmath $x$}}}
\def\hypfi{\varphi}

\begin{document}

\title{Study of $n+\tri$, $p+\het$, $p+\tri$, and $n+\het$ scattering with the HH
  method}

\author{M. Viviani$^1$, L. Girlanda$^{2,3}$, A. Kievsky$^1$, 
and L.E. Marcucci$^{1,4}$}

\affiliation{
$^1$ Istituto Nazionale di Fisica Nucleare, Sezione di Pisa, 
  Largo B. Pontecorvo 3, I-56127, Pisa, Italy \\
  $^2$ Department of Mathematics and Physics, University of Salento,
  Via Arnesano, I-73100 Lecce, Italy \\
$^3$ INFN-Lecce,  Via Arnesano, I-73100 Lecce, Italy \\  
$^4$ Department of Physics ``E. Fermi'', University of Pisa, 
Largo B. Pontecorvo 3, I-56127, Pisa, Italy }

\begin{abstract}

The $n+\tri$, $p+\het$, $p+\tri$, and $n+\het$ elastic and
charge exchange reactions at low energies are studied by means of
the hyperspherical harmonic method. The considered nuclear Hamiltonians
include modern two- and three-nucleon interactions, in particular
results are reported in case of chiral two-nucleon potentials,
with and without the inclusion of chiral three-nucleon (3N) interactions.
A detailed study of the convergence and numerical stability of
the method is presented. We have found that the effect the 
3N force is in general tiny except for $p+\tri$ scattering below
the opening of the $n+\het$ channel. In such a case, the effect of 3N forces
is appreciable and a clear dependence on the cutoff used to regularize the
high-momentum tail of the interactions is observed. Such a dependence
is related to the presence of the poorly known sharp $0^+$ resonance,
considered to be the first excited state of $\heq$. 
\end{abstract}

\pacs{13.75.Cs 21.45.+v 21.30.-x 24.70.+s 25.10.+s 25.40.Cm 25.55.Ci 26.65.+t
      27.10.+h}
%\keywords{}

\maketitle


\section{Introduction}
\label{sec:intro}

The four nucleon (4N) system has been object of intense studies
in recent years. In first place, this system is particularly interesting as a
``theoretical laboratory" to test the accuracy of our present
knowledge of the nucleon--nucleon (NN) and three nucleon (3N) interactions. 
In particular, the effects of the NN P-waves and of the 3N forces  are 
believed to be larger than in the $A=2$ or $3$ systems. Moreover,
it is the simplest system where the 3N interaction can be studied in
channels of total isospin $T=3/2$. There is a number
of reactions involving 4Ns which are of extreme importance for
astrophysics, energy production, and studies of fundamental
symmetries. As an example, the  $p + \tri \rightarrow \heq + e^+ +
e^-$ reaction is currently exploited as a tool for the discovery of an
unknown particle~\cite{Kras19}. 

Nowadays, the 4N bound state problem can be numerically solved with
good accuracy. For example, in Ref.~\cite{Kea01} the binding energies and
other properties of the $\alpha$-particle were studied using the
AV8$'$~\cite{AV18+} NN interaction; several different techniques produced
results in very close agreement with each other (at the level of, or less than,
1\%). More recently, the same agreement has also been obtained  considering
different realistic NN+3N interactions~\cite{Nogga03,Lazaus04,Viv05,Wea00}. 

In recent years, there has been a rapid advance in solving the 4N
scattering problem with realistic Hamiltonians. Accurate calculations of four-body
scattering observables have been achieved in the framework of the
Faddeev-Yakubovsky (FY) equations~\cite{DF07,DFS08,DF12,DF14,DF15,DF17}, solved in momentum
space, where the long-range Coulomb interaction is treated using the 
screening-renormalization me\-thod \cite{Alt78,DFS05}.
Solutions of the FY equations in configuration space~\cite{Cie98,Lea05,Lazaus09,Lazaus12,LC20}
and several calculations using the resonating group model (RGM)~\cite{HH97,PHH01,HH03}
were also reported. The application of the RGM together with the no-core
shell model (NCSM) technique is being vigorously pursued~\cite{Sofia08,Sofia10},
and the possibility of calculations of scattering observables using the Green
Function Monte Carlo method has been explored, too~\cite{Wiringapc}.

In this contribution, the four-body scattering problem is solved using the 
Kohn variational principle and expanding the ``core'' part of the wave function
(namely,  the part which describes the system where the particles
are close to each other) in terms of the hyperspherical harmonic (HH) functions (for a review, 
see Ref. \cite{rep08,fip19}). Preliminary applications of this method were
reported in Refs.~\cite{VKR98,Vea01,Lea05} for local potentials, as the
Argonne $v_{18}$ (AV18)~\cite{AV18} NN potential, and in Refs.~\cite{Vea06,Mea09,Vea09}
for non-local potentials.
Accurate benchmarks between FY (solved in momentum and configuration space)
and HH results were reported in Ref.~\cite{bm11} for $n+\tri$ and
$p+\het$ elastic scattering, and in Ref.~\cite{bm16} for $p+\tri$ and
$n+\het$ elastic scattering and charge-exchange reactions.
These calculations were limited to energies below the threshold
for three-body breakup. The good agreement found between the results
obtained by the different methods attested the high accuracy
reached in solving the 4N scattering problem.

In the present paper, the application of the HH method to study these
reactions is presented in full detail, focusing in particular on the selection
of the basis and the techniques used to evaluate the matrix
elements. A discussion of the convergence of the calculated
observables is also reported. Finally, we critically compare the results
of this first campaign of calculations with the available experimental data.

The potentials used in this study are the chiral interactions derived
at next-to-next-to-next-to-leading order (N3LO) by Entem and
Machleidt~\cite{EM03,ME11}, with cutoff $\Lambda=500$ and $600$ MeV.
In some selected cases, we have also performed calculations
using the new interactions derived at
next-to-next-to-next-to-next-to-leading order (N4LO) in
Ref.~\cite{MEN17}. In this case, the adopted values of the cutoff
parameter are $\Lambda=450$,  $500$, and
$550$ MeV. The calculations performed using one of these
chiral potentials are labeled as N3LO500, etc., i.e. specifying the
value of the cutoff. We present the convergence of the calculated
phase-shifts for the case of the AV18 interaction, too.

We have also performed calculations including
the chiral 3N interaction derived at next-to-next-to leading order (N2LO)
in Refs.~\cite{Eea02,N07}. The two free parameters in this
N2LO 3N potential, denoted usually as  $c_D$ and $c_E$, have been
fixed in order to reproduce the experimental values of the $A=3$ binding energies
and the Gamow-Teller matrix element (GTME) of the tritium $\beta$ decay~\cite{GP06,GQN09}.
Note that these parameters have been recently redetermined~\cite{Mea12,Bea18,Mea18}
after finding (and correcting) an inconsistency in the relation
between the 3N parameter $c_D$  and the axial current used so far~\cite{Schiavilla}.

The cutoff in the 3N interaction has been chosen to be consistent with
the corresponding value of the NN interaction. The development of a 3N interaction
including N3LO and N4LO contributions is still under
progress~\cite{Bea07,Krebs12,Hebeler15}. In some cases, we
have considered the Urbana IX (UIX) and Illinois 7 (IL7) 3N potentials~\cite{Pudliner95,Illinois}, 
used with the AV18 potential. The calculation including
both NN and 3N interactions have been labeled as N3LO500/N2LO500, AV18/IL7, etc.

The four-body studies performed so far have highlighted several discrepancies between
the theoretical predictions and experimental data. Let us consider
first the $n+\tri$ elastic scattering. Calculations based on NN 
interactions disagree~\cite{VKR98,Cie98,Fon99,PHH01,Lea05,DF07}
rather sizably with the measured total cross section~\cite{PBS80}, both at
zero energy and in the ``peak'' region ($E_n\approx 3.5$ MeV). Such an
observable is found to be very sensitive to the NN interaction
model~\cite{DF07}. At low energy, the discrepancy is removed by including a 3N
force fixed to reproduce the triton binding energy~\cite{VKR98,Cie98,PHH01}, but it
remains in the peak region. In Ref.~\cite{Vea09}, in a preliminary calculation,
we observed that this disagreement is noticeably reduced using the N3LO500/N2LO500
interaction. In the present paper, we will report the results of more refined calculations
performed with the N3LO500/N2LO500 and N3LO600/N2LO600 interactions, confirming
the results of Ref.~\cite{Vea09}.

Regarding the $p+\het$ elastic scattering, several accurate measurements
of both the unpolarized cross section~\cite{Fam54,Mcdon64,Fisher06} and
the proton analyzing power $A_{y0}$~\cite{All93,Vea01,Fisher06} can be found
in literature, allowing for a detailed study. The calculations performed so far with a 
variety of NN interactions have shown a glaring  
discrepancy between theory and experiment for
$A_{y0}$~\cite{Fon99,Vea01,PHH01,Fisher06,DF07}. This discrepancy is very 
similar to the well known ``$A_y$  Puzzle''  in $N+d$ scattering. This is a
fairly old problem, already reported about 30 years ago~\cite{KH86,WGC88} in
the case of $n+d$ and later confirmed also in the $p+d$ case~\cite{Kie96}.
The inclusion of usual models of the 3N force has little effect on
these $A=3$ observables.  To solve this puzzle, speculations about
the deficiency of the NN potentials in ${}^3P_J$ waves (where the
spectroscopic notation $^{2S+1}L_J$ has been adopted) have been
advanced. More recently, the effects of the contact terms appearing at
N4LO in the 3N force have been explored in order to explain this
puzzle~\cite{Girla11,Girla19}. The situation of other $p+\het$
observables (the $\het$ analyzing power $A_{0y}$ and some spin
correlation observables as $A_{yy}$, $A_{xx}$, etc.) is less clear due
to the lack of equally accurate measurements. About ten years
ago~\cite{Dan10}, at the Triangle University National Laboratory
(TUNL), a new set of accurate  measurements of various spin
correlation coefficients were obtained at 
$E_p=1.60$, $2.25$, $4$ and $5.54$ MeV, allowing for a phase-shift 
analysis (PSA). The aim of this paper is to compare the results of
the theoretical calculations to these data. The effect of the
inclusion of the chiral 3N force in $p+\het$ has been already reported
in Ref.~\cite{Vea13}, where we have shown that the inclusion of the 
chiral 3N interaction improves the agreement with the experimental
data, in particular, for the proton vector analyzing power. This
result is confirmed by the present study.

Regarding $p+\tri$ and $n+\het$ below the $d+d$ threshold only a few
accurate calculations exist~\cite{DF07,DFS08,Lazaus09,DF15}. From the
experimental point of view, for this range of energies there exist
several measurements of the $p+\tri$ elastic differential cross
section~\cite{Hemme49,Clas51,Balas65,Jarmie59,Brol64,Mandu68,Iva68,Kanko76},
$n+\het$ elastic cross section~\cite{Sea60,Say61}
and total cross section~\cite{Sea60,Say61,Alfi81,Haes83}, and various $n+\het$ elastic
polarization observables~\cite{Holla72,Sinra76,Klage85,Jany88,Este13}.
Regarding the $n+\het \rightarrow  p+\tri$ charge exchange reaction,
there exist measurements of the total 
cross section~\cite{Coon50,Batch55,Gibb59,Say61,Als64,Mack65,Cost70,Borza82,Haes83},
of the differential cross section~\cite{Jarvis50,Will53,Jarvis56,Drosg80}
and polarization observables~\cite{Cra71,Doy81,Tornow81,Walst98,Wilburn98}.
Preliminary results obtained for these observables with the HH method were already
presented in Ref.~\cite{bm16}.

Here, we complete those previous
studies and, in particular, we study the effect of the inclusion of the
3N interaction in the low-energy ${}^1S_0$ phase shifts in order to
extract the resonance energy and width of the first excited state of
$\heq$. Such a state of $\heq$ is of particular interest.
Its energy is slightly above the threshold for
$p+\tri$ breakup, but below that of $n+\het$~\cite{W70}. For the description
of this resonance therefore the Coulomb potential plays a very
important role. The nature of such a
resonance is still a puzzle after many years of studies. Electron
scattering can give directly information on the transition form factor~\cite{W70,Fea65,Kea83}
\begin{equation}
  {\cal S}_{\cal M}(q,\omega)= \sum_n |\langle n|{\cal M}(q)| 0\rangle|^2\delta(\omega-E_n+E_0)
  \ , \label{eq:tff}
\end{equation}
where $| 0\rangle$, $| n\rangle$, $E_0$, $E_n$ are eigenfunctions and
eigenvalues of the 4N Hamiltonian $H$, respectively,
${\cal M}(q)$ is the isoscalar monopole operator, and $\omega$, $q$
the energy and three-momentum transferred by the external probe.
At low values of $\omega$,  ${\cal S}_{\cal M}(q,\omega)$ is dominated by
the contribution of the first $0^+$ excited state of $\heq$, which therefore
can be studied theoretically and experimentally. The interpretation of this excited state 
as a collective breathing mode or a particle-hole-like excitation
is still to be clarified. Recently, two theoretical studies of
${\cal S}_{\cal M}(q,\omega)$ were performed.
In Ref.~\cite{Hiyama04}, ${\cal S}_{\cal M}(q,\omega)$ was calculated using a
bound state technique, i.e. expanding the wave function over a 
Gaussian basis. In Refs.~\cite{Bacca13,Bacca14} a calculation
using the Lorentz Integral Transform method to sum implicitly all the
intermediate states was performed. The calculated
transition form factors differ by a factor two and the
origin of this discrepancy has not yet been clarified. 
In this contribution, we  estimate the
position and width of the resonance directly from
the calculated phase-shifts. 

This paper is organized as follows. In Section~\ref{sec:theory}, a 
description of the method is reported, while in Sec.~\ref{sec:conv} a detailed 
discussion of the convergence and numerical stability of the calculated
phase-shifts is presented. The results are reported in Section~\ref{sec:res}, and
compared with the available experimental data.  The conclusions and the perspectives of this
approach will be given in Section~\ref{sec:conc}. Some details
regarding the regularization of the irregular Coulomb functions are given in the Appendix.

\section{The HH Technique for Scattering States}
\label{sec:theory}

This Section is divided into five subsections. First, we discuss the
asymptotic part of the wave function, and then the HH expansion of the
core part. In Subsections~\ref{sec:kvp} and~\ref{sec:theory3}, we discuss the
application of the Kohn variational principle and then give some details of the
calculations. Finally, in Subsection~\ref{sec:theory4} we discuss the choice of
the subset of HH functions considered in the calculation.

\subsection{Asymptotic functions}
\label{sec:theory1}

In this paper we limit ourselves to consider asymptotic states with
two clusters in the initial/final states, denoted generally as
$A+B$. For the sake of simplicity, a specific clusterization $A+B$ will be
denoted by the index $\gamma$. More specifically, $\gamma=1,\ldots,4$ will
correspond to the following clusterizations: $n+\tri$, $p+\het$,
$p+\tri$, and $n+\het$, respectively. Depending on the total charge (and on the energy), some of
these asymptotic states may enter or not in the wave function. 

Let us consider a scattering state with total angular momentum quantum number
$JJ_z$, and parity $\pi$ (the dependence on the wave function and other
quantities on $JJ_z\pi$ will be understood in the following).
The wave function $\Psi^{\gamma L S}$ describing incoming clusters $\gamma$
with relative orbital angular momentum $L$ and channel spin
$S$ can be written as
\begin{equation}
    \Psi^{\gamma LS}=\Psi_C^{\gamma LS}+\Psi_A^{\gamma LS} \ ,
    \label{eq:psica}
\end{equation}
where the core part $\Psi_C^{\gamma LS}$ vanishes in the limit of large inter-cluster
separations, and hence describes the system where the particles
are close to each other and their mutual interactions are strong. On the other
hand, $\Psi_A^{\gamma LS}$ describes the relative motion of the two clusters in
the asymptotic regions, where the mutual interaction is
negligible (except for the long-range Coulomb interaction). In the
asymptotic region the wave functions $\Psi^{\gamma LS}$ reduces to
$\Psi_{A}^{\gamma LS}$, which must therefore be the appropriate asymptotic
solution of the Schr\"odinger equation. $\Psi_{A}^{\gamma LS}$ can be decomposed
as a linear combination of the following functions
\begin{eqnarray}
  \Omega_{\gamma LS}^\pm &=& D_{\gamma } {\cal A}\biggl\{
  \Bigl [ Y_{L}(\hat{\bm y}_\gamma) \otimes  [ \phi_A \otimes \phi_B]_{S} 
   \Bigr ]_{JJ_z} \nonumber\\
  &&\times \left ( {\frac{\widetilde G_{L}(\eta_\gamma,q_\gamma y_\gamma)}{q_\gamma y_\gamma}
          \pm {\rm i} {\frac{F_L(\eta_\gamma,q_\gamma y_\gamma)}{q_\gamma y_\gamma}}} \right )\biggr\} \ ,
  \label{eq:psiom}
\end{eqnarray}
where $D_\gamma$ are appropriate normalization factors (see below), 
$y_\gamma$ is the distance between the center-of-mass (c.m.) of clusters $A$
and $B$, $q_\gamma$ is the magnitude of the relative momentum between the
two clusters, and $\phi_A$ and $\phi_B$ bound state wave functions. 
In the present work, the trinucleon bound state wave functions (for both $\het$ and $\tri$)
are calculated very accurately by means of the HH method~\cite{Nogga03,Vea06}
using the corresponding $A=3$ Hamiltonian. For a single nucleon,
$\phi$ reduces to the spin-isospin state. The channel spin $S$ is
obtained coupling the angular momentum of the two clusters. In our
case, clearly $S=0,1$. The symbol ${\cal A}$ means that the expression
between the curly braces has to be properly antisymmetrized.

The total energy of the scattering state in the c.m. system is 
\begin{equation}
  E=-B_A-B_B+T_{r}\, \label{eq:energy}
\end{equation}
where 
\begin{equation}
  T_{r}={q_\gamma^2\over 2\mu_\gamma}\ , \qquad
  {1\over \mu_\gamma} = {1\over M_A}+{1\over M_B} 
   \ ,\label{eq:tcm}
\end{equation}
and $M_X$ ($B_X$) is the mass (binding energy) of the cluster $X$.
Clearly, in the case of a single nucleon $M_X=M_N$, where $M_N$ is the
nucleon mass, and $B_X=0$. 

In Eq.~(\ref{eq:psiom}), the functions $F_L$ and
$\widetilde G_{L}$ describe the asymptotic radial motion of the
clusters $A$ and $B$. If the two clusters are composed of $Z_A$ and $Z_B$
protons, respectively, the parameter $\eta_\gamma$ is defined as
$\eta_\gamma=\mu_\gamma Z_A Z_B e^2/q_\gamma$, where $e^2\approx 1.44$
MeV fm. The function $F_L(\eta,qy)$ is
the regular Coulomb function, while $\widetilde G_{L}(\eta,qy)$ is a
``regularized'' version of the irregular Coulomb function
$G_{L}(\eta,qy)$.  In this work, we  
have used two different methods of regularization, namely
\begin{eqnarray}
  1)\quad {\widetilde G_{L}(\eta,qy)\over qy} & = & {G_{L}(\eta,qy)\over qy}
    -{ f_L(y)\over y^{L+1}} \exp(-\beta y) \ ,\label{eq:reg1} \\
  2) \quad {\widetilde G_{L}(\eta,qy)\over qy}  & = & {G_{L}(\eta,qy)\over qy}
    [1-\exp(-\beta y)]^{2 L+1}\ .\label{eq:reg2}
\end{eqnarray}
where $f_L(y)$ is chosen so that both functions $\widetilde
G_{L}(\eta,qy)/ qy$  and 
\begin{eqnarray}
  \overline G_{L}(\eta,qy) &=& \biggl[ {d^2\over dy^2}+ {2\over y} {d\over dy}
    -{L(L+1)\over y^2} \nonumber \\
   && \quad -{2\eta q \over y} + q^2\biggr] 
   {\widetilde  G_{L}(\eta,qy)\over qy} 
   \ ,\label{eq:ovrl}
\end{eqnarray}
be regular for $y\rightarrow0$. The functions $f_L$ are in general given as
\begin{eqnarray}
  f_L(y) &=& a_0 + a_1 y+ a_2 y^2 + \cdots + a_N y^N \nonumber\\
      &&     + ( b_1 y+ b_2 y^2 + \cdots + b_M y^M)\log(2qy)\ ,
   \label{eq:flreg}
\end{eqnarray}
where $N$, $M$ are positive integers and the coefficients $a_i$, $b_i$ can be
determined considering the analytic behavior of functions
$G_{L}(\eta,qy)/qy$ and $\overline G_{L}(\eta,qy)$ for $y\rightarrow 0$. The
expressions of the functions $f_L(y)$ are given 
in the Appendix. Method 2 is simpler. However, regularizing using method 1 has
the following advantage: In computing $(H-E)\Omega_{\gamma LS}^{G}$, we are left
(between others) with a term proportional to $\overline G_L$. Note
that the Coulomb functions (both the regular and the irregular) are the
solution of the equation
\begin{equation}
     \left[ {d^2\over dy^2}+ {2\over y} {d\over dy}
    -{L(L+1)\over y^2} -{2\eta q \over y} + q^2\right]
     {X_{L}(\eta,qy)\over qy} =0
   \ ,\label{eq:couleq}
\end{equation}
therefore, using method 1, we have
\begin{eqnarray}
 \overline G_L&=& -\biggl\{ f_L''-\biggl(2\beta+{2L\over y}\biggr)
 f_L' \nonumber \\
  &&\quad  +\biggl(\beta^2+2{\beta L-\eta q\over y } +q^2\biggr) f_L\biggr\}
     {e^{-\beta y}\over y^{L+1}} \ ,\label{eq:ovrlgg}
\end{eqnarray}
where $f'=df/dy$, etc. As discussed in the Appendix, the functions $f_L(y)$
are constructed so that $\overline G_L$ be regular at the origin. 
Therefore, the resulting function $\overline G_L$ is a
smooth function, not having the oscillatory behavior of $G_L$. For this
reason, using method 1, the matrix elements $\langle \Psi| H-E|
\Omega_{\gamma LS}^G\rangle$ are (slightly) less problematic from the
numerical point of view than using method 2 of regularization.

Note that using both methods, the functions $\overline G_{L}$
vanish exponentially as $y\rightarrow\infty$. Moreover, $\widetilde
G_{L}(\eta,qy)\rightarrow G_{L}(\eta,qy)$ when $y\gg 1/\beta$,
thus not affecting the asymptotic behavior of $\Psi_{A}^{\gamma LS}$, namely
\begin{equation}
  \widetilde G_{L}(\eta,qy)\pm {\rm i} F_L(\eta,qy) \rightarrow
  e^{\pm {\rm i} \bigl (q y-L\pi/2-\eta\ln(2qy)+\sigma_L\bigr ) }\ ,
\end{equation}
where $\sigma_L$ is the Coulomb phase shift.
Therefore, $\Omega_{\gamma LS}^+$  ($\Omega_{\gamma LS}^-$) describes the
outgoing (ingoing) relative motion of the clusters specified by $\gamma$. 

If one of the clusters is a neutron (cases $\gamma=1$ or $4$), then $\eta=0$ and
the functions $F_L$ and $G_L$ reduce to
\begin{equation}
 {F_L(\eta,qy)\over qy } \rightarrow j_L(qy)\ , \qquad
 {G_L(\eta,qy)\over qy } \rightarrow - y_L(qy)\ , 
 \label{eq:eta0}
\end{equation}
where $j_L$ and $y_L$ are the regular and irregular spherical Bessel functions
defined, for example, in Ref.~\cite{abra}. The corresponding regularizing
function $f_L(y)$ defined in Eq.~(\ref{eq:flreg}) can be obtained by taking the
expression of the coefficients $a_i$, $b_i$ for $\eta\rightarrow 0$ (note that
in this case all  $b_i\rightarrow 0$, see the Appendix). 

For example, the $p+\tri$ asymptotic states are (in our notation, this
corresponds to the clusterization $\gamma=3$)
\begin{eqnarray}
  \Omega_{3LS}^\pm &=& D_{3} \sum_{l=1}^4
  \Bigl [ Y_{L}(\hat{\bm y}_l) \otimes  [ \phi_3^{t}(ijk) \otimes \chi_l \xi_l^{p}]_{S} 
   \Bigr ]_{JJ_z} \nonumber\\
  &&\times \left ( {\frac{\widetilde G_{L}(\eta_3,q_3y_l)}{q_3 y_l}}
          \pm {\rm i} {\frac{F_L(\eta_3,q_3y_l)}{q_3 y_l}} \right ) \ ,
  \label{eq:psiom2}
\end{eqnarray}
where $y_l$ is the distance between the proton (particle $l$) and $\tri$
(particles $ijk$), $q_3$ is defined via Eqs.~(\ref{eq:energy})
and~(\ref{eq:tcm}), and
\begin{equation}
  \eta_3={\mu_3 e^2\over q_3}\ , \qquad
  \mu_3\approx {3\over 4}M_N \ . \label{eq:etamu}
\end{equation}
Moreover, $\phi_3^{t}$ is the
$\tri$ wave function (with the  $z$-component of isospin $T_z=-{1\over2}$), $\chi_l$
($\xi_l^p$) the spin (isospin) state of the
free proton. Note that we do not couple the isospin states. Therefore
$\Omega_{3LS}^\pm$ are superpositions of states with total isospin
$T=0$, $1$. The antisymmetry operator ${\cal A}$ in
this case reduces simply to the sum over the four possible $1+3$
partitions of the particles, assuming $\phi_3^t(ijk)$ to be completely
antisymmetric with respect to the exchange of particles $i$, $j$, and $k$.

In this paper, we consider only $1+3$ clusterizations, and 
the normalization factors $D_{\gamma}$ can be conveniently chosen to be 
\begin{equation}
   D_{\gamma}= \sqrt{1\over 4} \sqrt{2\mu_\gamma q_\gamma\over(\kappa_{\gamma})^3}\ , \qquad
   \kappa_{\gamma}=\sqrt{3\over2}\ , \qquad \gamma=1,\ldots,4\ . \label{eq:D31}
\end{equation}
The parameter $\kappa_{\gamma}$ is the coefficient of proportionality
between the Jacobi vector $\jacb_1$ and the distance between the two
clusters $\y$, namely $\jacb_1=\kappa_\gamma \y$, see next subsection.
Finally, the general expression of $\Psi_A^{\gamma LS}$ entering
Eq.~(\ref{eq:psica}) is
\begin{equation}
  \Psi_A^{\gamma LS}= \sum_{\gamma' L^\prime S^\prime}
 \bigg[\delta_{\gamma,\gamma'}\delta_{L L^\prime} \delta_{S S^\prime}
   \Omega_{\gamma' L'S'}^-
  - {\cal S}^{\gamma,\gamma'}_{LS,L^\prime S^\prime}(E)
     \Omega_{\gamma' L^\prime S^\prime }^+ \bigg] \ ,
  \label{eq:psia}
\end{equation}
where the parameters ${\cal S}^{\gamma,\gamma'}_{LS,L^\prime
  S^\prime}(E)$ are $S$-matrix elements.  Of course, the sum over
$L^\prime$ and 
$S^\prime$ is over  all values compatible with the given $J$ and
parity $\pi$. In particular, the sum over $L^\prime$ 
is limited to include either even or odd values such that
$(-1)^{L^\prime}=\pi$. The sum over $\gamma'$ is over the possible
final clusters compatible with the conservation of the total charge.
Clearly, the parameters ${\cal S}^{\gamma,\gamma'}_{LS,L^\prime
  S^\prime}(E)$ with $\gamma\neq\gamma'$ are related to the cross
section of the reaction $\gamma\rightarrow\gamma'$, while
${\cal S}^{\gamma,\gamma}_{LS,L^\prime  S^\prime}(E)$ to an elastic
scattering process.

\subsection{The hyperspherical harmonic functions}
\label{sec:theory2}
The core wave function $\Psi^{\gamma LS}_C$ has been here
expanded using the HH basis. The superscript $\gamma LS$ means
that $\Psi^{\gamma LS}_C$  is the core part of the wave
function given in Eq.~(\ref{eq:psica}) describing a process where there are 
two incoming clusters specified by $\gamma$ having a relative 
orbital angular momentum $L$ and channel spin $S$. For four equal mass
particles, a suitable choice of the Jacobi vectors is
\begin{eqnarray}
   \jacb_{1p}& = & \sqrt{\frac{3}{2}} 
    \left ({\bm r}_l - \frac{ {\bm r}_i+{\bm r}_j +{\bm r}_k}{3} \right )\ , \nonumber\\
   \jacb_{2p} & = & \sqrt{\frac{4}{3}}
    \left ({\bm r}_k-  \frac{ {\bm r}_i+{\bm r}_j}{2} \right )\ , \label{eq:JcbV}\\
   \jacb_{3p} & =& {\bm r}_j-{\bm r}_i\ , \nonumber
\end{eqnarray}
where $p$ specifies a given permutation corresponding to the order $i$, $j$,
$k$ and $l$ of the particles. By definition, the permutation $p=1$ is chosen
to correspond  to the order $1$, $2$, $3$ and $4$. In terms of
the Jacobi vectors, the kinetic energy $T$ is written as
\begin{equation}
  T=-{1\over M_N}\Bigl( \nabla^2_{\jacb_{1p}}
  +\nabla^2_{\jacb_{2p}}+\nabla^2_{\jacb_{3p}}\biggr)\ .
\end{equation} 
The other possible choice of the Jacobi vectors is
\begin{eqnarray}
   \y_{1p}& = & {\bm r}_l - {\bm r}_k \ , \nonumber\\
   \y_{2p} & = & {1\over \sqrt{2} }
    \left ({\bm r}_l + {\bm r}_k - {\bm r}_i-{\bm r}_j\right )\ , \label{eq:JcbVH}\\
   \y_{3p} & =& {\bm r}_j-{\bm r}_i\ . \nonumber
\end{eqnarray}
In the following, we are going to use only the HH functions constructed
with the Jacobi vectors given in Eq.~(\ref{eq:JcbV}). In fact, the HH
functions are essentially harmonics polynomials and those constructed
with the Jacobi vectors given in Eq.~(\ref{eq:JcbVH})  are just
linear combinations of the HH functions constructed with the
Jacobi vectors of Eq.~(\ref{eq:JcbV}). On the other hand, HH functions constructed
for different choices of the particle permutation $p$ are needed in order
to construct wave functions with the correct permutational symmetry.

For a given choice of the Jacobi vectors, the hyperspherical coordinates are
given by the so-called hyperradius $\rho$, defined by
\begin{equation}
   \rho=\sqrt{\jac_{1p}^2+\jac_{2p}^2+\jac_{3p}^2}\ ,\quad ({\rm independent\
    of\ }p)\ ,
    \label{eq:rho}
\end{equation}
and by a set of angular variables which in the Zernike and
Brinkman~\cite{zerni,F83} representation are (i) the polar angles $\hat
\jacb_{ip}\equiv (\theta_{ip},\phi_{ip})$  of each Jacobi vector, and (ii) the
two additional ``hyperspherical'' angles $\hypfi_{2p}$ and $\hypfi_{3p}$
defined as
\begin{equation}
    \cos\phi_{2p} = \frac{ \jac_{2p} }{\sqrt{\jac_{1p}^2+\jac_{2p}^2}}\ ,
    \quad
    \cos\phi_{3p} = \frac{ \jac_{3p} }{\sqrt{\jac_{1p}^2+\jac_{2p}^2+\jac_{3p}^2}}\ ,
     \label{eq:phi}
\end{equation}
where $\jac_{jp}$ is the modulus of the Jacobi vector $\jacb_{jp}$. The set of angular
variables $\hat \jacb_{1p}, \hat \jacb_{2p}, \hat \jacb_{3p}, \phi_{2p}$, and $\phi_{3p}$ is
denoted  hereafter as $\Omega_p$. The expression of a generic HH
function is
\begin{eqnarray}
 \lefteqn{ {\cal H}^{K,\Lambda, M}_{\ell_1,\ell_2,\ell_3, L_2 ,n_2,
     n_3}(\Omega_p) =\qquad\qquad} &&  \nonumber \\
  && {\cal N}^{\ell_1,\ell_2,\ell_3}_{ n_2, n_3} 
      \left [ \Bigl ( Y_{\ell_1}(\hat \jacb_{1p})
    Y_{\ell_2}(\hat \jacb_{2p}) \Bigr )_{L_2}  Y_{\ell_3}(\hat \jacb_{3p}) \right
    ]_{\Lambda M}  \nonumber \\
  && 
   \times (\sin\phi_{2p})^{\ell_1 }    (\cos\phi_{2p})^{\ell_2}
   (\sin\phi_{3p})^{\ell_1+\ell_2+2n_2}
      (\cos\phi_{3p})^{\ell_3}  \nonumber \\
   && \times
      P^{\ell_1+\frac{1}{2}, \ell_2+\frac{1}{2}}_{n_2}(\cos2\phi_{2p})
      \nonumber\\
   &&\times   P^{\ell_1+\ell_2+2n_2+2, \ell_3+\frac{1}{2}}_{n_3}(\cos2\phi_{3p})\ ,
      \label{eq:hh4P}
\end{eqnarray}
where $P^{a,b}_n$ are Jacobi polynomials and the coefficients ${\cal
N}^{\ell_1,\ell_2,\ell_3}_{ n_2, n_3}$ normalization factors. The quantity 
$K=\ell_1+\ell_2+\ell_3+2(n_2+n_3) $ is the grand angular quantum
number.  The HH functions are the eigenfunctions of the hyperangular part of
the kinetic energy operator. Furthermore, 
$\rho^K   {\cal  H}^{K,\Lambda,M}_{\ell_1,\ell_2,\ell_3, L_2 ,n_2,
n_3}(\Omega_p)$ are homogeneous polynomials of the particle coordinates of
degree $K$.

A set of antisymmetric hyperangular--spin--isospin states of 
grand angular quantum number $K$, total orbital angular momentum $\Lambda$,
total spin $\Sigma$, and total isospin $T$  (for given values of
total angular momentum $J$ and parity $\pi$) can be constructed as follows:
\begin{equation}
  \Psi_{\mu}^{K\Lambda\Sigma T} = \sum_{p=1}^{12}
  \Phi_\mu^{K\Lambda\Sigma T}(i,j,k,l)\ ,
  \label{eq:PSI}
\end{equation}
where the sum is over the $12$ even permutations $p\equiv i,j,k,l$, and
\begin{eqnarray}
 \lefteqn{  \Phi^{K\Lambda\Sigma T}_{\mu}(i,j;k;l)
   =\qquad\qquad} &&  \nonumber \\
  && \biggl \{
   {\cal H}^{K,\Lambda}_{\ell_1,\ell_2,\ell_3, L_2 ,n_2, n_3}(\Omega_p)
      \biggl [\Bigl[\bigl( s_i s_j \bigr)_{S_a}
      s_k\Bigr]_{S_b} s_l  \biggr]_{\Sigma} \biggr \}_{JJ_z}
     \nonumber \\
  && \times \biggl [\Bigl[\bigl( t_i t_j \bigr)_{T_a}
      t_k\Bigr]_{T_b} t_l  \biggr]_{TT_z}\ .
     \label{eq:PHI}
\end{eqnarray}
Here, ${\cal H}^{K,\Lambda}_{\ell_1,\ell_2,\ell_3, L_2 ,n_2, n_3}(\Omega_p)$ is the
HH state defined in Eq.~(\ref{eq:hh4P}), and $s_i$ ($t_i$) denotes the spin 
(isospin) function of particle $i$. The total orbital angular  momentum $\Lambda$ of
the HH function is coupled to the total spin $\Sigma$ to give the total angular
momentum $JJ_z$, whereas $\pi=(-1)^{\ell_1+\ell_2+\ell_3} $. The
quantum number $T$ specifies the total isospin of the state. The
integer index $\mu$ labels the possible choices of hyperangular, spin and
isospin quantum numbers, namely
\begin{equation}
   \mu \equiv \{ \ell_1,\ell_2,\ell_3, L_2 ,n_2, n_3, S_a,S_b, T_a,T_b
   \}\ ,\label{eq:mu}
\end{equation}
compatibles with the given values of $K$, $\Lambda$, $\Sigma$, 
$T$, $J$ and $\pi$. Another
important classification of the states is to group them in ``channels'': states
belonging to the same channel have the same values of angular
$\ell_1,\ell_2,\ell_3, L_2 ,\Lambda$, spin $S_a,S_b,\Sigma$, 
isospin $T_a,T_b,T$ quantum
numbers but different values of $n_2$, $n_3$.

Each state  $\Psi^{K\Lambda\Sigma T}_\mu$ entering 
the expansion of the 4N wave function must 
be antisymmetric under the exchange of any pair of particles. To this aim 
it is sufficient to consider states such that
\begin{equation}
    \Phi^{K\Lambda\Sigma T}_\mu(i,j;k;l)= 
    -\Phi^{K\Lambda\Sigma T}_\mu(j,i;k;l)\ ,
     \label{eq:exij}
\end{equation}
which is fulfilled when the condition
\begin{equation} 
    \ell_3+S_a+T_a = {\rm odd}\ , \label{eq:lsa}
\end{equation}
is satisfied.

The number $M_{K\Lambda\Sigma T}$ of  antisymmetric functions 
$\Psi^{K\Lambda\Sigma T}_\mu$
having given values of $K$, $\Lambda$, $\Sigma$, and $T$ but different
combinations of quantum numbers $\mu$ (see Eq.(\ref{eq:mu})) is in general very
large.  In addition to the degeneracy of the HH basis, the four
spins (isospins) can be coupled in different ways to $\Sigma$ ($T$). However, many
of the states $\Psi^{K\Lambda\Sigma T}_\mu$, 
$\mu=1,\ldots,M_{K\Lambda\Sigma T}$ are linearly
dependent between themselves. In the expansion of $\Psi^{\gamma LS}_C$ it is
necessary to include only the subset of linearly independent states, whose
number  is fortunately noticeably smaller than the corresponding value of
$M_{K\Lambda\Sigma T}$.

The core part of the  wave function can be finally written as
\begin{equation}\label{eq:PSI3}
  \Psi^{\gamma LS}_C= \sum_{K\Lambda\Sigma T}\sum_{\mu} 
    u^{\gamma LS}_{K\Lambda\Sigma T\mu}(\rho)
    \Psi_{\mu}^{K\Lambda\Sigma T}\ ,
\end{equation}
where the sum is restricted only to the linearly independent states. 
We have found convenient to expand the ``hyperradial'' functions
$u^{\gamma  LS}_{K\Lambda\Sigma T\mu}(\rho)$ in a 
complete set of functions, namely
\begin{equation}
     u^{\gamma LS}_{K\Lambda\Sigma T\mu}(\rho)=\sum_{m=0}^{M-1} 
      c^{\gamma LS}_{K\Lambda\Sigma T\mu m} \; g_m(\rho)
      \ ,     \label{eq:fllag}
\end{equation}
and we have chosen 
\begin{equation}
   g_m(\rho)= 
     \sqrt{b^{9}\frac{m!}{(m+8)!}}\,\,\,  
     L^{(8)}_m(b\rho)\,\,{\rm e}^{-\frac{b}{2}\rho} \ ,
      \label{eq:fllag2}
\end{equation}
where $L^{(8)}_l(b\rho)$ are Laguerre polynomials~\cite{abra} and 
$b$ is a parameter to be variationally optimized.

Using the expansion given in Eq.~(\ref{eq:fllag}), finally the
core part can be written as
\begin{equation}
  \Psi^{\gamma LS}_C= \sum_{K\Lambda\Sigma T\mu m} c^{\gamma
    LS}_{K\Lambda\Sigma T\mu m}\; 
  \Psi_{\mu}^{K\Lambda\Sigma T} g_m(\rho)    \ . \label{eq:psic}
\end{equation}
    
\subsection{The Kohn variational principle}
\label{sec:kvp}
The $S$-matrix elements ${\cal S}^{\gamma,\gamma'}_{LS,L^\prime S^\prime}(E)$ of
Eq.~(\ref{eq:psia}) and the coefficients $c^{\gamma LS}_{K\Lambda\Sigma
  T\mu,m}$ occurring in the expansion of $\Psi^{\gamma LS}_C$
are determined using the Kohn variational principle (KVP). Recalling
Eqs.~(\ref{eq:psica}), (\ref{eq:psia}), and~(\ref{eq:psic}), the wave
function can be written in a compact way as  
\begin{equation}
  \Psi^{\gamma LS}\equiv \Psi_\nu = \Omega_\nu^- - \sum_{\nu'} {\cal S}_{\nu,\nu'} \Omega_{\nu'}^+ +
  \sum_k c_{\nu k} \Psi_k\ , \label{eq:kvp1}
\end{equation}
where hereafter we use the notation $\nu\equiv\{\gamma LS\}$,
$k\equiv\{K\Lambda\Sigma T\mu m\}$, and
\begin{equation}
  {\cal S}_{\nu,\nu'}\equiv {\cal S}^{\gamma,\gamma'}_{LS,L^\prime
    S^\prime}  \ ,\quad
  c_{\nu k}\equiv c^{\gamma LS}_{K\Lambda\Sigma T\mu m}
  \ , \quad
  \Psi_k = \Psi_{\mu}^{K\Lambda\Sigma T} g_m(\rho) \ .\label{eq:kvp2}
\end{equation}
In practice, the index $\nu$ specifies the possible asymptotic waves and
the index $k$ runs over all the terms used to expand the core part.
To use the KVP for the $S$-matrix, we need also the related wave function
\begin{equation}
  \widetilde\Psi_\nu = \Omega_\nu^+ - \sum_{\nu'} {\cal S}^*_{\nu,\nu'} \Omega_{\nu'}^- +
  \sum_k c_{\nu k}^* \Psi_k\ , \label{eq:kvp3}
\end{equation}
where the asterisk denotes the complex conjugate.  In particular, we can
define $ \widetilde\Psi_\nu^{J,J_z}= (-)^{L+J+J_z} {\cal T}
\Psi_\nu^{J,-J_z}$, where here we have shown explicitly
the dependence on the total angular momentum and ${\cal T}$ is the time-reversal
operator. Since $H$ commutes with ${\cal T}$, then both
$\Psi_\nu^{J,J_z}$ and $ \widetilde\Psi_\nu^{J,J_z}$ are eigenstates
of $H$ with the same eigenvalue $E$.  The KVP states that
the coefficients ${\cal S}_{\nu,\nu'}$ and $c_{\nu k}$ are determined
by making the functional
\begin{eqnarray}
   [{\cal S}_{\nu,\nu'}]&=&
     { {\cal S}_{\nu,\nu'}+{\cal S}_{\nu',\nu}\over 2}\nonumber\\
     &-& {
     \langle \widetilde\Psi_{\nu'} | H-E | \Psi_\nu\rangle
       + \langle \widetilde\Psi_\nu | H-E | \Psi_{\nu'}\rangle
       \over 4{\rm i}} \ , \label{eq:kohn}
\end{eqnarray}
stationary with respect to variations of them~\cite{Kohn48,Del72,K97}.
The expression above is obtained when the normalization factors
$D_\gamma$ are chosen as in Eq.~(\ref{eq:D31}). After the variations
of the functional, a linear set of equations for ${\cal S}_{\nu,\nu'}$
and $c_{\nu k}$ is obtained. For example, let us consider the
functional for the diagonal case $\nu=\nu'=\nu_0$. Then
\begin{eqnarray}
  [{\cal S}_{\nu_0,\nu_0}] &=& {\cal S}_{\nu_0,\nu_0} -{1\over 2
    {\rm i} }\biggl[A_{\nu_0,\nu_0}^{-,-} - \sum_\nu {\cal S}_{\nu_0,\nu}
    (A_{\nu,\nu_0}^{+,-}+A_{\nu_0,\nu}^{-,+})\nonumber \\
    &+&  \sum_{\nu,\nu'}
    {\cal S}_{\nu_0,\nu} {\cal S}_{\nu_0,\nu'}
    A_{\nu,\nu'}^{+,+}
    \!+\! \sum_k c_{\nu_0,k} 2 B_{k,\nu_0}^-\nonumber \\
    &-&  \sum_{k,\nu} c_{\nu_0,k} {\cal S}_{\nu_0,\nu} 2 B_{k,\nu}^+
    \!+\!\sum_{k,k'} c_{\nu_0,k} c_{\nu_0,k'} C_{k,k'}\biggr]\,\!, \label{eq:kvp4}
\end{eqnarray}
where
\begin{eqnarray}
  A_{\nu,\nu'}^{\lambda,\lambda'}= \langle \Omega_\nu^{-\lambda}|H-E|
  \Omega_{\nu'}^{\lambda'} \rangle \ , \label{eq:kspa}\\
  B_{k,\nu}^{\lambda}= \langle \Psi_k |H-E|
  \Omega_\nu^\lambda\rangle  \ , \label{eq:kspb}\\
  C_{k,k'}           = \langle \Psi_k |H-E| \Psi_{k'}\rangle
  \ , \label{eq:kspc}
\end{eqnarray}
and $\lambda,\lambda'\equiv\pm$. Note the definition of the matrix
elements $ A_{\nu,\nu'}^{\lambda,\lambda'}$ in
Eq.~(\ref{eq:kspa}) which takes into account that
$(\Omega_\nu^\lambda)^\dag = \Omega_\nu^{-\lambda}$.
Moreover, $\langle \Psi_k |H-E|\Omega_\nu^\lambda\rangle
= \langle \Omega_\nu^{-\lambda} |H-E|  \Psi_k\rangle$ since
the wave functions $\Psi_k$ are square integrables. On the other hand,
$A_{\nu,\nu'}^{\lambda,\lambda'}\neq
A_{\nu',\nu}^{\lambda',\lambda}$. With the normalization factors
$D_\gamma$ chosen as in Eq.~(\ref{eq:D31}), it can be proved that
\begin{equation}
  {1\over 2{\rm i}} (A_{\nu,\nu}^{+,-}-A_{\nu,\nu}^{-,+})=1 \ .
  \label{eq:one}
\end{equation}
This relation can be used to test the numerical accuracy of the
calculated matrix elements. 
From the variation of the expression given in Eq.~(\ref{eq:kvp4}), we
can determine the  $S$-matrix elements ${\cal S}_{\nu_0,\nu}$ and
the coefficients $c_{\nu_0,k}$. In the following, we refer to
${\cal S}_{\nu_0,\nu}$ determined in this way as the ``first-order''
$S$-matrix elements. Explicitly, one obtains the following linear
system 
\begin{eqnarray}
\left(\begin{array}{cc} 
         C_{k,k'} & -B_{k,\nu'}^+ \\
         -B_{k',\nu}^+ & {1\over
           2}(A_{\nu,\nu'}^{+,+}+A_{\nu',\nu}^{+,+}) \\
      \end{array}\right)
     \left(\begin{array}{c} 
         c_{\nu_0,k'}\\
         {\cal S}_{\nu_0,\nu'}
     \end{array}\right) &&
     \nonumber\\
   =   \left(\begin{array}{c} 
         -B_{k,\nu_0}^-\\
         {\rm i}\delta_{\nu_0,\nu} +{1\over 2} (A_{\nu,\nu_0}^{+,-}+A_{\nu_0,\nu}^{-,+}) \\
     \end{array}\right)\ . &&  \label{eq:lisy}  
\end{eqnarray}
This linear system is solved using the Lanczos algorithm. A typical calculation
involves the expansion of the core part with $10,000$ HH functions
and $16$ functions $g_m(\rho)$. So the matrix elements $ C_{k,k'}$ form a matrix
of dimension $160,000\times160,000$. This part does not depend on the energy and
can be calculated only once. The possible $\nu$ values are much less. In this work
at maximum we can have $4$ combinations, for the $p+\tri$ and $n+\het$ scattering
in the $J>0$ waves. For example, for this process and the wave
$J^\pi=1^-$, we may have the
combinations $\nu\equiv\{\gamma L S\}=\{3\,1\,0\},\{3\,1\,1\},\{4\,1\,0\},\{4\,1\,1\}$.
Clearly the matrix elements $B_{k,\nu'}^\lambda$ and
$A_{\nu,\nu}^{\lambda,\lambda'}$ depend on the energy and have to be calculated every time
from the beginning. However, their number is much less than that of
the $C$ matrix elements. Moreover, the matrix on the left hand side of
Eq.~(\ref{eq:lisy}) does not depend on $\nu_0$ and
therefore can be inverted only once for all $\nu_0$.

The calculation has to be performed for each values of $J^\pi$
and for all the different types of interaction of interest. Finally, the
procedure has to be repeated separately for the $T_z=-1$ ($n+\tri$
scattering), $T_z=+1$ ($p+\het$ scattering), and  $T_z=0$ ($p+\tri$
and $n+\het$ scattering) cases. 

The KVP also states~\cite{Kohn48,Del72,K97} that the quantities
$[{\cal S}_{\nu,\nu'}]$ are a variational approximation to the exact
$S$-matrix elements ${\cal S}^{\rm exact}_{\nu,\nu'}$. To clarify
better this assertion, let us write $\Psi_\nu=\Psi_\nu^{\rm
  exact}+\epsilon_\nu$, where $\Psi_\nu$ are the wave functions
determined as discussed above, $\Psi_\nu^{\rm exact}$ the exact wave
functions, and $\epsilon_\nu$ the corresponding ``error''
functions. Then, the KVP assures that $|[{\cal S}_{\nu,\nu'}]-{\cal
  S}^{\rm exact}_{\nu,\nu'}|\propto \epsilon^2$. Therefore, the
convergence of the quantities $[{\cal S}_{\nu,\nu'}]$ to the 
exact $S$-matrix elements is quadratic in the error functions and
consequently much faster than the convergence of the first-order
estimates ${\cal S}_{\nu,\nu'}$. Usually, the quantities
$[{\cal S}_{\nu,\nu'}]$ are called the ``second-order'' $S$-matrix
elements. We note also that the quantities $[{\cal S}_{\nu,\nu'}]$
automatically verify the condition $[{\cal S}_{\nu,\nu'}]=[{\cal S}_{\nu',\nu}]$
(principle of detailed balance).
On the other hand, for the $S$-matrix elements calculated solving the linear system
given in Eq.~(\ref{eq:lisy}), this property is not guaranteed. Only
after the inclusion of a sufficient number of terms in the expansion of the
core part in Eq.~(\ref{eq:kvp1}), the symmetry property for
${\cal S}_{\nu,\nu'}$ is approximately verified.

\subsection{Details of the calculation}
\label{sec:theory3}

Let us now consider the problem of the computation of the matrix elements
of the Hamiltonian, and in particular of the NN and 3N interactions.
First, let us consider the ``core-core'' matrix elements, which
explicitly read
\begin{equation}
  C_{k,k'}=\langle \Psi_{\mu}^{K\Lambda\Sigma T} g_m(\rho)| H-E |
  \Psi_{\mu'}^{K'\Lambda'\Sigma' T'} g_{m'}(\rho) \rangle\ ,\label{eq:ccme}
\end{equation}
where $\Psi_{\mu}^{K\Lambda\Sigma T}$ are given in Eqs.~(\ref{eq:PSI})
and~(\ref{eq:PHI}). This calculation is considerably simplified using 
the following property of the functions given in Eq.~(\ref{eq:PHI}):
\begin{equation}\label{eq:arare}
  \Phi^{K\Lambda\Sigma T}_{\mu}(i,j,k,l) =
  \sum_{\mu'}  a^{K\Lambda\Sigma T}_{\mu,\mu'}(p) 
   \Phi^{K\Lambda\Sigma T}_{\mu'}(1,2,3,4)\ ,
\end{equation}
where the coefficients $a^{K\Lambda\Sigma T}_{\mu,\mu'}(p)$ have been obtained
using the techniques described in Ref.~\cite{V98}. In this way,  we
can write
\begin{equation}
   \Psi_{\mu}^{K\Lambda\Sigma T} g_m(\rho) = 
   g_m(\rho) \sum_{\mu'} \tilde a^{K\Lambda\Sigma T}_{\mu,\mu'}
   \Phi^{K\Lambda\Sigma T}_{\mu'}(1,2,3,4)\ ,\label{eq:PSI4}
\end{equation}
where
\begin{equation}
    \tilde a^{K\Lambda\Sigma T}_{\mu,\mu'}=  
   \sum_{p=1}^{12} a^{K\Lambda\Sigma T}_{\mu,\mu'}(p)  \ .
   \label{eq:sumperm}
\end{equation}
The sum over the permutations enters only in the construction
of the coefficients $\tilde a$, and it can be performed beforehand.
With a wave function written in this way, most of the
integrations needed to compute $C_{k,k'}$ can be performed analytically.
The remaining low-dimensional integrations can therefore be easily calculated with 
sufficiently dense grids to obtain relative errors $\le 10^{-6}$.
The adopted procedure is the same as described
in Ref.~\cite{rep08}.  

Second, let us consider the ``core-asymptotic'' and
``asymptotic-asymptotic'' matrix elements,
\begin{eqnarray}
  B_{k,\nu}^\lambda&=&\langle  \Psi_{\mu}^{K\Lambda\Sigma T} g_m(\rho)| H-E
   |  \Omega_{\gamma LS}^\lambda \rangle \ ,\label{eq:came}\\
  A_{\nu,\nu'}^{\lambda,\lambda'} & = & \langle \Omega_{\gamma LS}^{-\lambda}
  |  H-E | \Omega_{\gamma' L' S'}^{\lambda'} \rangle
  \ .\label{eq:aame}
\end{eqnarray}
The computation of $(H-E) \Omega_{\gamma LS}^\lambda$ can be
simplified as follows. We refer specifically to the $p+\tri$
asymptotic state given in Eq.~(\ref{eq:psiom2}). Then
\begin{eqnarray}
 (H-E) \Omega_{3LS}^\pm &=& D_{3} \sum_{l=1}^4
  \biggl(H_3(ijk)+V_{il}+V_{jl}+V_{kl}\nonumber\\
  && +W_{ijl}+W_{ikl}+W_{jkl}\nonumber\\
  && + {e^2_{il}\over r_{il}}
     +{e^2_{jl}\over r_{jl}}+{e^2_{kl}\over r_{kl}} -{\nabla_{\jacb_{1p}}^2\over
    M_N} +B_3-{q_3^2\over 2\mu_3}\biggr) \nonumber\\
  &&\times  \Bigl [ Y_{L}(\hat{\bm y}_l) \otimes  [ \phi_3^{t}(ijk) \otimes \chi_l \xi_l^p]_{S} 
    \Bigr ]_{JJ_z} \nonumber \\
  &&\times \left ( {\frac{\widetilde G_{L}(\eta_3,q_3y_l)}{q_3 y_l}}
          \pm {\rm i} {\frac{F_L(\eta_3,q_3y_l)}{q_3 y_l}} \right)\,\!,
  \label{eq:hpsia1}
\end{eqnarray}
where $H_3(ijk)$ is the Hamiltonian of the three-body subsystems
formed by particles $ijk$, $\phi_3^{t}$ the $\tri$ bound state wave function,
$B_3$ the corresponding binding energy, $V_{il}$ ($W_{ijl}$) the NN (3N) potential acting on the pair 
(triplet) of particles $il$ ($ijl$), and
\begin{equation}
  {e^2_{il}\over r_{il}} \equiv {e^2\over r_{il}}  {1+\tau_z(i)\over
    2}{1+\tau_z(l)\over 2}\ , \label{eq:coulil}
\end{equation}
is the point-Coulomb potential between particles $i$ and $l$ including the
isospin projection over the proton states (eventual additional electromagnetic
interactions are lumped in $V$). Above $\tau_z(i)$ is the isospin Pauli
matrix acting on particle $i$. Using
$H_3(ijk)\phi_3^{t}(ijk)=-B_3\phi_3^{t}(ijk)$ and that
$\jacb_{1p}=\kappa_3\y_l$, see Eqs.~(\ref{eq:D31}) and~(\ref{eq:JcbV}),
one obtains
\begin{eqnarray}
 (H-E) \Omega_{3LS}^\pm  &=&
  \Omega_{3LS}^\pm(T)+\Omega_{3LS}^\pm(V)\ , \label{eq:hpsia2}\\
  \Omega_{3LS}^\pm(T) &=& -{D_3 \over 2\mu_3}  \sum_{l=1}^4
  \Bigl [ Y_{L}(\hat{\bm y}_l) \otimes[ \phi_3^{t}(ijk)\! \otimes\! \chi_l \xi_l^p]_{S} 
   \Bigr ]_{JJ_z} \nonumber\\
    & & \times \overline G_{L}(\eta_3,q_3y_l) \,, \label{eq:hpsiak}\\
 \Omega_{3LS}^\pm(V) &=& D_3 \sum_{l=1}^4
 \biggl(V_{il}+V_{jl}+V_{kl} \nonumber\\
 && +W_{ijl}+W_{ikl}+W_{jkl}\nonumber\\
 && + {e^2_{il}\over r_{il}}
      +{e^2_{jl}\over r_{jl}}+{e^2_{kl}\over r_{kl}}-{e^2\over y_l}\biggr) \nonumber\\
  &&\times  \Bigl [ Y_{L}(\hat{\bm y}_l) \otimes  [ \phi_3^{t}(ijk) \otimes \chi_l \xi_l^p]_{S} 
   \Bigr ]_{JJ_z} \nonumber\\
  &&\times\!\! \left (\!\! {\frac{\widetilde G_{L}(\eta_3,q_3y_l)}{q_3 y_l}}
          \!\pm\! {\rm i} {\frac{F_L(\eta_3,q_3y_l)}{q_3 y_l}}\!\! \right )\,.\label{eq:hpsiav}
\end{eqnarray}
We have divided the expression of $ (H-E) \Omega_{3LS}^\pm$ in a kinetic energy part plus a
potential energy part. Note in the kinetic energy part the
appearance of the function $\overline G_{L}$ defined in
Eq.~(\ref{eq:ovrl}). Moreover,  $2\mu_3=\kappa_3^2 M_N$, see Eq.~(\ref{eq:D31}), and
$2\eta_3 q_3= e^2 \kappa_3^2 M_N$, see Eq.~(\ref{eq:etamu}). For the 
potential part, since $\phi_3^{t}$ is anti-symmetric with
respect to the exchange of the particles $i$, $j$, and $k$, 
in the matrix elements defined in  Eqs.~(\ref{eq:came}) and~(\ref{eq:aame}) one
can also take $V_{il}+V_{jl}+V_{kl}\rightarrow 3 V_{il}$, etc.

We note that the functions $\Omega_{3LS}^\pm(T)$ and
$\Omega_{3LS}^\pm(V)$ now vanish asymptotically at least as $1/(y_l)^2$.  In fact,
due to the presence of the bound state wave function
$\phi_3^{t}(ijk)$, the particles $i$, $j$, and
$k$ must be close. Then, we need only to discuss what happens for $y_l\rightarrow\infty$.
In this limit, the function
$\overline G_{L}$ goes to zero exponentially as discussed in
Subsect.~\ref{sec:theory1}. For $\Omega_{3LS}^\pm(V)$, when $y_l\rightarrow\infty$, all
distances $r_{il}$, $r_{jl}$, and $r_{kl}$ go to $\infty$ and therefore
all the NN and 3N potential terms $V$ and $W$ rapidly
vanish. Regarding the Coulomb term, it can be re-written as
\begin{eqnarray}  
  \left[
    {e^2_{il}\over r_{il}} +{e^2_{jl}\over r_{jl}}+{e^2_{kl}\over r_{kl}}
    -{e^2\over y_l}\right]\phi_3^{t}(ijk) &&\nonumber \\
  = \biggl[e^2_{il}\Bigl({1\over r_{il}}-{1\over y_l}\Bigr)
    +e^2_{jl}\Bigl({1\over r_{jl}}-{1\over y_l}\Bigr)
    &&\nonumber\\
    +e^2_{kl}\Bigl({1\over r_{kl}}-{1\over y_l}\Bigr)\biggr]
    \phi_3^{t}(ijk) 
      \ , &&
\end{eqnarray}
since in the $\tri$ wave function $\phi_3^{t}(ijk)$
only one of the particle is a proton and always
$(e^2_{il}+e^2_{jl}+e^2_{kl})\phi_3^{t}=e^2\phi_3^{t}$.
Therefore, for $y_l\rightarrow\infty$ we have
${1\over r_{il}}-{1\over y_l}\sim O(1/y_l^2) \rightarrow 0$, etc.
A similar analysis can be performed for all the asymptotic states
$\Omega_{\gamma LS}^\pm$ with the other values of $\gamma$ ($\gamma=1$, $2$, and $4$).
Clearly, when the particle $l$ is a neutron, the Coulomb term is missing. 
Therefore, also for the matrix elements $A_{\nu,\nu'}^{\lambda,\lambda'}$
given in Eq.~(\ref{eq:aame}), the integrands are always short-ranged and
their calculation does not present any singular behavior asymptotically. 

As a final remark, we note that the relation
$H_3(ijk) \phi_3^{t}(ijk)=-B_3\phi_3^{t}(ijk)$ is not exactly verified in our
calculation, as we construct variationally $\phi_3^{t}$ in terms of an
expansion over the three-body HH functions. However, as discussed in
Ref.~\cite{Del72}, this inaccuracy  contributes at the end to increase
the error function $\epsilon_\nu$ and the full procedure maintains its
validity (for example, the quantities $[{\cal S}_{\nu,\nu'}]$ are still
a variational approximation of the exact ones). 
As discussed later, we control this potential source of inaccuracy
by increasing the number of terms included in the expansion of $\phi_3^{t}$.
We can anticipate that the error related to this approximation
is well under control. 

Let us now resume the discussion of the matrix elements
given in Eqs.~(\ref{eq:came}) and~(\ref{eq:aame}). 
Their calculations is simplified  by ``projecting'' the states
$\Omega_{\gamma LS}^\pm$ and also $\Omega_{\gamma LS}^\pm(T)$ over a
complete set of angular-spin-isospin states, constructed in terms of
the Jacobi vectors $\jacb_i$ corresponding to the 
particle order $1,2,3,4$. For example:
\begin{equation}
   \Omega_{\gamma LS}^\pm = \sum_\alpha F_\alpha^{\gamma LS\pm}(\jac_1,\jac_2,\jac_3)
   {\cal Y}_\alpha(\hat\jacb_1,\hat\jacb_2,\hat\jacb_3)
  \ ,\label{eq:proj}
\end{equation}
where
\begin{eqnarray}
\lefteqn{   {\cal Y}_\alpha(\hat\jacb_1,\hat\jacb_2,\hat\jacb_3) 
             =\qquad\qquad} \nonumber\\
  &=&  \biggl\{ \biggl[\Bigl(Y_{\ell_3}(\hat\jacb_3)(s_1 s_2)_{S_2}\Bigr)_{j_3}
   \Bigl(Y_{\ell_2}(\hat\jacb_2)s_3 \Bigr)_{j_2}\biggr]_{J_2}
  \nonumber \\
  &&   \Bigl(Y_{\ell_1}(\hat\jacb_1)s_4\Bigr)_{j_1}\biggr\}_{JJ_z}
       \Bigl[ \bigl[(t_1 t_2)_{T_2} t_3\bigr]_{T_3} t_4\Bigr]_{TT_z},
       \label{eq:proj2}
\end{eqnarray}
and $\alpha=\{\ell_1,\ell_2,\ell_3,j_1,j_2,j_3,J_2,S_2,
T_2,T_3,T\}$. Note that due to the antisymmetry of the wave function, we must
have $\ell_3+S_2+T_2=$ odd. This ``partial wave expansion'' is performed 
including all states $\alpha$ such that $\ell_i\le \ell_{\rm max}$. The
functions $F_\alpha^{\gamma LS\pm}$ can be obtained very accurately by direct
integration 
\begin{equation}
   F_\alpha^{\gamma LS\pm}(\jac_1,\jac_2,\jac_3) = 
  \int d\hat\jacb_1 d\hat\jacb_2 d\hat\jacb_3\; \Bigl[{\cal
    Y}_\alpha(\hat\jacb_1,\hat\jacb_2,\hat\jacb_3)\Bigr]^\dag 
  \Omega_{\gamma LS}^\pm\ .\label{eq:proj3}
\end{equation}
This six dimensional integrals can be reduced to a three dimensional integral by
performing the analytical integration over three Euler angles. Then, we are
left with the integration over the ``internal'' angles, or in other words
over the variables
$\mu_{12}=\hat\jacb_1\cdot\hat\jacb_2$, $\mu_{13}=\hat\jacb_1\cdot\hat\jacb_3$, and
$\mu_{23}=\hat\jacb_2\cdot\hat\jacb_3$. This integration is performed
using a Gauss-Legendre quadrature technique over $n_\mu$ points (see
Subsect.~\ref{sec:theory6}). 

Finally, using the transformation given in Eq.~(\ref{eq:arare}) and
the partial wave expansion given above, all these terms, $\Psi_k$, $\Omega_{\gamma LS}^\pm $,  and
$\Omega_{\gamma LS}^\pm(T) $, can be rewritten as 
\begin{equation}
  \Psi^X = \sum_\alpha {\cal
    F}^X_\alpha(\jac_1,\jac_2,\jac_3)  
     {\cal Y}_\alpha(\hat\jacb_1,\hat\jacb_2,\hat\jacb_3) 
    \ ,\label{eq:proj4}
\end{equation}
where $\Psi^X$ stands for $\Psi_k$, $\Omega_{\gamma LS}^\pm $, or $\Omega_{\gamma LS}^\pm(T)$.
Above, ${\cal F}$ is either a combinations of Jacobi
polynomials of the hyperangles and functions $g_m(\rho)$, see
Eq.~(\ref{eq:PSI4}) for $\Psi_k$, or corresponds to a function $F^{\gamma
  LS\pm}_\alpha$ for the asymptotic parts.

Then, the matrix elements of a two-body potential can be evaluated as
explained in the following. 
Permuting the particles in either the ``bra'' and in the ``ket'', and
using the antisymmetry properties of $\Psi_k$, $\Omega_{\gamma LS}^\pm $,
$\phi_3^{t}(ijk)$, etc., it is always possible to reduce these
matrix elements to
\begin{equation}
   \langle \Psi^{X}| V_{12} | \Psi^{X'}\rangle 
         \ . \label{eq:me}
\end{equation}
These integrals are easily calculated using the decomposition given in Eq.~(\ref{eq:proj4}). 
Here we have developed two different procedures depending if the potential is
local or non-local. 

\subsubsection{Local potentials}
In this case Eq.~(\ref{eq:me}) is given explicitly by
\begin{eqnarray}
 \lefteqn{\langle \Psi^{X}| V_{12} |
 \Psi^{X'}\rangle =\qquad\qquad} \nonumber\\
 &&= \int d^3\jacb_1 d^3\jacb_2 d^3\jacb_3\;
     \Bigl(\Psi^{X}(\jacb_1,\jacb_2,\jacb_3) \Bigr)^\dag
    \nonumber\\
  && \qquad\qquad \times  V(\jacb_3)
      \Psi^{X'}(\jacb_1,\jacb_2,\jacb_3)
      \ . \label{eq:me1}
\end{eqnarray}
The calculation of the above integral is performed in two steps. First, the
spin-isospin-angular matrix elements 
\begin{eqnarray}
 \lefteqn{ \int d\hat\jacb_1 d\hat\jacb_2 d\hat\jacb_3 \; 
     {\cal Y}_{\alpha}(\hat\jacb_1,\hat\jacb_2,\hat\jacb_3)^\dag \;
     \qquad\qquad}&&  \nonumber \\ 
  && \qquad\qquad\qquad     
      \times V(\jacb_3)\; 
      {\cal Y}_{\alpha'}(\hat\jacb_1,\hat\jacb_2,\hat\jacb_3) 
     \nonumber\\
  &&  \quad =v^{j_3,T_3,T,T_3',T'}_{\ell_3,S_2,\ell_3',S_2'}(\jac_3)
   \delta_{j_3,j_3'} \delta_{j_2,j_2'} \delta_{j_1,j_1'}
   \delta_{\ell_2,\ell_2'} \delta_{\ell_1,\ell_1'}\ ,
\end{eqnarray}
are computed analytically, and, second, the integration
over the moduli of the Jacobi vectors,
\begin{eqnarray}
\lefteqn{  \int_0^\infty d\jac_1 d\jac_2 d\jac_3 \; \jac_1^2\jac_2^2\jac_3^2
   \Bigl({\cal F}^{X}_\alpha(\jac_1,\jac_2,\jac_3)\Bigr)^*
  \qquad\qquad}\nonumber \\ 
  && \qquad\qquad \times
   v^{j_3,T_3,T,T_3',T'}_{\ell_3,S_2,\ell_3',S_2'}(\jac_3)\;
 {\cal F}^{X'}_{\alpha'}(\jac_1,\jac_2,\jac_3)\ ,
\end{eqnarray}
is obtained  in the following way:
\begin{eqnarray}
 \lefteqn{ \int_0^\infty d\jac_1 d\jac_2 d\jac_3  \; \jac_1^2\jac_2^2\jac_3^2
   =\qquad\qquad} &&  \nonumber\\
   && \int_0^\infty \rho^8 d\rho \int_0^{\pi\over 2}   d\hypfi_3  (\cos\hypfi_3)^2
   (\sin\hypfi_3)^5 \nonumber\\
   && \times \int_0^{\pi\over 2}   d\hypfi_2
   (\cos\hypfi_2)^2 (\sin\hypfi_2)^2  
   \ ,\label{eq:int1}
\end{eqnarray}
where the hyperspherical angles $\hypfi_2$ and $\hypfi_3$ are defined in
Eq.~(\ref{eq:phi}). The integration over $\rho$ is performed on a ``scaled''
grid, using the new variable $0\le t\le 1$
\begin{equation}
   \rho\equiv \rho(t)=h {\alpha_s^{n_\rho t}-1\over \alpha_s -1}\ .\label{eq:scaled}
\end{equation}
The parameters $h$, $\alpha_s$, and $n_\rho$ are chosen  to optimize the
integration. For example, most of the calculation performed in the present
work have been performed with the choice $h=0.04$ fm, $\alpha_s=1.05$, and
$n_\rho=96$. Note that $\rho(t=1)\approx 90$ fm in this case.

The integration over $\hypfi_2$ is performed using the variable
$x=\cos2\hypfi_2$ so that 
\begin{equation}
  \int_0^{\pi\over 2}
   d\hypfi_2  (\cos\hypfi_2)^2 (\sin\hypfi_2)^2
   = {1\over 8} \int_{-1}^{+1} dx \; \sqrt{1-x^2} 
   \ , \label{eq:int2}
\end{equation}
and the integration over $x$ is then performed using $n_x$ Gauss-Chebyshev
points~\cite{abra}.  Finally, the integration over $\hypfi_3$ is performed 
in a similar way, namely using the variable $z=\cos2\hypfi_3$, 
\begin{equation}
  \int_0^{\pi\over 2}
   d\hypfi_3  (\cos\hypfi_3)^2 (\sin\hypfi_3)^5
   = {1\over 16\sqrt{2}} \int_{-1}^{+1} dz \; \sqrt{1+z} (1-z)^2 \ , \label{eq:int3}
\end{equation}
using $n_z$ Gauss-Legendre points related to the zeros of the $P_{2
  n_z+1}$ Legendre polynomial~\cite{abra}. 

In summary, for local-potentials the accuracy of the matrix elements, and 
consequently of the phase-shifts, depends on the following parameters:
\begin{enumerate}
\item $\ell_{\rm max}$, the  maximum value of the orbital angular momentum used
to truncate the expansion of Eq.~(\ref{eq:proj}). Values $\ell_{\rm max}=5$ 
or $6$ have been found appropriate to obtain a sufficient numerical accuracy.
\item The numbers $n_x$, $n_z$, and  $n_\mu$ (the latter is used to perform the
projection given in Eq.~(\ref{eq:proj3})) of Gauss-Chebyshev and
Gauss-Legendre  points used to perform the integrations of
Eqs.~(\ref{eq:proj3}),~(\ref{eq:int2}), and~(\ref{eq:int3}).
Typical used values are $n_z=50$, $n_x=30$, and $n_\mu=16$. 
\item The values of the parameters $h$, $\alpha_s$, and $n_\rho$ used to perform
  the integration over the hyperradius.
\item  The number $N_3$ of three-body HH functions used to construct the trinucleon
bound state wave function $\phi_3(ijk)$ entering the asymptotic
functions $\Omega_{\gamma LS}^\pm$, see Eq.~(\ref{eq:psiom}). 
\item The number $M$ of Laguerre polynomials used to expand the hyperradial
functions $u^{\gamma LS}_{K\Lambda\Sigma T\mu}(\rho)$, as given in
Eqs.~(\ref{eq:fllag}) and~(\ref{eq:fllag2}). This expansion depends also on
the parameter $b$, and therefore one has also to check the dependence of the results on
this (non-linear) parameter. 
\end{enumerate}
In Subsec.~\ref{sec:theory6}, we report a study of the
dependence of the calculated phase shifts on these parameters.

\subsubsection{Non-local NN potentials}
In this case Eq.~(\ref{eq:me}) is calculated in a slightly different way.
Now we have
\begin{eqnarray}
 \lefteqn{\langle \Psi^{X}| V_{12} |
 \Psi^{X'}\rangle =\qquad\qquad} \nonumber\\
 &&= \int d^3\jacb_1 d^3\jacb_2 d^3\jacb_3 d^3\jacb_3'\;
     \Bigl(\Psi^{X}(\jacb_1,\jacb_2,\jacb_3) \Bigr)^\dag
    \nonumber\\
  && \qquad\qquad \times  V(\jacb_3,\jacb_3')
      \Psi^{X'}(\jacb_1,\jacb_2,\jacb_3')
      \ . \label{eq:me2}
\end{eqnarray}
The calculation of the above integral is performed in two steps. First, the
spin-isospin-angular matrix elements 
\begin{eqnarray}
 \lefteqn{ \int d\hat\jacb_1 d\hat\jacb_2 d\hat\jacb_3 d\hat\jacb_3'\; 
     {\cal Y}_{\alpha}(\hat\jacb_1,\hat\jacb_2,\hat\jacb_3)^\dag \;
     \qquad\qquad}&&  \nonumber \\ 
  && \qquad\qquad\qquad     
      \times V(\jacb_3,\jacb_3')\; 
      {\cal Y}_{\alpha'}(\hat\jacb_1,\hat\jacb_2,\hat\jacb_3') 
     \nonumber\\
  &=&  v^{j_3,T_3,T,T_3',T'}_{\ell_3,S_2,\ell_3',S_2'}(\jac_3,\jac_3')
   \delta_{j_3,j_3'} \delta_{j_2,j_2'} \delta_{j_1,j_1'}
   \delta_{\ell_2,\ell_2'} \delta_{\ell_1,\ell_1'}\ ,
\end{eqnarray}
are computed analytically, and, second, the integration
over the moduli of the Jacobi vectors,
\begin{eqnarray}
\lefteqn{  \int_0^\infty d\jac_1 d\jac_2 d\jac_3 d\jac_3'\; \jac_1^2\jac_2^2\jac_3^2
  \jac_3^{\prime 2} \Bigl({\cal F}^{X}_\alpha(\jac_1,\jac_2,\jac_3)\Bigr)^*
  \qquad\qquad}\nonumber \\ 
  && \qquad \times
   v^{j_3,T_3,T,T_3',T'}_{\ell_3,S_2,\ell_3',S_2'}(\jac_3,\jac_3')\;
 {\cal F}^{X'}_{\alpha'}(\jac_1,\jac_2,\jac_3')\ ,
\end{eqnarray}
is obtained by using Gauss quadrature methods, in the following way:
\begin{eqnarray}
 \lefteqn{ \int_0^\infty d\jac_1 d\jac_2 d\jac_3 d\jac_3' \; \jac_1^2\jac_2^2\jac_3^2
  (\jac_3')^2
  \qquad\qquad}\nonumber \\ 
 && =\int_0^\infty d\rho_2  d\jac_3 d\jac_3' \; 
   (\rho_2)^5 \jac_3^2   (\jac_3')^2 \nonumber\\
   && \quad \int_0^{\pi\over 2}
   d\phi_2  (\cos\phi_2)^2 (\sin\phi_2)^2  \ ,\label{eq:int4}
\end{eqnarray}
where $\jac_2=\rho_2\cos\phi_2$ and $\jac_1=\rho_2\sin\phi_2$. The integration 
over $\phi_2$ is performed as specified in Eq.~(\ref{eq:int2}). Moreover, 
\begin{eqnarray}
 \lefteqn{  \int_0^{\infty} d\rho_2  (\rho_2)^5 F(\rho_2) =\qquad\qquad} \nonumber\\
 &=& \int_0^{\infty} d\rho_2  (\rho_2)^5 e^{-a_y \rho_2}  e^{a_y \rho_2}
  F(\rho_2) \nonumber \\
 &=& {1\over (a_y)^6} \int_0^{\infty} dy \Bigl(y^5 e^{-y}\Bigr)  e^{y} F(y/a_y)
   \ , \label{eq:int5}
\end{eqnarray}
where $y=a_y \rho_2$, and $a_y$ is a parameter. The integration over $y$
is performed using $n_y$ Gauss points $y_i$ generated from the weight function $y^5
e^{-y}$. The parameter $a_y$ is then chosen in order to achieve accurate
integrals with as small as possible values of $n_y$. Finally, the integration
of $\jac_3$ (and $\jac_3'$) is performed in a similar way, namely
\begin{eqnarray}
 \lefteqn{  \int_0^{\infty} d\jac_3\;  (\jac_3)^2 F(\jac_3)
   =\qquad\qquad} \nonumber\\
 &=&  \int_0^{\infty} d\jac_3\;  (\jac_3)^2 e^{-a_z \jac_3}  e^{a_z
    \jac_3} F(\jac_3)\nonumber \\
   &=&
 {1\over (a_z)^3} \int_0^{\infty} dz\; \Bigl(z^2 e^{-z}\Bigr)  e^{z} F(z/a_z)
   \ , \label{eq:int6}
\end{eqnarray}
where $z=a_z \jac_3$, $a_z$ being a free parameter. The integration over $z$
is performed using $n_z$ Gauss points $z_i$ generated from the weight
function
$z^2 e^{-z}$. The parameter $a_z$ is then chosen in order to achieve accurate
integrals with as small as possible values of $n_z$.  

In this case, the accuracy of the matrix elements, and consequently also of
the calculated phase-shifts, depends on the following parameters:
\begin{enumerate}
\item $\ell_{\rm max}$, the  maximum value of the orbital angular momentum used
to truncate the expansion of Eq.~(\ref{eq:proj}). Values $\ell_{\rm max}=5$ 
or $6$ have been found appropriate to obtain a sufficient numerical
accuracy also in this case.
\item The values of the number of points used to perform the integrations,
namely $n_z,n_y,n_x$, and $n_\mu$ (as before, the latter is used to perform the
projection given in Eq.~(\ref{eq:proj3})). Typical used values are $n_z=30$,
$n_y=50$, $n_x=20$, and $n_\mu=16$. The precision of the integrals depends
also on the parameters $a_y$ and $a_z$. However, once values large enough
of $n_y$ and $n_z$ are used, the dependence on these two parameters is
negligible, and therefore in this work we consider $a_y=a_z=7$ fm${}^{-1}$
without commenting anymore on their impact on the calculation.
\item  The number $N_3$ of three-body HH functions used to construct the trinucleon
bound state wave function $\phi_3(ijk)$.
\item The number $M$ of Laguerre polynomials used to expand the hyperradial
functions $u^{\gamma LS}_{K\Lambda\Sigma T\mu}(\rho)$, as given in
Eqs.~(\ref{eq:fllag}) and~(\ref{eq:fllag2}). This expansion depends also on
$b$, and therefore one has also to check the dependence of the results on
this (non-linear) parameter. 
\end{enumerate}
In Subsec.~\ref{sec:theory6}, we'll report a study of the
dependence of the calculated phase shifts on these parameters.

\subsubsection{Matrix elements of the 3N potential}
\label{sec:detail3n}

The matrix elements of a three-body potential $W_{ijk}$ can be evaluated in a similar
way. In this work, we have taken into account only local 3N
potentials. Consider the matrix element of $W_{123}$. This
operator is completely symmetric under the exchange of particles
$1,2,3$ and depends only on the Jacobi vectors $\jacb_3$ and
$\jacb_2$. By decomposing the wave function as in
Eq.~(\ref{eq:proj4}), we have explicitly
\begin{eqnarray}
 \lefteqn{\langle \Psi^{X}| W_{123} |
 \Psi^{X'}\rangle =\qquad\qquad} \nonumber\\
 &&= \int d^3\jacb_1 d^3\jacb_2 d^3\jacb_3 \;
     \Bigl(\Psi^{X}(\jacb_1,\jacb_2,\jacb_3) \Bigr)^\dag
    \nonumber\\
  && \qquad\qquad \times  W(\jacb_2,\jacb_3)
      \Psi^{X'}(\jacb_1,\jacb_2,\jacb_3)
      \ , \label{eq:me3N1}
\end{eqnarray}  
where the dependence of $W(\jacb_2,\jacb_3)$ on
spin-isospin operators is understood. 
The calculation of the above integral is performed in two steps. First, the
spin-isospin-angular matrix elements 
\begin{eqnarray}
 \lefteqn{ \int d\hat\jacb_1 d\hat\jacb_2 d\hat\jacb_3 \; 
     {\cal Y}_{\alpha}(\hat\jacb_1,\hat\jacb_2,\hat\jacb_3)^\dag \;
     \qquad\qquad}&&  \nonumber \\ 
  && \qquad\qquad\qquad     
      \times W(\jacb_2,\jacb_3)\; 
      {\cal Y}_{\alpha'}(\hat\jacb_1,\hat\jacb_2,\hat\jacb_3) 
     \nonumber\\
  &&  \quad =w^{J_2,T_3,T,T_3',T'}_{\ell_3S_2j_3\ell_2j_2,
       \ell_3'S_2'j_3'\ell_2'j_2'}(\jac_2,\jac_3)
        \delta_{j_1,j_1'} \delta_{\ell_1,\ell_1'}
        \delta_{J_2,J_2'} \ ,\label{eq:me3N2}
\end{eqnarray}
are computed mostly analytically (we are left with an one-dimensional integration
with respect to $\hat \jacb_2\cdot\hat \jacb_3$, which can be readily obtained). 
We have prepared a code which for given values of $\jac_2,\jac_3$ computes
efficiently those matrix elements for all forms of  3N potentials considered
so far, namely Tucson-Melbourne, Brazil, Urbana, Illinois, and chiral N2LO.
The calculation is completed by  the integration
over the moduli of the Jacobi vectors,
\begin{eqnarray}
\lefteqn{  \int_0^\infty d\jac_1 d\jac_2 d\jac_3 \; \jac_1^2\jac_2^2\jac_3^2
   \Bigl({\cal F}^{X}_\alpha(\jac_1,\jac_2,\jac_3)\Bigr)^*
  \qquad\qquad}\nonumber \\ 
  && \times
   w^{J_2,T_3,T,T_3',T'}_{\ell_3S_2j_3\ell_2j_2,
       \ell_3'S_2'j_3'\ell_2'j_2'}(\jac_2,\jac_3)
     {\cal F}^{X'}_{\alpha'}(\jac_1,\jac_2,\jac_3)\ ,\label{eq:me3N3}
\end{eqnarray}
again obtained by using Gauss quadrature method, as discussed
previously. In order to speed up the calculation, we have imposed the following
truncation to the 3N matrix elements:
\begin{eqnarray}
 \lefteqn{   w^{J_2,T_3,T,T_3',T'}_{\ell_3S_2j_3\ell_2j_2,
       \ell_3'S_2'j_3'\ell_2'j_2'}(\jac_2,\jac_3)=0
  \qquad\qquad}\nonumber \\ 
  && \qquad \ {\rm for}\ 
   \ell>\ell_{\rm max}^{3N}\ ,\ J_2>J_{\rm max}^{3N}\ , \ K>K_{\rm
     max}^{3N}\ , 
   \label{eq:me3N4}
\end{eqnarray}
where $\ell$ can be any of $\ell_3$, $\ell_2$, $\ell_3'$, and
$\ell_2'$. In our calculation, $w$ has been taken to vanish when acting on HH
functions (either on the right or on the left) having a grand angular
quantum number $K>K_{\rm max}^{3N}$. This truncation can be
justified since the 3N potentials under consideration are rather smooth
at short interparticle distances, and the contribution of components
of large $\ell$, $J_2$, and $K$ has been found very small. This has
been verified numerically increasing the values of $\ell_{\rm  max}^{3N}$, $J_{\rm
  max}^{3N}$, and $K_{\rm max}^{3N}$ until the calculated phase-shifts
were found rather insensitive to further changes. Examples of the
dependence of the results on these parameters will be discussed in 
Subsection~\ref{sec:theory7}.

\subsubsection{Parameterization of the $S$-matrix}
\label{sec:smatrix}
The scattering observables can be obtained directly from the
$S$-matrix elements. In the following, we present also the results for a selected set of 
$S$-matrix elements in order to check the convergence and compare
with the results of the PSA of Ref.~\cite{Dan10} for $p+\het$ scattering. 
Always we calculate the $S$-matrix elements (and the observables) via Eq.~(\ref{eq:kohn}),
namely using the ``second-order'' estimates given by the quantities
$[{\cal S}^{\gamma,\gamma}_{LS,L'S'}]$ (we simply call 
them ${\cal S}^{\gamma,\gamma}_{LS,L'S'}$ from now on). 
Moreover, these $S$-matrix elements are
parameterized as follows.

For $n+\tri$, $p+\het$, and $p+\tri$ scattering below the $n+\het$ threshold
($T_{r}\lesssim 0.73$ MeV), the number of open asymptotic clusterizations is one. Then, 
for $J=0$ there is only one $LS$ combination in the sum  over $L'S'$ of
Eq.~(\ref{eq:psia}), namely $L'=0$, $S'=0$ ($L'=1$, $S'=1$) for the even (odd)
parity state.  Consequently, for these cases the $S$-matrix  reduces to one parameter 
which is parameterized as usual as ${\cal S}^{\gamma,\gamma'}_{LS,LS}= \eta_{J\pi} \exp(2{\rm i}
\delta_{J\pi})$. For $J>0$, there are always two possible $LS$ combinations, and
correspondingly the $S$-matrix has been parameterized as~\cite{BB52}
\begin{widetext}
\begin{equation}
   {\cal S}^{\gamma,\gamma}_{LS,L'S'}= \left( \begin{array}{cc}
            \cos\epsilon_{J\pi} & -\sin\epsilon_{J\pi} \\
            \sin\epsilon_{J\pi} & \cos\epsilon_{J\pi}  \\
             \end{array}
   \right)
   \left( \begin{array}{cc}
            \eta_1 \exp(2i\delta^1_{J_\pi}) & 0 \\
             0  & \eta_2 \exp(2i\delta^2_{J\pi})  \\
             \end{array}
   \right)
   \left( \begin{array}{cc}
            \cos\epsilon_{J\pi} & \sin\epsilon_{J\pi} \\
            -\sin\epsilon_{J\pi} & \cos\epsilon_{J\pi}  \\
             \end{array}
   \right) \ .\label{eq:smpar}
\end{equation}
\end{widetext}
In this case we define $\eta_{J\pi}=\sqrt{(\eta_1^2+\eta_2^2)/2}$. 
As it is well known, the $S$-matrix should be unitary. However,
in the application of the Kohn principle given in
Eq.~(\ref{eq:kohn}), the value $\eta_{J\pi}=1$ is not imposed: it is achieved only
when the corresponding core part $\Psi_C^{\gamma LS}$  is well described by
the HH basis. We can use the value of $\eta_{J\pi}$ as a test of the convergence of
the HH expansion. In cases of poor convergence, $\eta_{J\pi}$
is found to depend very much also on the choice of 
$f_L(y_i)$, the function used to
regularize the Coulomb function $G_L$. This function depends
on the non-linear parameter $\beta$, and thus another test of the convergence is
performed by analyzing the dependence of $\eta_{J\pi}$ vs. the parameter $\beta$. At
the beginning of the calculation, when the number of HH functions is not enough
to get convergence, $\eta_{J\pi}$ will be extremely dependent on the value of $\beta$ (the
phase shifts depend less critically on $\beta$). By increasing the
number of HH components in the core wave function, we observe that
$\eta_{J\pi}\rightarrow 1$ and the dependence on $\beta$ becomes negligible. Note
that the convergence rate has been found to depend on the value of $\beta$,
and there exist some critical values of this parameter where the convergence
can be very slow. However, it is not difficult to find regions of values of
$\beta$ where the convergence is fast and smooth and the final results are
independent of $\beta$.  Since we are here interested in the study of the
convergence of the HH function, we have chosen $\beta$ in one of the
``favorable'' region, where the convergence is achieved in a smooth and fast
way. A detailed study on this subject is reported in Subsec~\ref{sec:theory5}.

For $p+\tri$ above the $n+\het$ threshold ($T_{r}\gtrsim 0.73$ MeV)
and $n+\het$ scattering, there are two open asymptotic clusterizations
$\gamma=3,4$. Now
the dimension of the $S$-matrix is doubled with respect to the cases discussed above.
Then, it is more convenient to presents the results directly in terms
of the matrix elements parameterized as
\begin{equation}
   {\cal S}^{\gamma,\gamma'}_{LS,L^\prime S^\prime}=
    \eta^{\gamma,\gamma'}_{LS,L^\prime S^\prime}\ 
    \exp\Bigl[2i\delta^{\gamma,\gamma'}_{LS,L^\prime S^\prime}\Bigr]\ .
\label{eq:etadelta}
\end{equation}
The parameters $\eta^{\gamma,\gamma'}_{LS,L^\prime S^\prime}$ are
always $\le 1$.

\subsection{Choice of the basis}
\label{sec:theory4}

The main difficulty of the application of the HH technique is the slow
convergence of  the basis with respect to the grand angular quantum number
$K$. This problem has been overcome by dividing the HH basis in \textit{classes},
depending on the value of $\mathcal{L} =
\ell_1 + \ell_2 + \ell_3$, total spin $\Sigma$, and $n_2$, $n_3$.   
The calculation is started  by including in the expansion of
the wave function the HH states of the first class (class ``C1'') having grand angular
quantum number $K\le K_1$ and studying the  convergence of a 
quantity of interest (for example, the phase-shifts) increasing the value of
$K_1$. Once a satisfactory value of $K_1 = K_{1\rm max}$ is reached, the states
of the second class (class ``C2'') with $K\le K_2$ are added in the expansion, keeping all the
states of the class C1 with $K_1 \le  K_{1\rm max}$. Then $K_2$ is increased until
the desired convergence is achieved and so on. 

Note that in the case of
$p+\het$ or $n+\tri$ scattering, the $z$-component of the total isospin
is $|T_z|=1$, and therefore, only channels with total isospin $T=1$ or $2$ have
to be included in the expansion. The contribution of the $T=2$ channels is
expected to be quite tiny, and in this paper  they have been disregarded.
On the other hand, for $p+\tri$ and $n+\het$ scattering, the
$z$-component of the total isospin is $T_z=0$, and therefore we have
to include in the HH expansion channels with total isospin $T=0$, $1$, and
$2$. However, also in this case we have disregarded the contribution of the
$T=2$ channels.

Let us now discuss the choice of the classes of HH states for the
various $J^\pi$ cases (in the following, we will use also the spectroscopic
notation). For example for $J^\pi=1^+$, both $^3S_1$ and $^3D_1$ 
components can be constructed by including a rather small number of
channels, since in this case the Pauli principle does not allow for
the overlaps between the 4Ns. As a consequence, the core
part is rather small and does not require a large number of channels
to be well described. The same happens for $L\ge 2$ waves
($J^\pi=2^+$, $3^\pm$, $4^\pm$ and so on), where the centrifugal 
barrier prevents the two clusters to come close to each other.

On the other hand, it is well known that there is a strong attraction
in P-waves~\cite{tilley92}. In fact, various R-matrix analyses have
shown the presence of resonances for the $J^\pi=0^-$, $1^-$, and $2^-$
waves. As a consequence, the convergence of the HH expansion in these cases
is much more problematic and, correspondingly, for these cases we have
organized differently the HH expansion as explained below.

Regarding the $J^\pi=0^+$ state, we have to distinguish between $T=0$
and $T=1$ states. For the $T=1$ states, the only needed
for the study of $n+\tri$ and $p+\het$ processes, the Pauli principle
prevents the overlaps between identical nucleons and
consequently  the core part does not require
a large number of channels to be well described. On the other hand,
in the $J^\pi=0^+$ $T=0$ wave, needed for the study of $p+\tri$ and
$n+\het$ processes, the potential is strongly attractive
and the construction of the wave function turns out to be more difficult.
In fact, in this wave there is the formation of the $\alpha$-particle with
binding energy of $28.3$ MeV. Moreover, just below the threshold of
$n+\het$ scattering, there is the first excited state of the $\alpha$
particle and therefore the $S$-matrix in vicinity of this resonance
will vary very fast with the energy.  As a 
consequence the convergence of the HH expansion for this state
will require a large number of channels.

Let us now define in detail the choice of the classes in the various
cases. For the less critical cases ($J=1^+$, $2^+$, $3^\pm$, $4^\pm$,
and so on), we have organized the HH expansion simply grouping the HH functions
in classes depending on the value of $\mathcal{L} = \ell_1 + \ell_2 +
\ell_3$. For example, for $J^\pi=1^+$, the first class includes all HH
functions with $\mathcal{L}=0$, the second one all HH functions with
$\mathcal{L}=2$, etc. So class C1 is composed by 3 $T=0$ channels and
4 $T=1$ channels with $\mathcal{L}=0$, class C2 by 51 $T=0$ channels
and 76 $T=1$ channels with $\mathcal{L}=2$, class C3 by 159 $T=0$
channels and 239 $T=1$ channels with $\mathcal{L}=4$, and so on. 
Clearly in the study of $n+\tri$ and $p+\het$ scattering, we need only
to include the channels with $T=1$.  As will be shown
below, the third class gives already a tiny contribution to the
$S$-matrix.  A similar procedure has been used for the other ``easy'' waves
$J^\pi=2^+$, $3^\pm$, etc. and also for the $J^\pi=0^+$ $T=1$ wave
in case of the study of $n+\tri$ and $p+\het$ scattering. 
Clearly, for negative parity states, the class C1 includes all HH
functions with $\mathcal{L}=1$, class C2 all HH
functions with $\mathcal{L}=3$, and so on. For these cases, in general,
the convergence is achieved, also with a strong repulsive potential
like the AV18, with fairly small values of grand angular quantum
number $K$ ($K\lesssim 30$). 

Regarding the expansion of the $J^\pi=0^+$ state for $p+\tri$ and
$n+\het$ scattering, we have already discussed how the construction of
the $T=0$ component of the wave function is more critical. In this
case, we need to include in the expansion HH functions with $K$ up to $60$ or
more. We have followed the same sub-division adopted for
the study of the ground state of the $\alpha$-particle~\cite{Viv05}.
First of all, we have seen that a very slow convergence is observed
for the particular sets of HH functions 
which incorporate ``two-body'' correlations. It is therefore appropriate to group
these HH functions in the first class and treat them with a particular
attention. In practice, in the first class we include the HH states with
$n_2=0$  belonging to the channels listed in Table~\ref{table:theory:chan0p0}.
Note that the corresponding radial part of the HH functions depends
only on $\cos\phi_{3p}=r_{ij}/\rho$ and thus these states take into account
two-body correlations (see Eq.~(\ref{eq:hh4P})). This is the part of the
wave function more difficult to construct due to the strong repulsions
between the particles at short distances. In the second class, we have 
included the HH functions belonging to the channels listed in
Table~\ref{table:theory:chan0p0} but with $n_2>0$. These HH functions
depend on $\cos\phi_{2p}$, which is proportional to the distance of particle $k$
from the center of mass of the pair $ij$. Therefore, these states
start to take into account three-body correlations. For the other classes
we have followed the procedure to group them depending on the values of
${\cal L}$ and $T$. In practice, class C3 includes the (remaining) channels with
$T=0$ and ${\cal L}=2$ and class C4 includes all $T=1$ channels with
${\cal L}\le 2$. Then class C5 (C6) includes both $T=0$ and $T=1$ channels
with  ${\cal L}=4$ (${\cal L}=6$), and so on. 

\begin{table}
\begin{ruledtabular}
    \caption[Table]{ Quantum numbers of the first channels considered in the
    expansion of the wave function of the $0^+$ state. See the text for details.}

    \begin{tabular}{r@{$\quad$}|@{$\quad$}r@{$\quad$}
    r@{$\quad$}r@{$\quad$}r@{$\quad$}r@{$\quad$}
    r@{$\quad$}r@{$\quad$}r@{$\quad$}r@{$\quad$}r@{$\quad$}r}
        $\alpha$ & $\ell_1$ & $\ell_2$ & $\ell_3$ & $L_2$ & $\Lambda$ &
        $S_a$ & $S_b$ & $\Sigma$ & $T_a$ & $T_b$ & $T$ \\ \hline
  1  &0 &0 &0 &0 &0  &1 &1/2 &0  &0 &1/2 &0 \\
  2  &0 &0 &0 &0 &0  &0 &1/2 &0  &1 &1/2 &0 \\
  3  &0 &0 &2 &0 &2  &1 &3/2 &2  &0 &1/2 &0 \\
    \end{tabular}
    \label{table:theory:chan0p0}
\end{ruledtabular}
\end{table}

Let us consider now the $J^\pi=0^-$, $1^-$, and $2^-$ waves. Note that, since
the waves under consideration are of negative 
parity, only HH functions with odd values of ${\cal 
L}=\ell_1+\ell_2+\ell_3$ (and $K$) have to be considered. Also in these cases
it is necessary to consider first the states that describe
``two-body'' correlations and group them in the first class.
The second class will contain HH functions describing
three-body correlations, and then we start to group them depending on the
values of ${\cal L}$. 
However, for these states, we have observed a quite different rate of convergence
with respect to the inclusion of HH functions belonging to channels with a
given total spin $\Sigma$. In particular, the channels with $\Sigma=1$ give a very
important contribution to the structure of the scattering state for these
values of $J^\pi$. On the contrary the channels with $\Sigma=0$ and $\Sigma=2$ are less
important. The final choice of the classes for the cases
$J^\pi=0^-$, $1^-$, and $2^-$ is detailed below.

\begin{table}
\begin{ruledtabular}
    \caption[Table]{ Quantum numbers of the first channels considered in the
    expansion of the wave function of the $0^-$ state. See the text for details.}

    \begin{tabular}{r@{$\quad$}|@{$\quad$}r@{$\quad$}
    r@{$\quad$}r@{$\quad$}r@{$\quad$}r@{$\quad$}
    r@{$\quad$}r@{$\quad$}r@{$\quad$}r@{$\quad$}r@{$\quad$}r}
        $\alpha$ & $\ell_1$ & $\ell_2$ & $\ell_3$ & $L_2$ & $\Lambda$ &
        $S_a$ & $S_b$ & $\Sigma$ & $T_a$ & $T_b$ & $T$ \\ \hline
  1 & 1 &0 &0 &1 &1  &1 &1/2 &1  &0 &1/2 &0 \\
  2 & 1 &0 &0 &1 &1  &1 &3/2 &1  &0 &1/2 &0 \\
  3 & 1 &0 &0 &1 &1  &0 &1/2 &1  &1 &1/2 &0 \\
  4 & 1 &0 &2 &1 &1  &1 &1/2 &1  &0 &1/2 &0 \\
  5 & 1 &0 &2 &1 &1  &1 &3/2 &1  &0 &1/2 &0 \\
  6 & 1 &0 &2 &1 &1  &0 &1/2 &1  &1 &1/2 &0 \\
\hline
         1 & 1 & 0 & 0 & 1 & 1 & 1 & 1/2 & 1 & 0 & 1/2 & 1\\
         2 & 1 & 0 & 0 & 1 & 1 & 1 & 3/2 & 1 & 0 & 1/2 & 1\\
         3 & 1 & 0 & 0 & 1 & 1 & 0 & 1/2 & 1 & 1 & 1/2 & 1\\
         4 & 1 & 0 & 0 & 1 & 1 & 0 & 1/2 & 1 & 1 & 3/2 & 1\\
         5 & 1 & 0 & 2 & 1 & 1 & 1 & 1/2 & 1 & 0 & 1/2 & 1\\
         6 & 1 & 0 & 2 & 1 & 1 & 1 & 3/2 & 1 & 0 & 1/2 & 1\\
         7 & 1 & 0 & 2 & 1 & 1 & 0 & 1/2 & 1 & 1 & 1/2 & 1\\
         8 & 1 & 0 & 2 & 1 & 1 & 0 & 1/2 & 1 & 1 & 3/2 & 1\\
    \end{tabular}
    \label{table:theory:chan0m1}
\end{ruledtabular}
\end{table}

\begin{table}
\begin{ruledtabular}
    \caption[Table]{ Quantum numbers of the first channels considered in the
    expansion of the wave function of the $1^-$ state. See the text for details.}

    \begin{tabular}{r@{$\quad$}|@{$\quad$}r@{$\quad$}
    r@{$\quad$}r@{$\quad$}r@{$\quad$}r@{$\quad$}
    r@{$\quad$}r@{$\quad$}r@{$\quad$}r@{$\quad$}r@{$\quad$}r}
        $\alpha$ & $\ell_1$ & $\ell_2$ & $\ell_3$ & $L_2$ & $\Lambda$ &
        $S_a$ & $S_b$ & $\Sigma$ & $T_a$ & $T_b$ & $T$ \\ \hline
  1  &1 &0 &0 &1 &1 & 1 &1/2 &0 & 0 &1/2 &0 \\
  2  &1 &0 &0 &1 &1 & 0 &1/2 &0 & 1 &1/2 &0 \\
  3  &1 &0 &0 &1 &1 & 1 &1/2 &1 & 0 &1/2 &0 \\
  4  &1 &0 &0 &1 &1 & 1 &3/2 &1 & 0 &1/2 &0 \\
  5  &1 &0 &0 &1 &1 & 0 &1/2 &1 & 1 &1/2 &0 \\
  6  &1 &0 &2 &1 &1 & 1 &1/2 &0 & 0 &1/2 &0 \\
  7  &1 &0 &2 &1 &1 & 0 &1/2 &0 & 1 &1/2 &0 \\
  8  &1 &0 &2 &1 &1 & 1 &1/2 &1 & 0 &1/2 &0 \\
  9  &1 &0 &2 &1 &2 & 1 &1/2 &1 & 0 &1/2 &0 \\
 10  &1 &0 &2 &1 &1 & 1 &3/2 &1 & 0 &1/2 &0 \\
 11  &1 &0 &2 &1 &2 & 1 &3/2 &1 & 0 &1/2 &0 \\
\hline
1  & 1 & 0 & 0 & 1 & 1 &  1 & 1/2 & 0 &  0 & 1/2 & 1 \\
2  & 1 & 0 & 0 & 1 & 1 &  0 & 1/2 & 0 &  1 & 1/2 & 1 \\
3  & 1 & 0 & 0 & 1 & 1 &  0 & 1/2 & 0 &  1 & 3/2 & 1 \\
4  & 1 & 0 & 0 & 1 & 1 &  1 & 1/2 & 1 &  0 & 1/2 & 1 \\
5  & 1 & 0 & 0 & 1 & 1 &  1 & 3/2 & 1 &  0 & 1/2 & 1 \\
6  & 1 & 0 & 0 & 1 & 1 &  0 & 1/2 & 1 &  1 & 1/2 & 1 \\
7  & 1 & 0 & 0 & 1 & 1 &  0 & 1/2 & 1 &  1 & 3/2 & 1 \\
8  & 1 & 0 & 2 & 1 & 1 &  1 & 1/2 & 0 &  0 & 1/2 & 1 \\
9  & 1 & 0 & 2 & 1 & 1 &  0 & 1/2 & 0 &  1 & 1/2 & 1 \\
10 & 1 & 0 & 2 & 1 & 1 &  0 & 1/2 & 0 &  1 & 3/2 & 1 \\
11 & 1 & 0 & 2 & 1 & 1 &  1 & 1/2 & 1 &  0 & 1/2 & 1 \\
12 & 1 & 0 & 2 & 1 & 2 &  1 & 1/2 & 1 &  0 & 1/2 & 1 \\
13 & 1 & 0 & 2 & 1 & 1 &  1 & 3/2 & 1 &  0 & 1/2 & 1 \\
14 & 1 & 0 & 2 & 1 & 2 &  1 & 3/2 & 1 &  0 & 1/2 & 1 \\
    \end{tabular}
    \label{table:theory:chan1m1}
\end{ruledtabular}
\end{table}

\begin{table}
\begin{ruledtabular}
    \caption[Table]{ Quantum numbers of the first channels considered in the
    expansion of the wave function of the $2^-$ state. See the text for details.}

    \begin{tabular}{r@{$\quad$}|@{$\quad$}r@{$\quad$}
    r@{$\quad$}r@{$\quad$}r@{$\quad$}r@{$\quad$}
    r@{$\quad$}r@{$\quad$}r@{$\quad$}r@{$\quad$}r@{$\quad$}r}
        $\alpha$ & $\ell_1$ & $\ell_2$ & $\ell_3$ & $L_2$ & $\Lambda$ &
        $S_a$ & $S_b$ & $\Sigma$ & $T_a$ & $T_b$ & $T$ \\ \hline
  1  &1 &0 &0 &1 &1  &1 &1/2 &1  &0 &1/2 &0 \\
  2  &1 &0 &0 &1 &1  &1 &3/2 &1  &0 &1/2 &0 \\
  3  &1 &0 &0 &1 &1  &0 &1/2 &1  &1 &1/2 &0 \\
  4  &1 &0 &2 &1 &1  &1 &1/2 &1  &0 &1/2 &0 \\
  5  &1 &0 &2 &1 &2  &1 &1/2 &1  &0 &1/2 &0 \\
  6  &1 &0 &2 &1 &3  &1 &1/2 &1  &0 &1/2 &0 \\
  7  &1 &0 &2 &1 &1  &1 &3/2 &1  &0 &1/2 &0 \\
  8  &1 &0 &2 &1 &2  &1 &3/2 &1  &0 &1/2 &0 \\
  9  &1 &0 &2 &1 &3  &1 &3/2 &1  &0 &1/2 &0 \\
 \hline
  1 &  1 & 0 & 0 & 1 & 1 &  1 & 1/2 & 1 &  0 & 1/2 & 1 \\
  2 &  1 & 0 & 0 & 1 & 1 &  1 & 3/2 & 1 &  0 & 1/2 & 1 \\
  3 &  1 & 0 & 0 & 1 & 1 &  0 & 1/2 & 1 &  1 & 1/2 & 1 \\
  4 &  1 & 0 & 0 & 1 & 1 &  0 & 1/2 & 1 &  1 & 3/2 & 1 \\
  5 &  1 & 0 & 2 & 1 & 1 &  1 & 1/2 & 1 &  0 & 1/2 & 1 \\
  6 &  1 & 0 & 2 & 1 & 2 &  1 & 1/2 & 1 &  0 & 1/2 & 1 \\
  7 &  1 & 0 & 2 & 1 & 3 &  1 & 1/2 & 1 &  0 & 1/2 & 1 \\
  8 &  1 & 0 & 2 & 1 & 1 &  1 & 3/2 & 1 &  0 & 1/2 & 1 \\
  9 &  1 & 0 & 2 & 1 & 2 &  1 & 3/2 & 1 &  0 & 1/2 & 1 \\
 10 &  1 & 0 & 2 & 1 & 3 &  1 & 3/2 & 1 &  0 & 1/2 & 1 \\
    \end{tabular}
    \label{table:theory:chan2m1}
\end{ruledtabular}
\end{table}

\begin{enumerate}

   \item {Class C1}. In this class are included the HH states with
   $n_2=0$  belonging to the channels of Tables~\ref{table:theory:chan0m1},
   ~\ref{table:theory:chan1m1}, and~\ref{table:theory:chan2m1} for the waves
   with $J^\pi=0^-$, $1^-$, and $2^-$, respectively, for both $T=0$
   and $T=1$. As discussed above, these states take into
   account two-body correlations. 

  \item {Class C2}. This class includes HH functions belonging {\it (i)} to the same
  channels as for class C1, but with $n_2>0$ and {\it (ii)} to the rest of 
  channels with $\ell_1+\ell_2+\ell_3=1$. The HH functions of type {\it (i)}
  take into account the three--body correlations.

  \item {Class C3}. This class includes the HH functions belonging to the
  remaining channels with $\ell_1+\ell_2+\ell_3=3$ and $\Sigma=1$. 

  \item {Class C4}. This class includes the HH functions  belonging to the
  channels with $\ell_1+\ell_2+\ell_3=3$ and $\Sigma=0$ and $2$. 

  \item {Class C5}. This class includes the HH functions belonging to the
  channels with $\ell_1+\ell_2+\ell_3=5$.

\end{enumerate}

We remark again that the classification related to the total spin is important since we have
observed that the component with $\Sigma=1$ requires more
states to be well accounted for, while the $\Sigma=2$ and $\Sigma=0$
components give only a tiny contribution to the phase shift (however, they are
important for achieving $\eta_{J\pi}=1$).
Some examples of convergence for the phase shifts, mixing angles, and
``elasticity parameter'' $\eta_{J\pi}$ will be given in the next section.

\section{Convergence and numerical stability}
\label{sec:conv}

In this section, an analysis of the convergence and
numerical stability of the results will be discussed.

\subsection{Study of the convergence for $n+\tri$ and $p+\het$ scattering}
\label{sec:theory5}

Let us first concentrate on  $n+\tri$ and $p+\het$ scattering. At
the energies considered here only one asymptotic state is open, and the $S$-matrix can
be conveniently decomposed in terms of a single phase-shift (for the
$J^\pi=0^\pm$ waves), or in terms of two phase-shifts and one mixing
parameter, as discussed in Subsect.~\ref{sec:smatrix}. We recall
that for these processes we need to include only $T=1$
channels in the HH expansion.

Here we have considered  the AV18 and N3LO500 potential models. Both
potentials represent the NN interaction in its full richness, with
short-range repulsion, 
tensor and other non-central components and charge symmetry breaking
terms, and both reproduce the NN scattering data with a $\chi^2/$datum very close to
1. The main difference is that the AV18 interaction is local and has a strong
repulsive part at short interparticle distances, while the N3LO500 potential is
non-local and has a somewhat less repulsive core.
In both cases, the electromagnetic interaction has been limited
to just the point-Coulomb potential. We have used $1/M_N=41.47108$ MeV
fm$^2$. The function $G_L$ has been regularized with method 1
corresponding to two values of $\beta$ as reported in the tables.

Since the convergence is similar for both $n+\tri$ and $p+\het$
phase-shifts, we will concentrate on the charged case, where the presence of
the long-range Coulomb potential can complicate the calculation. Moreover, the
convergence has been found to be slower as the proton energy
$E_p=(4/3) T_{r}$ increases, so the tests have been
performed for $E_p=5.54$ MeV, the larger $p+\het$ energy considered in
this paper. In this subsection, the matrix elements have been computed
using $\ell_{\rm max}=5$, $n_z=30$, $n_y=50$, $n_x=20$, and $n_\mu=16$. 
We have in all cases used $M=16$ and $b=4.0$ fm${}^{-1}$ in the expansion
of the hyperradial functions $u^{\gamma LS}_{K\Lambda\Sigma
  T\mu}(\rho)$.

Let us discuss first the convergence for the ``easy'' cases, namely for
$J^\pi=0^+$, $1^+$, $2^+$, $3^\pm$ etc. As an example, we consider here
only the $J^\pi=0^+$ case. As discussed previously, for $n+\tri$ and $p+\het$ scattering we have
only one possible clusterization and for  $J^\pi=0^+$,
$L,S=0,0$, the $S$-matrix is one-dimensional, parameterized as 
$\eta_{0+} \exp[2{\rm i} \delta({}^1S_0)]$. The results obtained for
the $p+\het$ $\eta_{0+}$ and $\delta({}^1S_0)$ at $E_p=5.54$ MeV (corresponding to
$T_{r}=4.15$  MeV) are reported in Table~\ref{table:conv1}.
As can be seen from the table, the convergence is similar for both potentials, since
in this wave the interaction between $p$ and $\het$ clusters is dominated by
the Pauli repulsion. The differences between the phase-shifts obtained by the
two potentials are related to the different $\het$ binding energy (and
radius). Including an appropriate 3N interaction, the phase-shifts calculated
with the different models becomes quite close to each other.
The inclusion of the first class (channels with $\mathcal{L}=0$),
already produces a very good estimate for the phase shift. The inclusion of
the second class (channels with $\mathcal{L}=2$) decreases the phases
shift by about $0.3$ ($0.7$) deg for the N3LO500 (AV18) case. Finally, the inclusion
of the third class (channels with $\mathcal{L}=4$), produces only tiny 
changes in the result, showing the rapid convergence with respect to ${\cal L}$.

\begin{table}[t]
\caption[Table]{\label{table:conv1}
Convergence of the  $J^\pi 0^+$ $p+\het$  inelasticity parameter $\eta_{0+}$ and 
phase-shift $\delta({}^1S_0)$ (deg) at $E_p=5.54$ MeV  corresponding
to the inclusion in the core part of the wave function
of three different classes in which the HH basis has been
subdivided. See the main text for more details.
The N3LO500 and AV18 potentials are considered here with the inclusion
of the point-Coulomb interaction. The function $G_L$ has been regularized with
method 1, using the two values of $\beta$.
}
\begin{tabular}{c@{$\quad$}c@{$\quad$}c@{$\quad\ $}
                @{$\ $} c@{$\ $}c@{$\ $}  
                @{$\ $} c@{$\ $}c@{$\ $} }
\hline
\hline
 &&& \multicolumn{4}{c}{N3LO500} \\
 &&& \multicolumn{2}{c}{$\beta=0.80$ fm${}^{-1}$}  
  & \multicolumn{2}{c}{$\beta=0.90$ fm${}^{-1}$}  \\
\hline
$K_1$ & $K_2$ & $K_3$ & 
     $\eta_{0+}$ & $\delta({}^1S_0)$ &
     $\eta_{0+}$ & $\delta({}^1S_0)$ \\
\hline
 32 &&& 1.00495 & -68.914 & 1.00027 & -68.909 \\
 36 &&& 1.00501 & -68.911 & 1.00029 & -68.906 \\
 40 &&& 1.00505 & -68.909 & 1.00029 & -68.904 \\
 44 &&& 1.00507 & -68.909 & 1.00030 & -68.903 \\
\hline
 44 &16&& 1.00025 & -68.581 & 1.00055 & -68.568 \\
 44 &20&& 1.00005 & -68.564 & 1.00026 & -68.558 \\
 44 &24&& 1.00001 & -68.557 & 1.00009 & -68.553 \\
 44 &28&& 1.00000 & -68.555 & 1.00002 & -68.549 \\
\hline
 44 &28&10& 1.00000 & -68.547 & 1.00002 & -68.544 \\
 44 &28&12& 1.00000 & -68.545 & 1.00002 & -68.540 \\
 44 &28&14& 1.00000 & -68.544 & 1.00001 & -68.538 \\
\hline
 &&& \multicolumn{4}{c}{AV18} \\
 &&& \multicolumn{2}{c}{$\beta=0.80$ fm${}^{-1}$}  
   & \multicolumn{2}{c}{$\beta=0.90$ fm${}^{-1}$}  \\
\hline 
$K_1$ & $K_2$ & $K_3$ & 
     $\eta_{0+}$ & $\delta({}^1S_0)$ &
     $\eta_{0+}$ & $\delta({}^1S_0)$ \\
\hline 
  32 &&& 1.00229 & -70.041 & 1.00415 & -70.511 \\
  36 &&& 1.00219 & -70.018 & 1.00421 & -70.488 \\
  40 &&& 1.00212 & -70.001 & 1.00429 & -70.472 \\
  44 &&& 1.00207 & -69.989 & 1.00436 & -70.459 \\
\hline
 44 & 16 && 1.00251 & -69.318 & 1.00093 & -69.318 \\
 44 & 20 && 1.00199 & -69.280 & 1.00083 & -69.271 \\
 44 & 24 && 1.00151 & -69.254 & 1.00076 & -69.246 \\
 44 & 28 && 1.00113 & -69.236 & 1.00070 & -69.231 \\
\hline
 44 & 28 & 10 & 1.00114 & -69.224 & 1.00068 & -69.217 \\
 44 & 28 & 12 & 1.00114 & -69.216 & 1.00067 & -69.206 \\
 44 & 28 & 14 & 1.00114 & -69.210 & 1.00065 & -69.198 \\
\hline
\hline
\end{tabular}
\end{table}

\begin{table}[h]
\caption[Table]{\label{table:conv1n}
Convergence of $1^-$ $p+\het$ inelasticity parameter $\eta_{1-}$, 
phase-shifts $\delta({}^1P_1)$ and $\delta({}^3P_1)$ (deg), and mixing angle
$\epsilon_{1^-}$ (deg) at $E_p=5.54$ MeV  corresponding
to the inclusion in the core part of the wave function
of the different classes C1--C5 in which the HH basis has been subdivided.
The N3LO500 is considered here with the inclusion
of the point-Coulomb interaction.The function $G_L$ has been regularized with
method 1, using $\beta=0.70$ fm${}^{-1}$. In the last row, the results
with $\beta=0.80$ fm${}^{-1}$ are also given.}
\begin{tabular}{c@{$ $}c@{$ $}
    c@{$ $}c@{$ $}c@{$\qquad$}
                @{$\ $} c@{$\ $}c@{$\ $}  
                @{$\ $} c@{$\ $}c@{$\ $} }
\hline
\hline
 $K_1$ & $K_2$ & $K_3$ & $K_4$ & $K_5$ & 
     $\eta_{1-}$ & $\delta({}^1P_1)$ & $\delta({}^3P_1)$ & $\epsilon_{1-}$ \\
\hline
 29&&&&&  1.00502 & 20.506 & 37.192 & 11.130  \\
 33&&&&&  1.00502 & 20.516 & 37.243 & 11.102  \\
 37&&&&&  1.00502 & 20.520 & 37.268 & 11.088  \\
 41&&&&&  1.00502 & 20.522 & 37.279 & 11.081  \\
\hline
 41&19&&&&  1.00525 & 21.774 & 42.115 & 10.104  \\
 41&23&&&&  1.00534 & 21.799 & 42.173 & 10.090  \\
 41&27&&&&  1.00541 & 21.814 & 42.201 & 10.083  \\
 41&31&&&&  1.00548 & 21.824 & 42.213 & 10.080  \\
\hline
 41&31&17&&&  1.00541 & 21.886 & 42.848 & 9.824  \\
 41&31&21&&&  1.00541 & 21.888 & 42.906 & 9.800  \\
 41&31&25&&&  1.00541 & 21.889 & 42.936 & 9.787  \\
\hline
 41&31&25&13&&  1.00075 & 22.755 & 43.859 & 9.393  \\
 41&31&25&17&&  1.00035 & 22.843 & 43.917 & 9.393  \\
 41&31&25&21&&  1.00023 & 22.883 & 43.928 & 9.399  \\
 \hline
 41&31&25&21&11&  1.00026 & 22.884 & 43.929 & 9.396  \\ 
 41&31&25&21&13&  1.00026 & 22.898 & 43.939 & 9.392  \\
 41&31&25&21&15&  1.00026 & 22.902 & 43.945 & 9.388  \\
 \hline
 &&&&& \multicolumn{4}{c}{$\beta=0.80$ fm$^{-1}$} \\
 41&31&25&21&15&  1.00006 & 22.887 & 43.951 & 9.378  \\ 
 \hline
 \hline
\end{tabular}
\end{table}

From the table, we can also observe the dependence of $\eta$ and
$\delta$ with respect to the chosen value of $\beta$. As it can be seen,
$\delta$ practically does not depend on $\beta$. On the contrary, the
inelasticity parameter is quite sensitive to $\beta$ when the 
number of HH functions included in the expansion is small. However,
including the second class with HH states up to $K_2\approx 28$, the
dependence on $\beta$ is noticeably reduced and $\eta\rightarrow 1$. Note that
the inclusion of the third class  has only a tiny effect on
$\eta$. For N3LO500, including the three classes, $\eta$ becomes $1$
with nearly five digits. For AV18, although the phase
shift has reached a good convergence, $\eta$ is slightly different
from unity. It appears necessary in this case to include more
states of the first and second class to reach an accuracy similar to
the N3LO500 case. An analogous behavior is observed using method 2 of regularization,
and for the other ``easy'' states $1^+$, $2^+$, $3^\pm$, etc.

Let us now concentrate on the ``difficult'' cases, namely on the waves
$J^\pi=0^-$, $1^-$, and $2^-$. As an example, let us show the results
for the state $J^\pi=1^-$. In this case we can have
$L,S=1,0$ and $1,1$ and the $S$-matrix has been decomposed as in
Eq.~(\ref{eq:smpar}), in terms of the parameters $\delta({}^1P_1)$,
$\delta({}^3P_1)$, $\epsilon_{1-}$ and two 
inelasticity parameters $\eta_{{}^1P_1}$ and $\eta_{{}^3P_1}$. 
The convergence for the various quantities obtained using the
N3LO500 potential is reported in
Table~\ref{table:conv1n}, where, for the sake
of simplicity, we have reported only the combination
$\eta_{1-}=\sqrt{ [(\eta_{ {}^1P_1 } )^2+(\eta_{ {}^3P_1 } )^2]/2 }$.
First of all, we notice the different rate of convergence for the two phase shifts,
$\delta({}^1P_1)$ and $\delta({}^3P_1)$ (this is true also for the
AV18 potential). 
For both potentials, the convergence of the class C1 is rather slow
and a fairly large values of $K_1$ have to be used.
The inclusion of the second and third classes increases
$\delta({}^1P_1)$ by about $1.5$  deg. The increase of  $\delta({}^3P_1)$
is more sizable, almost $6$ deg, and also the effect on $\epsilon_{1-}$ is
noticeable. Including the states with $\Sigma=0$ and $2$, first
appearing when the class C4 is considered, has the same effect on 
both $\delta({}^1P_1)$ and $\delta({}^3P_1)$, 
about $1$ deg, and its contribution is very important to obtain
$\eta_{1-}\rightarrow 1$. The contribution of class C5 (including the
channels with ${\cal L}=5$) is very small, and therefore we expect
that the contribution of the remaining HH states having ${\cal L}>5$
be negligible. 

The convergence of the $J^\pi=0^-$ and $2^-$ waves is similar
to that observed for the $J^\pi=1^-$ wave. In particular, the
convergence of the ${}^3P_0$ (${}^3P_2$) phase-shift follows a
similar pattern as the ${}^1P_1$ (${}^3P_1$) phase-shift. 
The convergence with the AV18 potential of all these parameters is
slower, see below.

\begin{figure} \centering
    \includegraphics[width=\columnwidth]{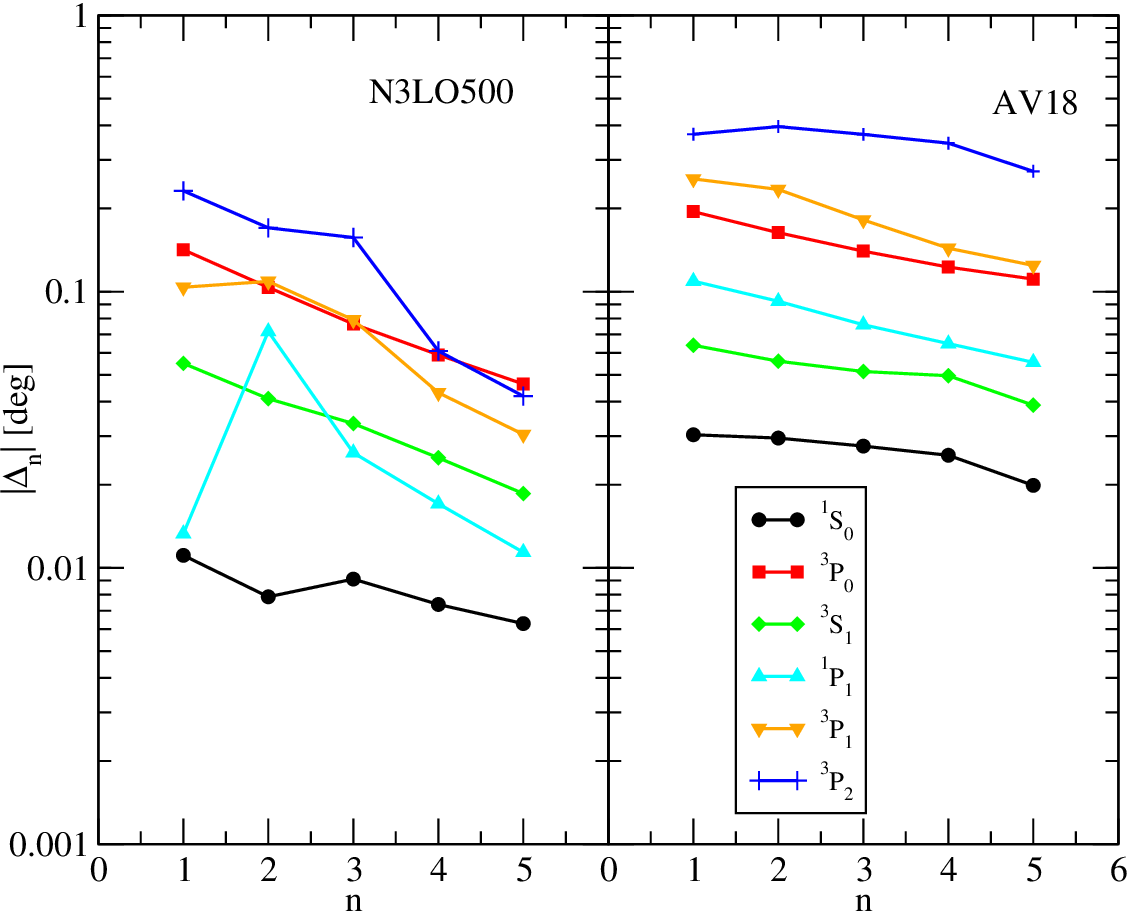}
    
    \caption{(color online) The values of $|\Delta_n(\overline K_1,\overline
             K_2,\ldots)|$ obtained for the N3LO500 (left panel) and
             AV18 (right panel) potentials and the
             ${}^1S_0$, ${}^3S_1$, ${}^3P_0$, ${}^1P_1$, ${}^3P_1$ and
             ${}^3P_2$ phase-shifts. The quantity $n$ is defined in
             Eq.~(\ref{eq:dbet}), i.e. $2n$ is the increase of the
             grand angular quantum number for each class starting from
             a given set $\overline K_1,\overline K_2,\ldots$. See the
             main text for more details.}
    \label{fig:deltan}
\end{figure}

In order to obtain a quantitative estimate of the ``missing''
phase-shift due to the truncation of the HH expansion of the various classes,
let us introduce $\delta(K_1,K_2,\ldots)$ as the phase-shift obtained
by including in the expansion all the HH states of the class C1 with $K\le
K_1$, all the HH states of the class C2 having $K\le K_2$, etc. 
Let us then consider $\overline{K}_1,\overline{K}_2,\ldots$ a given convenient
choice of the grand angular quantum number $K_i$ for each class $i$, and
define
\begin{eqnarray}
  \Delta_n(\overline K_1,\overline K_2,\ldots)&\!=\!& 
    \delta(\overline K_1+2n,\overline K_2+2n,\ldots)\nonumber \\
   &\!-\!& \delta(\overline K_1+2n\!-\!2,\overline K_2+2n\!-\!2,\ldots)
  \ .\label{eq:dbet}
\end{eqnarray}
Namely $\Delta_n$ is the difference of the phase shift computed by increasing
each $K_i$ by $2$ units. The values of $\Delta_n(\overline K_1,\overline
K_2,\ldots)$ obtained for the  N3LO500 and AV18 potentials and for the
${}^1S_0$, ${}^3S_1$, ${}^3P_0$, ${}^1P_1$, ${}^3P_1$, and ${}^3P_2$ phase-shifts
are shown in Fig.~\ref{fig:deltan}.

\begin{table}[tb]
\caption[Table]{\label{table:extra1}
Convergence of ${}^1S_0$ $p+\het$ phase-shift (deg) at $E_p=2.25$ and
$5.54$ MeV corresponding to the inclusion
in the core part of the wave function of the different subsets of HH
basis. The N3LO500 and AV18 potentials are considered here with the inclusion
of the point-Coulomb interaction. The corresponding values of  the missing
phase-shifts, as calculated with Eq.~(\ref{eq:extra}), are given by
the quantity $\Delta_M$. In the rows labeled ``EXT'', the extrapolated phase-shifts
computed as described in text have been reported (in all cases $x\approx0.8$).}
\begin{tabular}{lc@{$\;$}c@{$\;$}c@{$\;$}
                @{$\;$}c@{$\;\;$}c@{$\;\;$}c@{$\;\;$}c@{$\;$}}
\hline
\hline
 & & & & \multicolumn{2}{c}{$E_p=2.25$\ MeV} & \multicolumn{2}{c}{$E_p=5.54$\ MeV}\\
$n$ & $K_1$ & $K_2$ & $K_3$ & N3LO500 & AV18 & N3LO500 & AV18 \\
\hline
0 &34& 18&  4& -41.259 & -41.792 & -68.580 & -69.308  \\
1 &36& 20&  6& -41.255 & -41.767 & -68.572 & -69.278  \\
2 &38& 22&  8& -41.251 & -41.744 & -68.563 & -69.248  \\
3 &40& 24& 10& -41.248 & -41.722 & -68.555 & -69.220  \\
4 &42& 26& 12& -41.245 & -41.703 & -68.549 & -69.195  \\
5 &44& 28& 14& -41.243 & -41.687 & -68.544 & -69.175  \\
\hline
 &\multicolumn{3}{c}{$\Delta_M$} & -0.009 & -0.064 & -0.021  & -0.080 \\
\hline
 &\multicolumn{3}{c}{EXT} & -41.234 & -41.623 & -68.523 & -69.095 \\
 &\multicolumn{3}{c}{$x$} & 0.8  & 0.8  & 0.8 & 0.8  \\
\hline
\hline
\end{tabular}
\end{table}

Some explicit values of calculated $\delta(K_1,K_2,K_3,\ldots)$ are reported in
Tables~\ref{table:extra1} and~\ref{table:extra2}.
In Table~\ref{table:extra1}, we report the $p+\het$ phase-shifts
calculated using the N3LO500 potential at $E_p=5.54$ MeV, and also at lower energy $E_p=2.25$ MeV,
for the ``simple'' state ${}^1S_0$, while the corresponding 
phase-shifts for the ``difficult'' states  ${}^1P_1$ and
${}^3P_1$ are reported in Table~\ref{table:extra2}.
The values of the quantities $K_i$ reported in the rows corresponding
to the value ``$n=0$'' are just the $\overline K_i$ selected in
these cases. By inspecting these tables and Fig.~\ref{fig:deltan},
it is possible to observe in some cases an increase of $|\Delta_n|$.
This is due to the following fact. Since many HH states are linearly dependent and
have to be excluded from the expansion, sometimes states describing
important configurations appear only
for some $K\ge K_{min}$.  When the $K_i$'s are increased
and reach the value $K_{min}$, such configurations start
to be included in the expansion, and the corresponding phase-shift
has an abrupt change. For larger values of $K_i$, all important
configurations are already included and the values of $\Delta_n$ vary
smoothly. In particular, as it can be seen for $n\ge 3$, the
differences $\Delta_n$ start to decrease approximately linearly
in a logarithmic scale. Therefore we can extrapolate the behavior
$\Delta_n\propto x^n$, with $x\le 1$. From this simple behavior, we
can readily estimate the missing phase-shift due to the truncation
of the expansion to finite values of $n$. Suppose to have calculated
$\delta(\overline K_1+2n,\overline K_2+2n,\ldots)$
up to a given $n_{\rm max}$; then
$\Delta_{n_{\rm max}+1}= x \Delta_{n_{\rm max}}$, etc. Then the
missing phase-shift can be estimated as
\begin{eqnarray}
   \Delta_M&=&\sum_{n=n_{\rm max}+1}^\infty \Delta_n = x \Delta_{n_{\rm
       max}}+x^2 \Delta_{n_{\rm max}}+\cdots \nonumber\\
    &=& {x\over 1-x} \Delta_{n_{\rm max}}\ .
   \label{eq:extra}
\end{eqnarray}
Typical values for  $x$ are $\approx 0.8$.  The calculated missing
phase-shifts with Eq.~(\ref{eq:extra}) are reported in
Tables~\ref{table:extra1} and~\ref{table:extra2} in the rows  
labeled ``$\Delta_M$'', while in the rows denoted ``EXT'' we list the
extrapolated phase-shifts computed as
\begin{equation}
  \delta_{{\rm EXT}}=     \delta(\overline K_1+2n_{\rm max},\overline
  K_2+2n_{\rm max},\ldots)
  + \Delta_M\ .\label{eq:deltaext}
\end{equation}

\begin{table}[tb]
\caption[Table]{\label{table:extra2}
The same as in Table~\protect\ref{table:extra1} but for the ${}^1P_1$ and ${}^3P_1$
phase-shifts.}
\begin{tabular}{lc@{$\ $}c@{$\ $}c@{$\ $}c@{$\ $}c@{$\ $}
                @{$\ $}c@{$\ $}c@{$\ $}c@{$\ $}c@{$\ $}}
\hline
\hline
\multicolumn{10}{c}{N3LO500}\\
\hline
 &&&&&& \multicolumn{2}{c}{$\delta({}^1P_1)$\ [deg]} & \multicolumn{2}{c}{$\delta({}^3P_1)$\ [deg]}\\
\hline
$n$ & $K_1$ & $K_2$ & $K_3$ & $K_4$ & $K_5$ & $2.25$ &
$5.54$ & $2.25$ & $5.54$\\
&&&&&& MeV & MeV & MeV & MeV\\
\hline
0& 31& 21& 15& 13&  5& 10.301 &  22.771  & 16.800 & 43.609   \\
1& 33& 23& 17& 15&  7& 10.306 &  22.785  & 16.843 & 43.713   \\
2& 35& 25& 19& 17&  9& 10.334 &  22.857  & 16.891 & 43.823   \\
3& 37& 27& 21& 19& 11& 10.343 &  22.884  & 16.925 & 43.903   \\
4& 39& 29& 23& 21& 13& 10.350 &  22.901  & 16.944 & 43.947   \\
5& 41& 31& 25& 23& 15& 10.354 &  22.913  & 16.958 & 43.979   \\
\hline
&\multicolumn{5}{c}{$\Delta_M$} & 0.016 & 0.048 & 0.055  & 0.124 \\
\hline
&\multicolumn{5}{c}{EXT} & 10.370 & 22.961 & 17.013 &  44.103  \\
&\multicolumn{5}{c}{$x$} & 0.8  & 0.8 & 0.8  & 0.8 \\
\hline
\hline 
\multicolumn{10}{c}{AV18}\\
\hline
& &&&&& \multicolumn{2}{c}{$\delta({}^1P_1)$\ [deg]} & \multicolumn{2}{c}{$\delta({}^3P_1)$\ [deg]}\\
\hline
$n$ & $K_1$ & $K_2$ & $K_3$ & $K_4$ & $K_5$ & $2.25$ &
$5.54$ & $2.25$ & $5.54$\\
&&&&&& MeV & MeV & MeV & MeV\\
\hline
0& 51& 25& 21& 15&  1& 9.965  & 22.070 & 15.721 &  40.871  \\
1& 53& 27& 23& 17&  3& 9.999  & 22.179 & 15.809 &  41.127  \\
2& 55& 29& 25& 19&  5& 10.029 & 22.272 & 15.882 &  41.362  \\
3& 57& 31& 27& 21&  7& 10.055 & 22.347 & 15.945 &  41.544  \\
4& 59& 33& 29& 23&  9& 10.077 & 22.412 & 15.998 &  41.687  \\
5& 61& 35& 31& 25& 11& 10.098 & 22.468 & 16.055 &  41.812  \\
\hline
&\multicolumn{5}{c}{$\Delta_M$} & 0.119 & 0.417 & 0.323 & 0.708   \\
\hline
&\multicolumn{5}{c}{EXT} &  10.206 & 22.785 & 16.378 & 42.520   \\
&\multicolumn{5}{c}{$x$} & 0.85  & 0.85  & 0.85  & 0.85  \\
\hline
\hline
\end{tabular}
\end{table}

As it can be seen from Table~\ref{table:extra1}, the values of $\Delta_M$
are estimated to be rather small in all cases. For the ``difficult''
cases reported in Table~\ref{table:extra2}, the convergence seems to
be under control for N3LO500. On the contrary, for AV18 and
specifically at the largest energy, the values of $\Delta_M$ are
estimated to be sizable. In this case, higher values of $K_1\div K_5$
should be employed. We can see that the missing phase-shift is less
than 2\%. In any case, the extrapolation procedure affects
mainly the third digit of the phase-shifts, and this has no practical
consequences for the $p+\het$ observables.

The extrapolation of other phase-shifts is performed analogously.
Therefore, we can conclude saying that the convergence for the
N3LO500 potential is usually good. In this case, the extrapolation
factor $x\approx0.8$. The same is found with all other interactions
derived within chiral EFT. On the other hand, for the AV18 potential the
convergence is usually a bit slower. For this case, usually an
extrapolation factor $x\approx 0.85$ is found to be more appropriate.  

Finally, we mention that the convergence rate when including any type
of 3N interactions has been found similar to the cases when only the
NN interaction is considered. In fact, in general the 3N interactions
are rather soft at short interparticle distances, and therefore  
the convergence rate of the various classes does not change
appreciably.

\subsection{Convergence for $p+\tri$ and $n+\het$ scattering}
\label{sec:scatt0}

Let us now consider the convergence of the HH expansion for $p+\tri$ and
$n+\het$ scattering. Now, channels with
$T=0$ and $1$ have to be included in the expansion of the core
part. In general, we have observed similar convergence patterns 
as already discussed, except for the $J^\pi=0^+$ state. In fact,
it is well known that $\heq$ has a narrow resonance in the
$J^\pi=0^+$, $T=0$ wave just above the $p+\tri$ threshold. Therefore,
this wave has to be considered a ``difficult'' case, as the
description of the core part requires the inclusion of a large
number of HH states, those necessary to describe the $J^\pi=0^+$
resonance. In fact, for this case, the classes have been
organized in a slightly different method as discussed in
Subsect.~\ref{sec:theory4}.

Here we discuss only the convergence for the $J^\pi=0^+$ wave. For other
$J^\pi$ waves,  the convergence has a similar behavior as discussed
in the previous subsection.   Clearly, for $p+\tri$ scattering below the $n+\het$
threshold ($T_{r}\lesssim 0.73$ MeV) there is only one open asymptotic
clusterization and the $S$-matrix is one-dimensional. Above that
threshold, and for $n+\het$ scattering, there are two open
asymptotic clusterizations. For the $J=0^+$ wave, we have
again $LS=00$. The $S$-matrix is parameterized as in Eq.~(\ref{eq:etadelta})
${\cal S}^{\gamma,\gamma'}_{00,00}= \eta^{\gamma,\gamma'}_{00,00}
\exp\Bigl(2i\delta^{\gamma,\gamma'}_{00,00}\Bigr)$
and we recall that $\gamma=3$ ($4$) corresponds to the $p+\tri$
($n+\het$) clusterization.
The results obtained for $\eta^{3,3}_{00,00}$ and $\delta^{3,3}_{00,00}$
at $E_p=0.60$ MeV (corresponding to $T_{r}=0.45$  MeV) and  $E_p=2.0$
MeV (corresponding to $T_{r}=1.50$  MeV) are reported in
Table~\ref{table:convt0}.

\begin{table}[tb]
\caption[Table]{\label{table:convt0}
Convergence of $0^+$ $p+\tri$ parameters $\eta^{3,3}_{00,00}$ and 
$\delta^{3,3}_{00,00}$ (deg) defined in Eq.~(\ref{eq:etadelta}) at
$E_p=0.60$ and $2.0$ MeV  corresponding to the inclusion in the
core part of the wave function of the different classes C1--C6 in
which the HH basis has been subdivided. The N3LO500 potential is
considered here with the inclusion of the point-Coulomb
interaction. The function $G_L$ has been regularized with 
method 1, using $\beta=0.80$ fm${}^{-1}$.}
\begin{tabular}{c@{$\ $}c@{$\ $}
    c@{$\ $}c@{$\ $}c@{$\ $}c@{$\ $}
                @{$\ $} c@{$\ $}c@{$\ $} 
                @{$\ $} c@{$\ $}c@{$\ $} }
\hline
\hline
 &&&&&& \multicolumn{2}{c}{$E_p=0.60$ MeV}
      & \multicolumn{2}{c}{$E_p=2.0$ MeV}  \\  
\hline
$K_1$ & $K_2$ & $K_3$ & $K_4$ & $K_5$ & $K_6$ & 
     $\eta^{3,3}_{00,00}$ & $\delta^{3,3}_{00,00}$ &
     $\eta^{3,3}_{00,00}$ & $\delta^{3,3}_{00,00}$ \\
\hline
 38 &&&&&&  1.0588 & 2.89 & 0.3549 &  9.20 \\
 42 &&&&&&  1.0595 & 2.94 & 0.3553 &  9.27 \\
 46 &&&&&&  1.0598 & 2.97 & 0.3556 &  9.30 \\
 50 &&&&&&  1.0599 & 2.98 & 0.3556 &  9.32 \\
 \hline
 50 & 40 &&&&& 1.2895 &  14.26 & 0.1209 & 36.86\\
 50 & 42 &&&&& 1.3074 &  14.71 & 0.1204 & 36.73\\
 50 & 44 &&&&& 1.3239 &  15.11 & 0.1199 & 36.61\\
 50 & 46 &&&&& 1.3384 &  15.44 & 0.1194 & 36.51\\
 \hline
 50 & 46 & 32 &&&& 1.7115 &  26.90 & 0.2102 & 77.89\\
 50 & 46 & 34 &&&& 1.7240 &  27.09 & 0.2104 & 77.91\\
 50 & 46 & 36 &&&& 1.7325 &  27.21 & 0.2105 & 77.93\\
 \hline
 50 & 46 & 36 & 40 &&& 1.0117 &  44.84 & 0.2002 & 75.99\\
 50 & 46 & 36 & 42 &&& 1.0060 &  45.61 & 0.2002 & 75.99\\
 50 & 46 & 36 & 44 &&& 1.0025 &  46.30 & 0.2002 & 75.98\\
 \hline
 50 & 46 & 36 & 44 & 18 && 1.0030 & 51.79 & 0.2325 & 80.23 \\
 50 & 46 & 36 & 44 & 20 && 1.0032 & 51.89 & 0.2351 & 80.47 \\
 50 & 46 & 36 & 44 & 22 && 1.0016 & 52.08 & 0.2354 & 80.50 \\
 \hline
 50 & 46 & 36 & 44 & 22 & 14 & 1.0011 & 52.08 & 0.2354 & 80.56\\
 50 & 46 & 36 & 44 & 22 & 16 & 1.0010 & 52.16 & 0.2360 & 80.64\\
\hline 
\hline 
\end{tabular}
\end{table}

From the table, we can observe the large effect of class 2 (the
triplet basis). We observe also that for $E_p=0.60$ MeV, the inclusion
of the 4th class (states with isospin $T=1$) has a large effect, in
particular on the parameter $\eta^{3,3}_{00,00}$. Only after including this
class this parameter starts to approach the value $1$. At $E_p=2.0$ MeV,
the effect of the 4th class is less important. The inclusion of
the 5th class, including both $T=0$ and $1$ HH states with
$\ell_{\rm max}=4$ is still important, while the inclusion of 
HH states with $\ell_{\rm max}=6$ (6th class) is much less sizable.

Also for this case we can apply the extrapolation procedure discussed
previously and the results are reported in 
Table~\ref{table:extraT0}. We note that, also for the N3LO500
potential, the convergence is not well achieved and the values of
$\Delta_M$ are sizable, in particular at $E_p=0.60$ MeV, close to
the energy of the first excited state of $\heq$. Note that at this
energy, the $n+\het$ channel is closed and therefore one should find
$\eta^{3,3}_{00,00}=1$. At $E_p=2.0$ MeV, the convergence is less
problematic and we estimate $(\eta^{3,3}_{00,00})^2\approx 0.06$.
This means that at this energy the elastic process $p+\tri\rightarrow p+\tri$ 
will have a 6\% probability.

For the present case, the uncertainties connected to the
extrapolation formula given in Eq.~(\ref{eq:extra}) are more significant,
especially below the $n+\het$ threshold. Assuming to have an uncertainty
$\Delta x\approx 0.04$ for the factor $x$ (a rather conservative estimate), the
corresponding ``error'' in $\Delta_M$ is given by
${\Delta x\over (1-x)^2} |\Delta_{n_{\rm max}}|\approx |\Delta_{n_{\rm max}}|$
assuming $x\approx 0.8$. For example, for $E_p=0.60$ MeV, we obtain
$\delta^{3,3}_{00,00}(\textrm{EXT})=57.60 \pm 1.35$ deg, approximately a 2\% uncertainty.
This uncertainty will not spoil the comparison with the experimental data,
since the latter quantities are known with larger error bars.

\begin{table}[tb]
\caption[Table]{\label{table:extraT0}
  Convergence of the parameters $\eta^{3,3}_{00,00}$ and
  $\delta^{3,3}_{00,00}$ (deg) at $E_p=0.60$ and $2.00$ MeV
  corresponding to the inclusion in the core part of the wave
  function of the different classes of HH basis. The N3LO500 potential
  is considered here with the inclusion of the point-Coulomb
  interaction. The corresponding values of  the missing 
  phase-shifts, as calculated with Eq.~(\ref{eq:extra}), are given by
  $\Delta_M$. In the rows labeled ``EXT'', the extrapolated phase-shifts
  computed as described in text have been reported (in all cases $x\approx0.8$).}
\begin{tabular}{lc@{$\ $}c@{$\ $}c@{$\ $}c@{$\ $}c@{$\ $}c@{$\ $}
                c@{$\ $}c@{$\ $}c@{$\ $}c@{$\ $}}
\hline
\hline
& &&&&&& \multicolumn{2}{c}{$E_p=0.60$ MeV}
      & \multicolumn{2}{c}{$E_p=2.0$ MeV}  \\  
\hline
$n$ & $K_1$ & $K_2$ & $K_3$ & $K_4$ & $K_5$ & $K_6$ & 
     $\eta^{3,3}_{00,00}$ & $\delta^{3,3}_{00,00}$ &
     $\eta^{3,3}_{00,00}$ & $\delta^{3,3}_{00,00}$ \\
\hline
0& 40&36&26&34&12& 6& 1.0068 & 40.90 & 0.2156 & 78.11 \\ 
1& 42&38&28&36&14& 8& 1.0054 & 44.06 & 0.2219 & 78.93 \\ 
2& 44&40&30&38&16&10& 1.0040 & 46.78 & 0.2269 & 79.56 \\ 
3& 46&42&32&40&18&12& 1.0028 & 49.03 & 0.2309 & 80.03 \\ 
4& 48&44&34&42&20&14& 1.0017 & 50.81 & 0.2337 & 80.37 \\ 
5& 50&46&36&44&22&16& 1.0010 & 52.16 & 0.2360 & 80.64 \\ 
\hline
&\multicolumn{6}{c}{$\Delta_M$} & -0.0028 & 5.44 & 0.0092 & 1.08 \\ 
\hline
&\multicolumn{6}{c}{EXT} & 0.9982 & 57.60 & 0.2452 & 81.72 \\ 
&\multicolumn{6}{c}{$x$} & 0.8  & 0.8  & 0.8 & 0.8 \\ 
\hline
\hline
\end{tabular}
\end{table}

Again, the convergence for other chiral interactions with or without
the inclusion of the 3N forces are similar. Regarding the AV18
potential, the convergence of the $0^+$, $T=0$ phase-shift would
require the inclusion of HH functions with larger $K$ values. We have
not pursued such a calculation any longer in the present study.

\subsection{Numerical stability}
\label{sec:theory6}

In this subsection, we want to discuss the dependence of the results on
the grids used for the calculation of the matrix elements, the number of
three-body HH functions used to construct $\phi_3(i,j,k)$, and on the parameters
$M$ and $b$ entering the expansion of the hyperradial functions (see 
Eq.~(\ref{eq:fllag2})). We limit ourselves only to consider the
calculation of the ${}^1S_0$, ${}^3S_1$, ${}^3P_0$, ${}^1P_1$,
${}^3P_1$, and ${}^3P_2$ $p+\het$ phase shifts at $E_p=5.54$ MeV
with the N3LO500 potential. Similar results were obtained for
all the other cases considered in this paper.

\begin{table*}
  \caption[Table]{\label{table:grid_nl}
$p+\het$ phase-shifts (deg) at $E_p=5.54$ MeV
calculated for different values of the parameters used in the
calculation. The parameter $\ell_{\rm max}$ is the maximum value of
orbital angular momenta used to expand the asymptotic states, see
Eq.~(\protect\ref{eq:proj}). The number of grids points $n_z$, $n_y$,
$n_x$, and $n_\mu$ are used in the numerical integration of the 
potential matrix elements. $N_3$ is the number of three-body HH functions
used to construct the $\het$ wave function. Finally, the parameters
$b$ and $M$ are used in the expansion of the hyperradial functions in
terms of Laguerre polynomials, see Eq.~(\protect\ref{eq:fllag}). The
N3LO500 potential is considered here with the inclusion 
of the point-Coulomb interaction. The $\het$ binding energy obtained
with $N_3=390$ ($N_3=480$) three-body HH functions is $7.12869$
($7.12871$) MeV. See
Section~\protect\ref{sec:theory3} for more details.
The changed parameters with respect to ``case a'' are highlighted in
boldface. 
  }
\begin{tabular}{l@{$\ $}@{$\ $}c@{$\ $}c@{$\ $}c@{$\ $}c
                @{$\ $}c@{$\ $}c@{$\ $}c
                @{$\ $}c@{$\ $}c@{$\ $}c@{$\ $}c@{$\ $}c@{$\ $}c@{$\ $}c@{$\ $}}
\hline
\hline
case &  $\ell_{\rm max}$ & $n_z$ & $n_y$ & $n_x$ & $n_\mu$ & $N_3$ & $b$ & $M$ 
       & $\delta({}^1S_0)$ & $\delta({}^3S_1)$ & $\delta({}^3P_0)$  
       & $\delta({}^1P_1)$  & $\delta({}^3P_1)$  & $\delta({}^3P_2)$\\  
\hline
a & 5 & 30 & 50 & 20 & 16 & 365 & 4.0 & 16 & 
-68.518 & -60.104 & 25.065 & 22.955 &  44.091 & 47.811 \\ 
b & 5 & {\bf 40} & {\bf 60} & {\bf 30} & {\bf 18} & 365 & 4.0 & 16 & 
-68.533 & -60.111 & 25.042 & 22.960 & 44.102 & 47.815\\
c & {\bf 6} & 30 & 50 & 20 & 16 & 365 & 4.0 & 16 & 
-68.524 & -60.105 & 25.057 & 22.991 & 44.191 & 47.927 \\
d & 5 & 30 & 50 & 20 & 16 & {\bf 480} & 4.0 & 16 & 
-68.514 & -60.104 & 25.070 & 22.938 & 44.081 & 47.805 \\
e & 5 & 30 & 50 & 20 & 16 & 365 & {\bf 3.5} & 16 & 
-68.524 & -60.109 & 25.081 & 22.969 & 44.102 & 47.819 \\
f & 5 & 30 & 50 & 20 & 16 & 365 & {\bf 3.5} & {\bf 18} & 
-68.523 & -60.108 & 25.081 & 22.961 & 44.104 & 47.825 \\
\hline
\hline
\end{tabular}
\end{table*}

The values of the ``extrapolated'' (as discussed in the previous
subsection) phase shifts obtained for different values of the parameters
in case of the potential N3LO500 are reported in
Table~\ref{table:grid_nl}.
In the ``case a'' row, we have reported the phase shifts calculated using
the ``standard'' values of  $\ell_{\rm max}$, number of grids points,
number of three-body HH functions, and values of $M$ and
$b$ used so far. Increasing the values of grids points $n_z$, $n_y$,
$n_x$, $n_\mu$ used to compute the matrix elements (case b), the
calculated phase-shifts change only by approximately 0.1\%. The effect of
increasing $\ell_{\rm max}$ (case c) has a slightly larger effect, in
particular for the ${}^3P_1$ and ${}^3P_2$ phase-shifts.
The increasing of the number of three-body HH functions (case d) to describe the
$\het$ bound state produces negligible effects. The same using a different value of
$b$ (case e). Finally, the phase-shifts are rather
insensitive to the increase of the number $M$ of Laguerre
polynomials. Therefore, we can conclude that the calculated phase-shifts are
almost insensitive to the choice of the various parameters.

A similar analysis has been performed also for other potentials, in
particular for AV18 which has a stronger repulsion at short
inter-particle distances. We have found that the
calculated phase-shifts are almost insensitive to changes of the
various parameters also in this case. The greatest sensitivity is found again for the 
parameter $\ell_{\rm max}$. Increasing it by one unit, however, 
causes at most 0.5\% changes in the phase-shifts. 

\subsection{Numerical stability with the inclusion of the 3N potential}
\label{sec:theory7}

In this subsection, the numerical stability of the
results when the 3N potential is included is studied. The method now
involves the calculation of the 3N potential matrix elements
discussed in Subsec.~\ref{sec:detail3n}. Here we report the results of 
the inclusion of the N2LO500 3N interaction together with
the N3LO500 NN potential. 

In Table~\ref{table:grid_nl_3n}, the dependence of the usual
$p+\het$ phase-shifts at $E_p=5.54$ MeV on several
parameters is studied. Some of these quantities also enter the calculation of
the matrix elements of the NN potential, namely $\ell_{\rm max}$,
$n_z$, $n_y$, $n_x$, $n_\mu$, $N_3$, $b$, and $M$. 
In the cases reported below, we have used the same values of the
grid points $n_z$, $n_y$, and $n_x$ to calculate both NN and 3N
matrix elements, given in Eqs.~(\ref{eq:me2})
and~(\ref{eq:me3N3}), respectively. Moreover, adding the 3N force, the
calculations also depend on the values of the parameters $\ell_{\rm  max}^{3N}$,
$J_{\rm  max}^{3N}$, and $K_{\rm max}^{3N}$ used to truncate the
spin-isospin-angular matrix elements $w$ of the 3N force, given in
Eq.~(\ref{eq:me3N4}).

\begin{table*}
\caption[Table]{\label{table:grid_nl_3n}
$p+\het$ phase-shifts (deg) at $E_p=5.54$ MeV
calculated for different values of the parameters
 when a 3N interaction is included.
The parameters $\ell_{\rm max}$, $n_z$, $n_y$, $n_x$, $n_\mu$, $N_3$, 
$b$, and $M$ have the same meanings as in
Table~\protect\ref{table:grid_nl}. The parameters
$\ell_{\rm  max}^{3N}$, $J_{\rm  max}^{3N}$, and $K_{\rm max}^{3N}$
are used in the truncation of the spin-isospin-angular matrix elements
of the 3N force, see Sect.~\protect\ref{sec:detail3n} for more details.
The calculation are performed using the N3LO500/N2LO500 interaction
with the inclusion of the point-Coulomb potential.  
The $\het$ binding energy obtained with $N_3=390$ ($N_3=480$) three-body
HH functions is $7.72988$ ($7.72991$) MeV. The changed parameters with
respect to the ``case a'' are highlighted in boldface.
}
\begin{tabular}{l@{$\ $}@{$\ $}c@{$\ $}c@{$\ $}c@{$\ $}c
                @{$\ $}c@{$\ $}c@{$\ $}c @{$\ $}c @{$\ $}c @{$\ $}c @{$\ $}c
                @{$\ $}c@{$\ $}c@{$\ $}c@{$\ $}c@{$\ $}c@{$\ $}c@{$\ $}}
\hline
\hline
case &  $\ell_{\rm max}$ & $n_z$ & $n_y$ & $n_x$ & $n_\mu$ & $N_3$ &
        $b$ & $M$  & $\ell_{\rm max}^{3N}$ & $J_{\rm max}^{3N}$ & $K_{\rm max}^{3N}$
       & $\delta({}^1S_0)$ & $\delta({}^3S_1)$ & $\delta({}^3P_0)$  
       & $\delta({}^1P_1)$  & $\delta({}^3P_1)$  & $\delta({}^3P_2)$\\  
\hline
a & 5 & 30 & 50 & 20 & 16 & 365 & 4.0 & 16 & 5 & 15/2 & 16 &
-66.554 & -58.523 & 24.188 & 22.579 &  44.990 & 49.732 \\ 
b & 5 & 30 & 50 & 20 & 16 & 365 & 4.0 & 16 & {\bf 6} & {\bf 17/2} & 16 &
-66.554 & -58.524 & 24.188 & 22.580 &  44.993 & 49.734 \\ 
c & 5 & 30 & 50 & 20 & 16 & 365 & 4.0 & 16 & 5 & 15/2 & {\bf 18} &
-66.554 & -58.520 & 24.205 & 22.584 &  45.015 & 49.749 \\ 
d & {\bf 6} & {\bf 40} & {\bf 60} & {\bf 30} & {\bf 18} & {\bf 480} &
{\bf 3.5} & {\bf 18} & 5 & 15/2 & 16 &
-66.533 & -58.518 & 24.209 & 22.613 &  45.101 & 49.869\\
\hline
\hline
\end{tabular}
\end{table*}

As it can be seen from the table, the effect of the truncation of the
spin-isospin-angular matrix elements $w$ of the 3N force (cases b and
c) is rather well under control, since we observe only very tiny differences
between the phase-shifts. The use of denser grids, more
accurate $\het$ wave functions, and a larger number of Laguerre
polynomials (case d) produces at most changes of the order of 0.2\%. Therefore,
we can conclude that the numerical aspect of the inclusion of the 3N
interaction in the calculation of the $p+\het$ phase-shifts is well
under control. A similar degree of accuracy has been reached also for
other 3N interactions. 

\section{Results}
\label{sec:res}

In this section we report the results obtained for various scattering
observables. In the first subsection, a study of $n+\tri$ and $p+\het$
elastic scattering is presented, while the second subsection is
dedicated to the study of the resonant states of $\heq$ as
extracted from the $p+\tri$ phase-shifts. Finally, in the last subsection
we present an analysis of the
$p+\tri\rightarrow p+\tri$, $p+\tri\rightarrow n+\het$, and
$n+\het\rightarrow n+\het$ processes.

As stated before, in this section we report the results 
obtained mainly using the N3LO interaction derived 
by Entem and Machleidt~\cite{EM03,ME11}, corresponding to 
two different cutoff values ($\Lambda=500$ MeV and
$\Lambda=600$ MeV). These NN interactions are labeled,
respectively, N3LO500 and N3LO600.  In this way we can explore the
dependence on the cutoff value $\Lambda$ of the 4N observables. 
The 3N force considered here has been derived at N2LO in
Ref.~\cite{Eea02} (the 3N force at N3LO and N4LO are still under
construction but we plan in future to include them in the 4N calculations).
With the N3LO500 (N3LO600) NN interaction, we have considered the 3N
N2LO force labeled N2LO500 (N2LO600) with the parameters $c_D$ and $c_E$
fixed to reproduce the 3N binding energy and the tritium GTME. 
These values were recently redetermined in Ref.~\cite{Bea18}
after finding and correcting an inconsistency between the 3N force and the
axial current used so far~\cite{Schiavilla}.

In some cases, we have also considered the new potentials developed at
successive order (N4LO) in Ref.~\cite{MEN17} for three different 
cutoff values ($\Lambda=450$, $500$, and $550$ MeV). With such NN
interaction, we have used the same N2LO 3N force.  In this case,
however, the values for the $\pi N$ parameters $c_i$
entering the 3N N2LO force have been chosen as in the last column of
Table~IX of Ref.~\cite{MEN17}, taking into account in an effective
way part of the missing N3LO and N4LO 3N forces (the
two-pion-exchange contribution). In such a way, these N2LO 3N force
may be seen as effective N4LO 3N forces~\cite{MEN17}. The corresponding values for $c_D$
and $c_E$ have been fixed again by reproducing the 3N binding energy and
the tritium GTME~\cite{Mea18}. 

\begin{table*}[t]
\caption{\label{tab:par}
  NN+3N interaction models used in this work. In columns 2$-$4 the values of
  the cutoff parameter $\Lambda$ and the coefficients $c_D$ and $c_E$ entering
  the chiral 3N force are reported (the coefficients are
  adimensional). In the last columns we have reported the
  corresponding $\tri$, $\het$, and $\heq$ binding energies. The experimental
  values of the latter quantities are reported in the last line.}
\begin{center}
\begin{tabular}{lcccccc}
\hline 
\hline 
Model & $\Lambda$ [MeV]  & $c_D$ & $c_E$ & $B(\tri)$ [MeV] &
$B(\het)$ [MeV] & $B(\heq)$ [MeV] \\
\hline
N3LO500/N2LO500  & $500$ & $+0.945$ & $-0.0410$ & $8.471$ & $7.729$ & $28.34$ \\
N3LO600/N2LO600  & $600$ & $+1.145$ & $-0.6095$ & $8.467$ & $7.733$ & $28.59$ \\
\hline
N4LO450/N2LO450  & $450$ & $+0.560$ & $+0.460$ & $8.482$ & $7.714$ & $28.53$ \\
N4LO500/N2LO500  & $500$ & $-0.745$ & $-0.150$ & $8.473$ & $7.728$ & $28.15$ \\
N4LO550/N2LO550  & $550$ & $-1.030$ & $-0.570$ & $8.470$ & $7.731$ & $28.07$ \\
\hline
Expt.  & & & & $8.480$ & $28.30$ \\
\hline
\end{tabular}
\end{center}
\end{table*}

For the sake of clarity, the adopted
values of all employed parameters $c_D$ and $c_E$ are summarized in
Table~\ref{tab:par}, where we have also reported the corresponding
$\tri$, $\het$, and $\heq$ binding energies. As it can be seen, the calculated
$\heq$ binding energies are rather close to the experimental
value. Therefore, eventual 4N forces should be rather tiny and their
effect in $A=4$ scattering at low energy can be safely neglected.

For this study we have focused our attention to the effect of the 3N
interaction. For this reason we have restricted the electromagnetic interaction between
the nucleons to just the point-Coulomb interaction between the
protons. To be noticed that with the N3LO500 and N3LO600 NN
interactions, one should include only the effect of the two-photon
exchange, Darwin-Foldy term, and vacuum  polarization interactions in
the ${}^1S_0$ partial wave~\cite{ME11}. We have
disregarded them in this work. The effect of these additional electromagnetic
interactions is the subject of a forthcoming paper~\cite{newHH}.

\subsection{$p+\het$ and $n+\tri$ scattering}
\label{sec:rest1}

The $p+\het$ and $n+\tri$ observables are calculated at specific
values of the kinetic energy $E_{N}$ of the incident nucleon, related
to $T_{r}$ by  
\begin{equation}
  E_N={4\over 3} T_{r}\ .\label{eq:energy2}
\end{equation}
In the energy range considered here ($E_N\le 6$ MeV), the various
$n+\tri$ and $p+\het$ observables are dominated by $S$-wave and $P$-wave phase shifts
($D$-wave phase shifts give only a marginal contribution, and more
peripheral phase shifts are negligible).

Let us first discuss the results for $n+\tri$ zero--energy scattering. The
relevant quantities are the singlet $a_s$ and triplet $a_t$ scattering
lengths, the zero-energy total cross section $\sigma_T$, and the coherent scattering
length $a_c$, related as follows
\begin{equation} 
   \sigma_T= \pi (|a_s|^2+3|a_t|^2)\ , \quad 
    a_c={1\over4} a_s + {3\over4} a_t\ . \label{eq:def}
\end{equation}
The experimental accessible quantities are $\sigma_T$ and $a_c$. 
The $n+\tri$ cross section has been accurately measured over a wide
energy range and the extrapolation to zero energy does not present any
problems. The value obtained is $\sigma_T=1.70\pm 0.03$
b~\cite{PBS80}. The coherent scattering  length has been measured by
neutron--interferometry techniques. The most recent values reported in
the literature have been obtained by the same group; they are
$a_c=3.82\pm0.07$ fm~\cite{Rauch81} and  $a_c=3.59\pm0.02$
fm~\cite{Rauch85}, the latter value being obtained with a more
advanced experimental arrangement. Finally, the value
$a_c=3.607\pm0.017$ fm has been obtained from $p-\hel$ data  by using
an approximate Coulomb--corrected R--matrix theory~\cite{Hale90}.

\begin{table}[t]
\caption{\label{tab:nt0}
Total cross section $\sigma_T$ (b) and coherent  scattering
length  $a_c$ (fm) for $n+\tri$ zero energy scattering calculated with different
interactions. The last rows report the experimental values.}
\begin{center}
\begin{tabular}{lcc}
\hline
\hline
Interaction & $\sigma_T$ &  $a_c$ \\
\hline
AV18     & 1.85 & 3.83 \\
AV18/UIX & 1.73 & 3.71 \\
\hline
N3LO500 & 1.802 & 3.780 \\
N3LO600 & 1.797 & 3.775 \\
N3LO500/N2LO500 & 1.687 & 3.658 \\
N3LO600/N2LO600 & 1.693 & 3.663 \\
\hline
Expt.  & 1.70$\pm$0.03~\protect\cite{PBS80} & 3.82$\pm$0.07~\protect\cite{Rauch81} \\
       &  & 3.59$\pm$0.02~\protect\cite{Rauch85} \\
       &  & 3.607$\pm$0.017~\protect\cite{Hale90} \\
\hline
\end{tabular}
\label{tab:sigma}
\end{center}
\end{table}

The total cross section and coherent scattering length calculated
with the considered interactions are compared with the experimental values in
Table~\ref{tab:sigma}. It is well known that the
$n+\tri$ singlet and triplet scattering lengths are linearly
correlated with the $\tri$ binding energy $B_3$~\cite{VKR98}. Therefore,
only with the interactions including the 3N force (which well reproduce
$B_3$) the calculated $\sigma_T$ and $a_c$ are close to the
experimental values. From inspection of Table~\ref{tab:sigma}, it can
be concluded that there is a  satisfactory agreement between the
calculated and the measured value of $\sigma_T$. However, the calculated
coherent scattering lengths differ slightly from
the experimental value, in particular from the more
accurate one, $a_c=3.607\pm0.017$. Interestingly, the $a_c$ 
calculated using the $\chi$EFT interactions differ less from the
experimental value than the value calculated with AV18/UIX. It would
be interesting to study this observables with the most recent chiral
interactions of Ref.~\cite{MEN17}. Work in this direction is in progress.

\begin{figure} \centering
    \includegraphics[width=\columnwidth,clip]{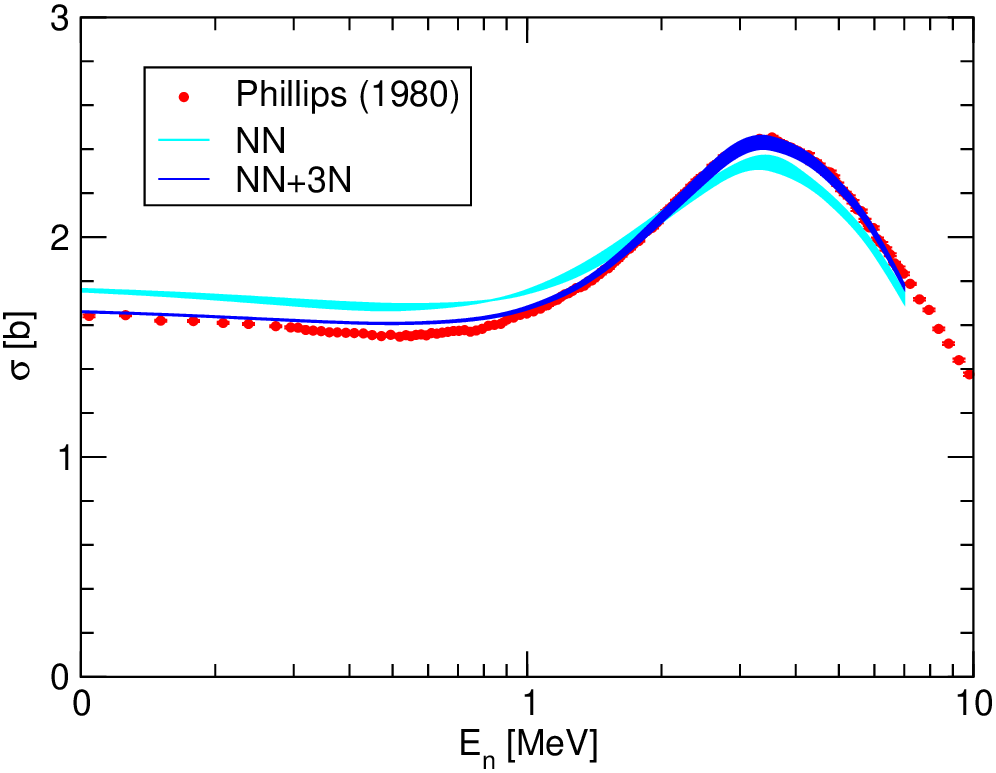}
    \caption{(color online) $n+\tri$ total cross section as function
      of the incoming neutron energy $E_n$ calculated
      with the NN N3LO interaction of
      Refs.~\protect\cite{EM03,ME11} (light cyan band) or including
      also the 3N N2LO interaction discussed in the text  (darker blue
      band). The width of the bands reflects the spread of theoretical
      results using $\Lambda=500$ or $600$ MeV cutoff values.
      See the main text for more details. The experimental
      values are taken from Ref.~\protect\cite{PBS80}.
    }
    \label{fig:ntcross}
\end{figure}

The $n+\tri$ total cross section as function of the incoming neutron energy
$E_n$ is shown in Fig.~\ref{fig:ntcross}.
The light cyan (darker blue) band shown in the figure collects the results
obtained using the N3LO500 and N3LO600 (N3LO500/N2LO500	and
N3LO600/N2LO600) interactions. Therefore the width of the bands
reflects the theoretical ``uncertainty'' connected to the 
use of interactions with two different cutoff values. As it can be seen by
inspecting the figure, the width of the bands is very tiny,
and a very good agreement with the experiment is observed, in
particular for the results obtained including the 3N force.  

\begin{figure} \centering
    \includegraphics[width=0.9\columnwidth]{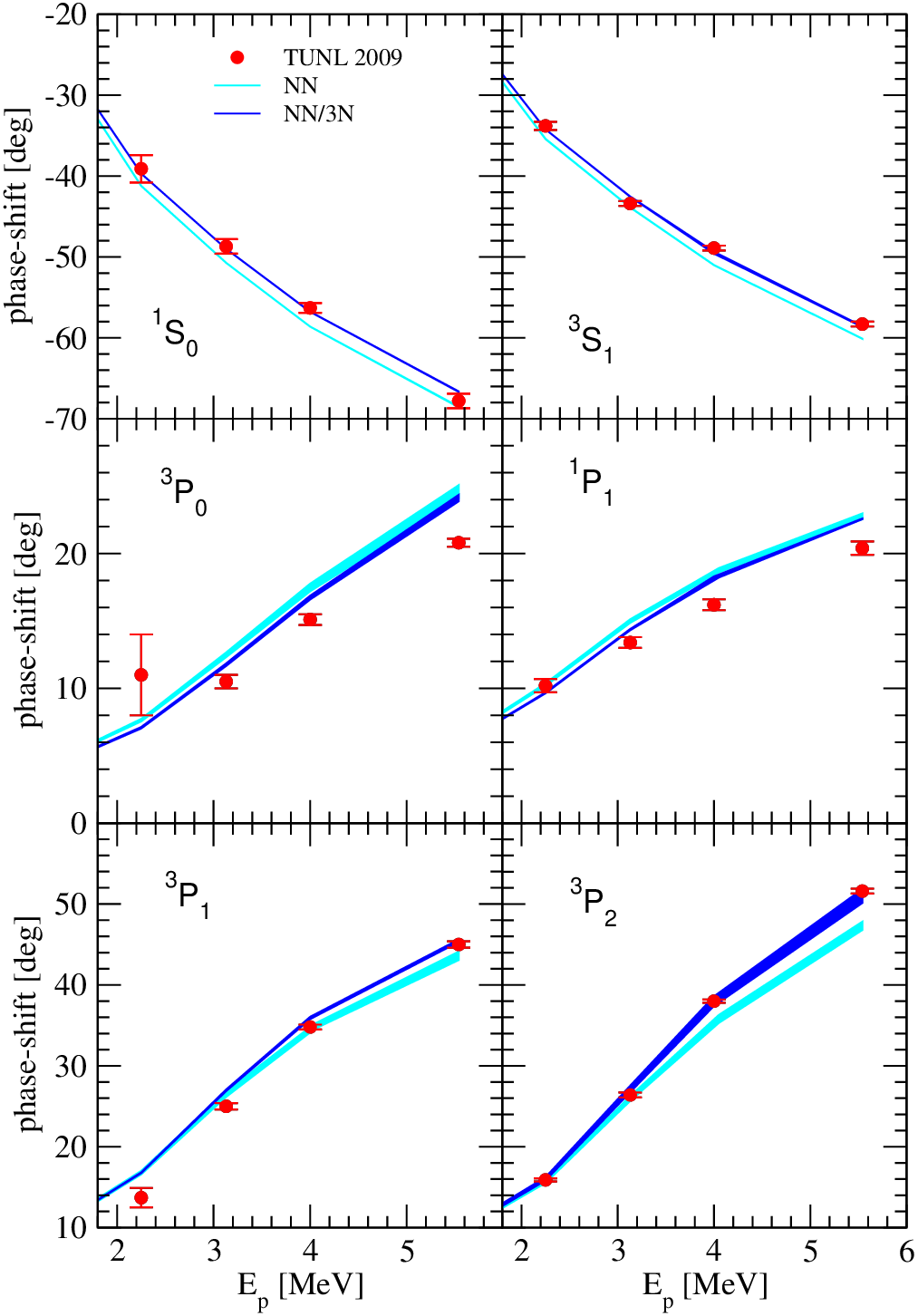}
    \caption{(color online)  $p+\het$ phase shifts as function
      of the incoming proton energy $E_p$ calculated
      with the NN N3LO interaction of Refs.~\protect\cite{EM03,ME11}
      (light cyan band) or including also the 3N N2LO interaction
      discussed in the text  (darker blue  band). 
      The results of the PSA performed at TUNL
      have been also reported~\protect\cite{Dan10}.} 
    \label{fig:phases}
\end{figure}

Let us discuss now $p+\het$ scattering. In this case, 
there exists an accurate PSA which has allowed for the extraction of 
phase-shifts and mixing parameters from the available experimental data~\cite{Dan10}.
A comparison of a selected set of calculated phase-shifts with those
obtained by this PSA is shown in Fig.~\ref{fig:phases}. Again, the
light cyan (darker blue) bands shown in the figure collect the results 
obtained using the N3LO500 and N3LO600 (N3LO500/N2LO500	and
N3LO600/N2LO600) interactions and the width of the bands
reflects the use of the two different cutoff values.
The inspection of the figure reveals that, using the interaction models
with only a NN potential, both $S$- and $P$-wave phase-shifts result 
to be at variance with the PSA. Including the 3N force, we observe a general 
improvement of the description of the phase shifts.
The decreasing (in absolute value) of the $S$ phase-shifts when the 3N
force is added is mainly due to the smaller dimension of the $\het$
nucleus following the increase of binding energy. These phase-shifts
are negative since the Pauli principle does not allow to have three
protons in $S$ wave. The $P$-waves are attractive. In particular, for
the ${}^3P_1$ and ${}^3P_2$ waves, the 3N interaction provides an
extra attraction; the resulting phase-shifts are in nice agreement
with the PSA. Regarding the ${}^1P_1$ and ${}^3P_0$ phase-shifts,
the 3N interaction reduces a little bit the disagreement with the
experimental ones, but the calculated values still overpredict the
experimental values at the largest energy.

\begin{figure*}
  \includegraphics[width=0.8\textwidth,clip,angle=0]{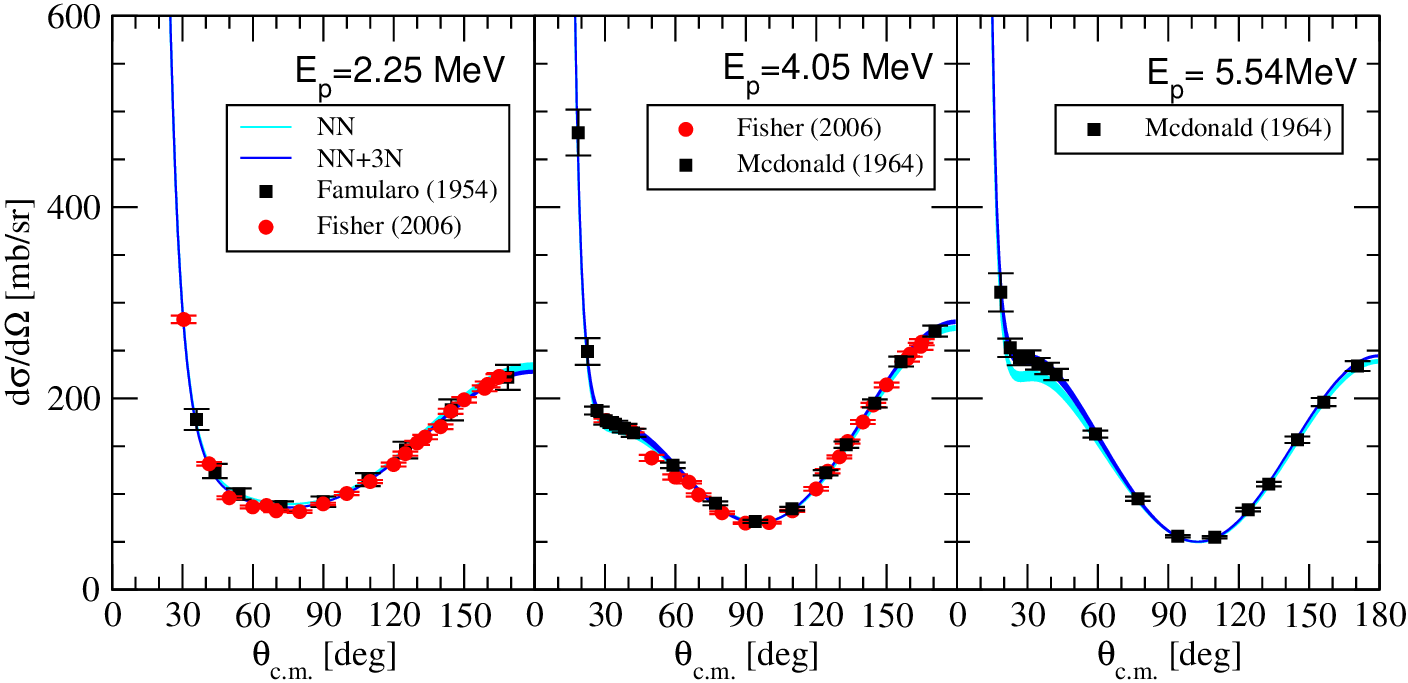}
  \caption{(color online) 
    $p+\het$ differential cross section as function of the c.m.
    scattering angle for three different proton energies $E_p$ calculated
    with the NN N3LO interactions of Refs.~\protect\cite{EM03,ME11}
    (light cyan band) or including also the 3N N2LO interactions
    (darker blue  band).
    The width of the bands reflects the use of two different cutoff
    values, $\Lambda=500$ and $600$ MeV.      
    The experimental data are from
    Refs.~\protect\cite{Fam54,Mcdon64,Fisher06}.} 
   \label{fig:xsu_ph}
\end{figure*}

Let us now compare the theoretical results directly with a
selected set of observables for which there are accurate experimental data.
We have reported the results for the $p+\het$ unpolarized differential
cross section in Fig.~\ref{fig:xsu_ph} for various  energies of the
incident proton. As usual, the results obtained with the NN (NN+3N) potentials
are shown as a light cyan (darker blue) band. As it can be seen by
inspecting the figure, the widths of the bands in this case are very tiny,
they can be appreciated only at energy $E_p=5.54$ MeV,
for $\theta_{c.m.}\approx 30$ deg. Furthermore, we observe a very good agreement
with the experimental values, in particular for the results obtained
including the 3N force.  

\begin{figure*}
  \includegraphics[width=0.8\textwidth,clip,angle=0]{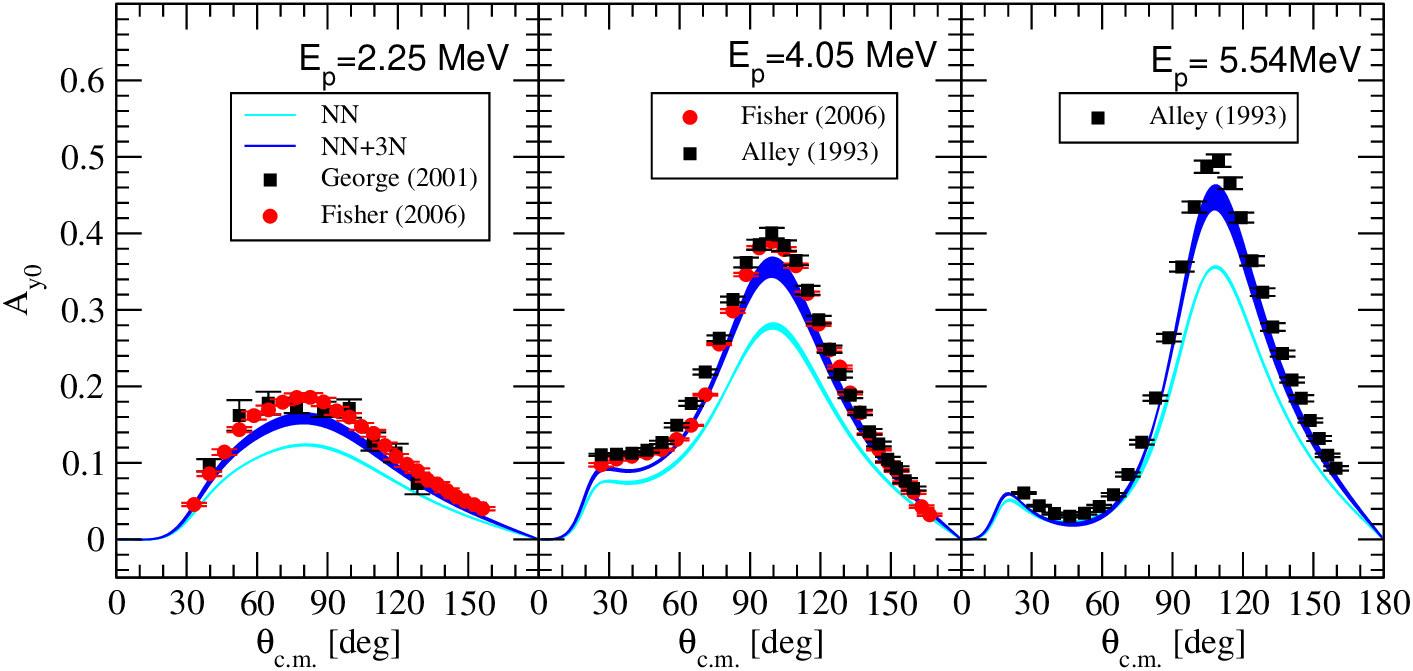}
  \caption{(color online) Same as in Fig.~\ref{fig:xsu_ph}, but for
    the $p+\het$  analyzing power $A_{y0}$.     
    The experimental data are from
    Refs.~\protect\cite{Vea01,All93,Fisher06}.} 
   \label{fig:ay0_ph}
\end{figure*}

On the contrary, for the proton analyzing power $A_{y0}$, shown in Fig.~\ref{fig:ay0_ph},
we note a large sensitivity to the inclusion of the 3N interaction.  
The calculations performed using N3LO500 and N3LO600, in fact, largely underpredict 
the experimental data, a fact already observed before also using other
interactions~\cite{Vea01,Fisher06,bm11}. A
sizable improvement is found by including the N2LO 3N interaction, as already found in Ref.~\cite{Vea13}
and recently confirmed by Ref.~\cite{LC20}.
The underprediction of the experimental data is now around 6-10\%. 

\begin{figure*}
  \includegraphics[width=0.8\textwidth,clip,angle=0]{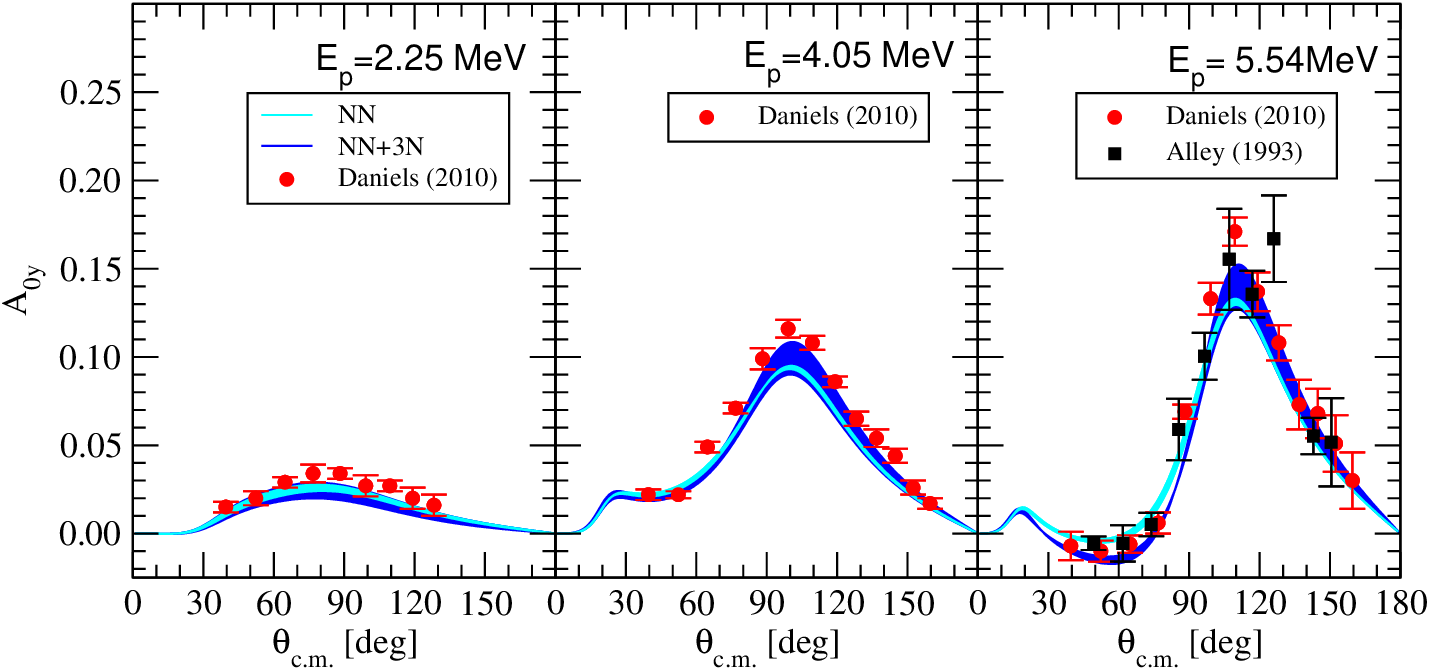}
  \caption{(color online) Same as in Fig.~\ref{fig:xsu_ph}, but for
    the  $p+\het$  analyzing power $A_{0y}$.    
    The experimental data are from
    Refs.~\protect\cite{All93,Dan10}.} 
   \label{fig:a0y_ph}
\end{figure*}

For the $\het$ analyzing power $A_{0y}$, shown in Fig.~\ref{fig:a0y_ph},
we note a smaller sensitivity to the inclusion of the 3N interaction.
However, the results obtained with the 3N force show a slightly
larger dependence on the cutoff. Here the experimental values have
larger errors, and therefore it is not possible to arrive to a
definite answer about the performance of the different interactions.

\begin{figure}
  \includegraphics[width=\columnwidth,clip]{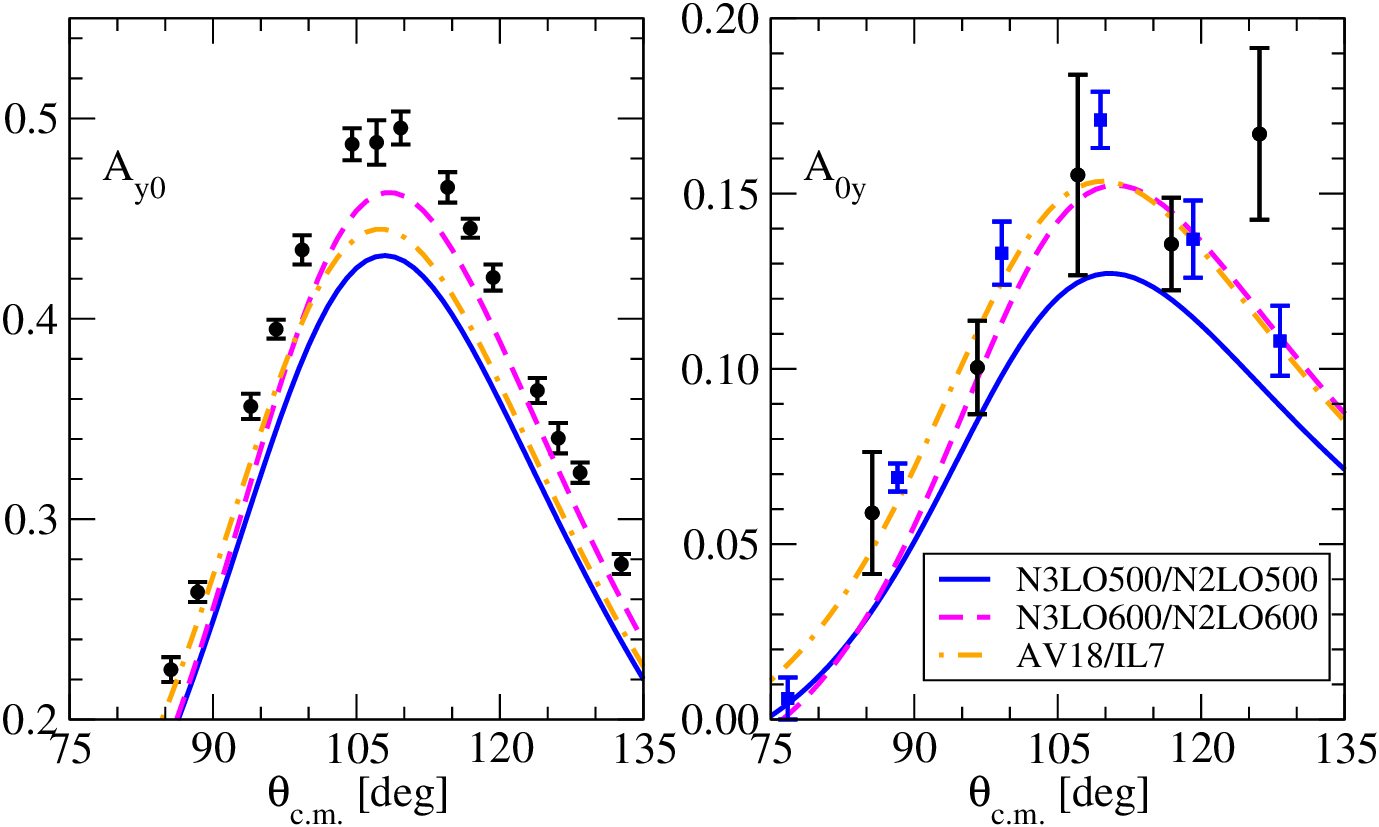}
  \caption{(color online) $p+\het$ analyzing powers at $E_p=5.54$ MeV 
   calculated with the N3LO500/N2LO500 (blue solid lines), N3LO600/N2LO600
   (dashed magenta lines), and AV18/IL7 (dot-dash red lines) interaction models. 
   The experimental data are from
    Refs.~\protect\cite{All93,Vea01,Fisher06}.}
\label{fig:det}
\end{figure}

To better point out the sensitivity to the particular interaction model,
in Fig.~\ref{fig:det} an enlargement of $A_{y0}$ and $A_{0y}$
at $E_p=5.54$ MeV in the peak region is shown. From the inspection of the
figure, we note that the observables are sensitive to the 
choice of the cutoff $\Lambda$, in particular $A_{y0}$ calculated with 
the $\Lambda=600$ MeV interaction model is slightly closer to the 
experimental data. Here, we have reported also the results obtained
using the AV18/IL7 phenomenological  interaction. We note
that $A_{y0}$ calculated with AV18/IL7 is very similar to the results obtained
with the chiral models, while $A_{0y}$ is in better agreement with the  
data than with N3LO500/N2LO500.

The previously observed large underprediction of the $p+\het$ $A_{y0}$
observable, when only NN forces were taken into account~\cite{Fon99,Vea01,Fisher06,bm11},
was considered to be due to some deficiencies of the
interaction in $P$-waves, as, for example, due to the appearance of a
unconventional ``spin-orbit'' interaction in $A>2$
systems~\cite{K99}. The IL7 model has been fitted to 
reproduce the $P$-shell nuclei spectra and, in particular, the two 
low-lying states in ${}^7$Li. This may explain the improvement in the
description of the $p+\het$ analyzing powers obtained with this interaction model.
Regarding the N2LO 3N force models, its two parameters have been
fitted to 3N observables (the 3N binding energy and the tritium GTME),
quantities which are more sensitive to $S$-waves. Therefore, its
capability to improve the description of the $p+\het$ analyzing powers
is not imposed but it is somewhat built-in.  

\begin{figure*}
  \includegraphics[width=0.8\textwidth,clip,angle=0]{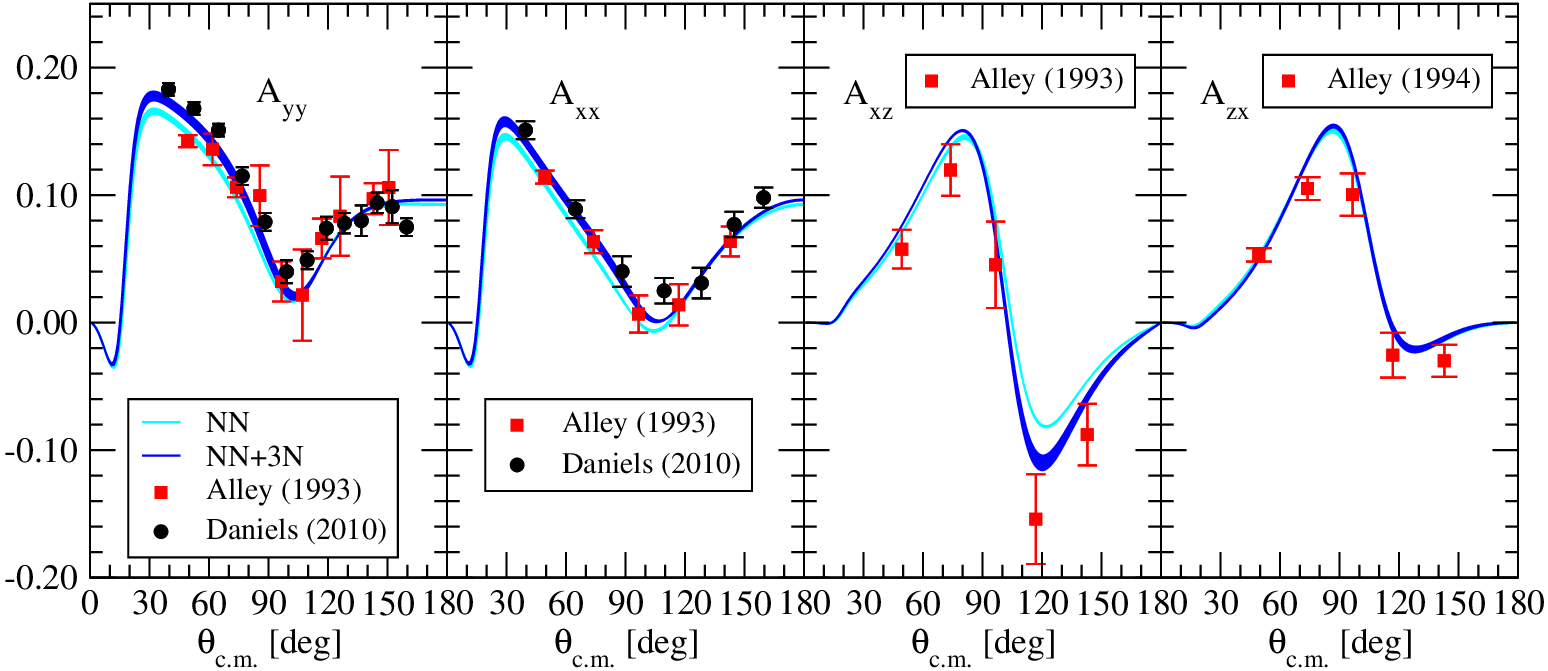}
  \caption{(color online) Same as in Fig.~\ref{fig:xsu_ph}, but for
    the $p+\het$ $A_{yy}$, $A_{xx}$, $A_{xz}$, and $A_{zx}$  spin polarization
    coefficients at $E_p=5.54$ MeV. The experimental data are from
    Refs.~\protect\cite{All93,Dan10}.} 
   \label{fig:4obs_ph}
\end{figure*}

In the literature, there are also measurements of spin polarization
coefficients. Unfortunately, these measurements have not the same
precision as for the unpolarized cross section and the proton
analyzing power. As an example, we report in Fig.~\ref{fig:4obs_ph}
$A_{yy}$, $A_{xx}$, $A_{xz}$, and $A_{zx}$ calculated at $E_p=5.54$
MeV compared with the available experimental data. As it can be seen, the
sensitivity to $\Lambda$ is small reflecting in the small widths
of the two bands. Also the effect of 3N force is tiny, and we observe a
good agreement between calculations and data.

\subsection{Resonances of $\heq$}
\label{sec:rest2}

Let us now consider the $p+\tri$ scattering. The incident energy of
the proton beam in the laboratory system is related to the c.m. kinetic
energy as $E_p={4\over 3} T_r$. We remember that for
$T_r> B(\tri)-B(\het)\equiv \Delta_3\approx0.72$ MeV,
the channel $n+\het$ is open. In this subsection, however, we
focus on the results obtained for the parameters
$\delta^{3,3}_{LS,L'S'}$ and $\eta^{3,3}_{LS,L'S'}$ describing the elastic
process $p+\tri\rightarrow p+\tri$. As usual, they are related to the
$S$-matrix as given in Eq.~(\ref{eq:etadelta}). For the sake of
simplicity, here we denote $\delta^{3,3}_{LS,L'S'}\equiv
\delta_{p+\tri}$ and we refer to it as the $p+\tri$ phase-shift.

\begin{figure}
  \begin{center}
  \includegraphics[width=0.8\columnwidth,clip,angle=0]{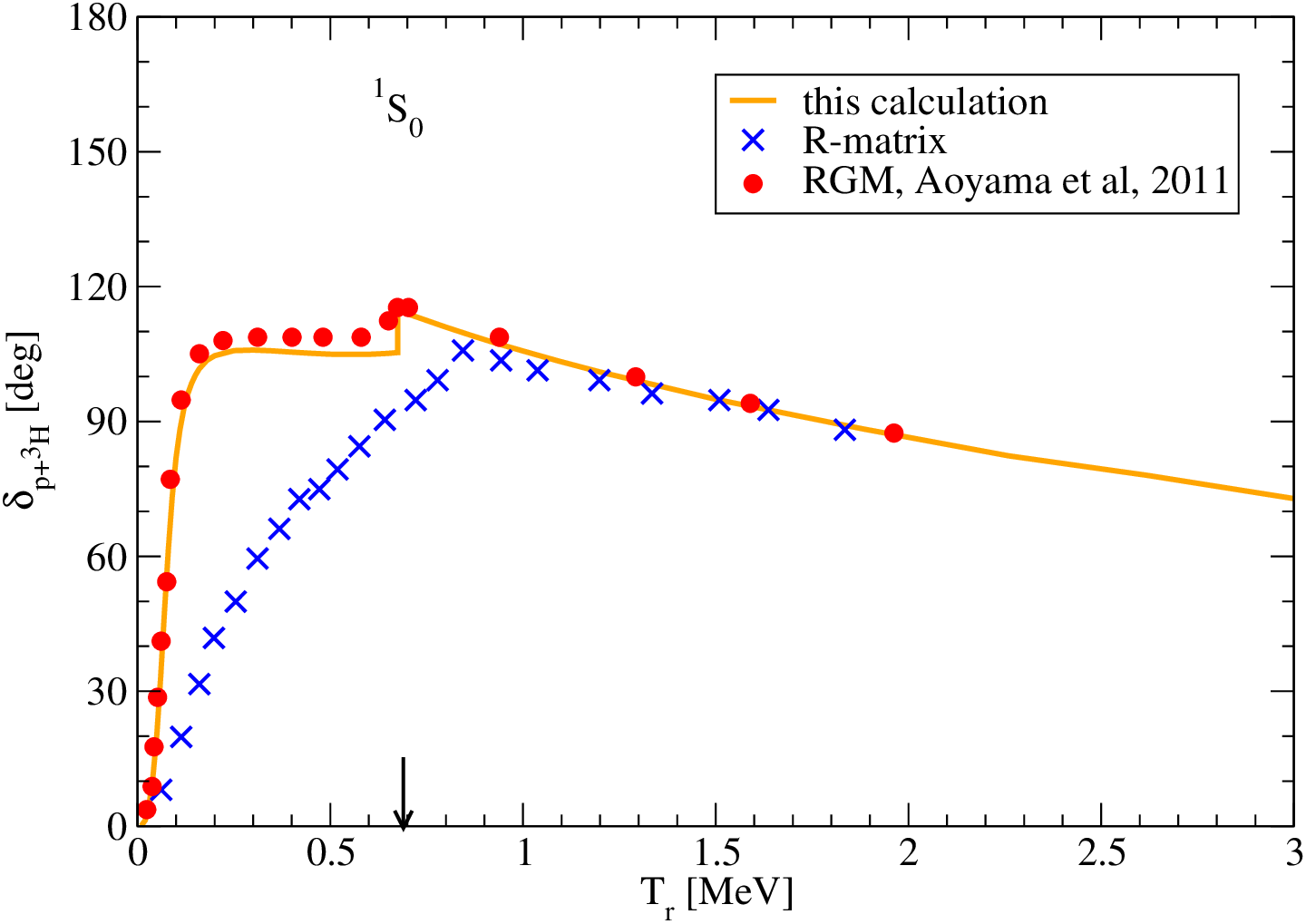}
  \end{center}
  \caption{(color online) $p+\tri$ ${}^1S_0$ phase-shift calculated with the Minnesota
    potential as function of the c.m. kinetic energy $T_r$. Solid line: present calculation; red dots: RGM
    calculation of Ref.~\protect\cite{Aoyama11}; crosses: phase-shift
    extracted from the R-matrix analysis~\cite{HH08}. The arrow
    denotes the energy of the $n+\het$ threshold.  } 
   \label{fig:minne}
\end{figure}

Let us present first of all a calculation performed for the $0^+$ wave
with the Minnesota (central) potential~\cite{minne}, in order to compare our results
with the accurate calculations performed in
Refs.~\cite{Aoyama11,Aoyama16}. 
We have reported the calculated values of $0^+$ phase-shift  $\delta_{p+\tri}$
(corresponding to the ${}^1S_0$ wave in spectroscopic notation) in 
Fig.~\ref{fig:minne}, together with the results of
Ref.~\cite{Aoyama11}. For the Minnesota potential, the $n+\het$
threshold is at $E=0.675$ MeV, shown in the figure by an arrow. For that energy
the phase-shift has a discontinuity. Probably, at energies just
below the opening of the $n+\het$ channel, it should be convenient to
include in the wave function also an asymptotic ``closed'' component like
\begin{equation}
  \Omega_{4 LS}^{c} = \sum_{l=1}^4\biggl\{
  \Bigl [ Y_{L}(\hat{\bm y}_l) \otimes  [ \phi_3^{h} \otimes \chi_l \xi_l^n]_{S} 
   \Bigr ]_{JJ_z} {\exp(-\beta_4 y_l)\over y_l}\ , 
  \label{eq:psiclosed}
\end{equation}
where $\phi_3^{h}$ is the $\het$ wave function,
$\xi_l^n$ the isospin state of the neutron (particle $l$), and $y_l$
the distance between the c.m. of $\het$ and the neutron.  Above,
we have specified that $\gamma=4$ and 
$\beta_4=\sqrt{2\mu_4(\Delta_3-T_r)}$, where $\Delta_3\approx0.72$ MeV
is the
difference between the $\tri$ and $\het$ binding energies.
When $T_r\rightarrow \Delta_3$, $\beta_4$ becomes rather small
and the component $\Omega_{4 LS}^{c}$ will have a long-range tail.
Configurations of this type are rather difficult to be
constructed in terms of the HH expansion, whence the
utility of explicitly including them in the variational wave function. 
Work in this direction is currently in progress. 

Returning to Fig.~\ref{fig:minne}, we note that the results of our
calculation and that of Ref.~\cite{Aoyama11} are very
close. The phase-shift has a ``resonant'' behavior, with a very sharp
increase followed by a plateau. In particular,  $\delta_{p+\tri}$
reaches the value of $90$ deg for $E\approx 0.12$ MeV. 

Now, let us try to extract the energy $E_R$ and width $\Gamma$ of the resonance using two
methods. In the first method, one can just estimate $E_R$ 
as the value of $T_{r}$  for which the first derivative $\delta_{p+\tri}^\prime$ has a maximum and
$\Gamma=2/\delta_{p+\tri}^\prime(E_R)$~\cite{TN09}.
Using the phase-shifts reported in Fig.~\ref{fig:minne},
we obtain  $E_R=0.064$ MeV and $\Gamma=0.088$ MeV.

Another method to determine $E_R$ and $\Gamma$ has been taken from
Ref.~\cite{RSE07}. The idea is to fit the calculated
$S$-matrix for various energies using a Pad\`e approximation, namely
\begin{equation}
   {\cal S}(k)={1+\sum_{n=1}^N a_n k^n \over 1+\sum_{n=1}^N (-)^n a_n
     k^n }\ , \label{eq:pade}
\end{equation}
where $k=\sqrt{2\mu T_{r}}$ and $\mu $ is the $p+\tri$ reduced
mass. This form is suggested by the general properties of the 
$S$-matrix, in particular that  ${\cal S}(-k)={\cal S}(k)^{-1}$ and
that ${\cal S}(k\to 0)\to 1$. Given a number $N$ of values ${\cal S}(k_i)$,
$i=1,\ldots,N$, the coefficients $a_n$
can be simply obtained solving the linear system
\begin{equation}
  \sum_{n=1}^N \Bigl[1+(-)^{n+1} {\cal S}(k_i)\Bigr]k_i^n a_n = 
          {\cal S}(k_i)-1\ . \label{eq:pades}
\end{equation}
The resonances are then calculated as the poles of the $S$-matrix,
namely the zeroes of the denominator of Eq.~(\ref{eq:pade}). The problem thus reduces
to find the zeros of the polynomial $1+\sum_{n=1}^N (-)^n a_n k^n$,
which can be readily obtained using the method described in Ref.~\cite{recipes}.
However, since the $S$-matrix is extracted using only a finite number of
energies, the procedure finds a number of spurious poles, in addition
to the ``true'' poles. To recognize the true poles, in Ref.~\cite{RSE07} it is
suggested to vary $N$, and to observe the position of the poles in the
plane $\Re(k),\Im(k)$: the position of the ``true'' poles should be
independent on $N$, while 
the spurious pole positions will vary considerably with $N$.

We have used this procedure using the phase-shifts calculated with
the Minnesota potential and selecting increasing values of
$N=4,6,\ldots$. We have found one stable pole, from which 
the values $E=E_R=0.067$ MeV and $\Gamma=0.070$ MeV are extracted,
in reasonable agreement with the values determined using the first
method. We note that in Ref.~\cite{Aoyama11}, the resonance energy is determined,
by a bound-state approximation, to be $E_R = 0.12$ MeV,
which corresponds to the energy $T_{r}$ for which
$\delta=90$ deg. In Ref.~\cite{Aoyama16}, the resonance is obtained by a complex
scaling method at $E_R=0.07$ MeV and $\Gamma=0.06$ MeV (with a numerical error
estimated to be several tens of keV), in good agreement with our results.

\begin{figure*}
  \begin{center}
  \includegraphics[width=0.8\textwidth,clip,angle=0]{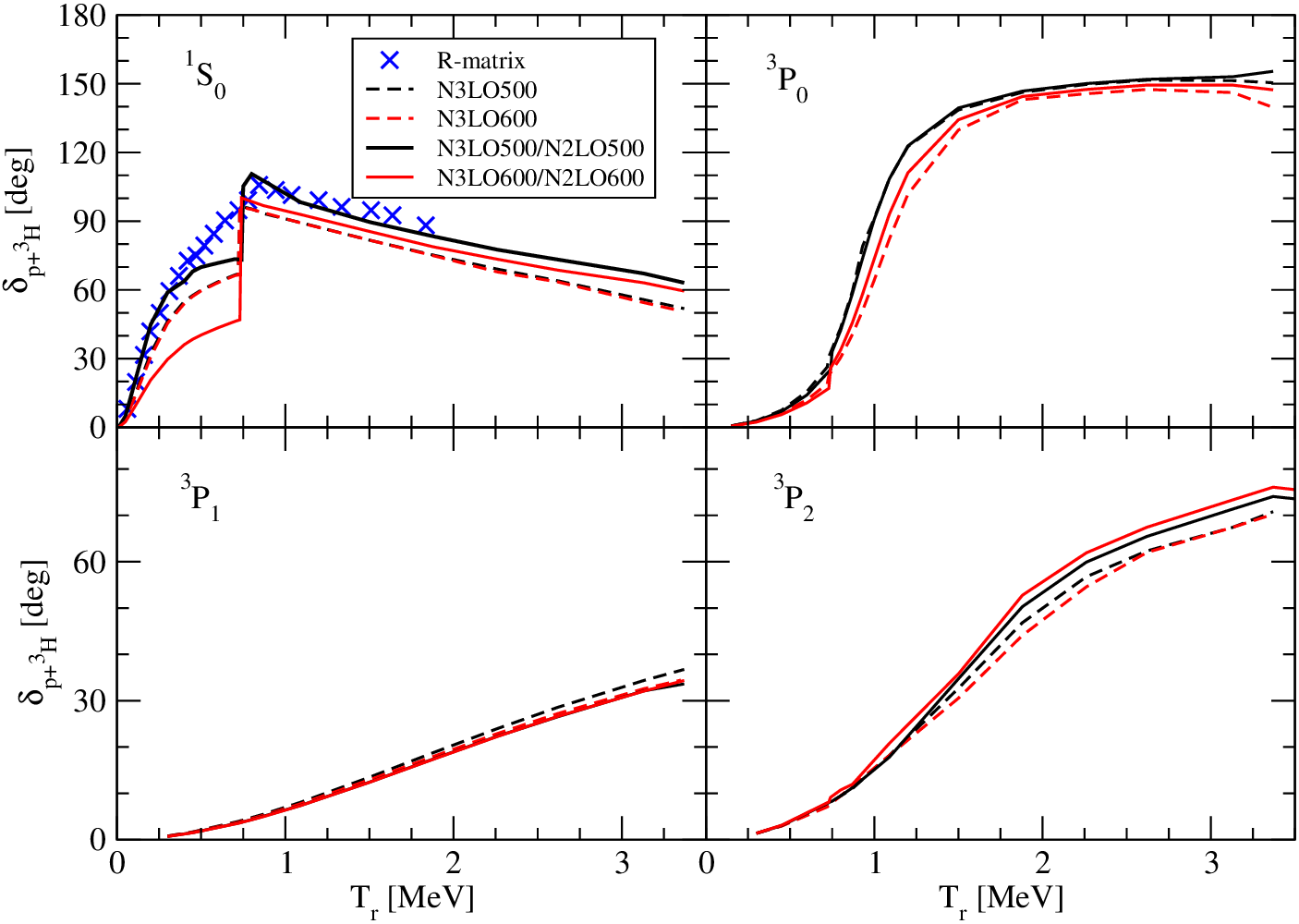}
  \end{center}
  \caption{(color online) $p+\tri$ phase-shifts as function of the c.m. kinetic energy $T_{r}$
    calculated with the N3LO500 (red dashed curves), N3LO600 (red dotted
    curves), N3LO500/N2LO500 (black solid curves), and
    N3LO600/N2LO600 (red dot-dashed curves) interactions.
    The experimental phase shifts have been extracted by an R-matrix
    analysis in  Ref.~\protect\cite{HH08}.} 
   \label{fig:phases_pt}
\end{figure*}

\begin{figure*}
  \begin{center}
  \includegraphics[width=0.8\textwidth,clip,angle=0]{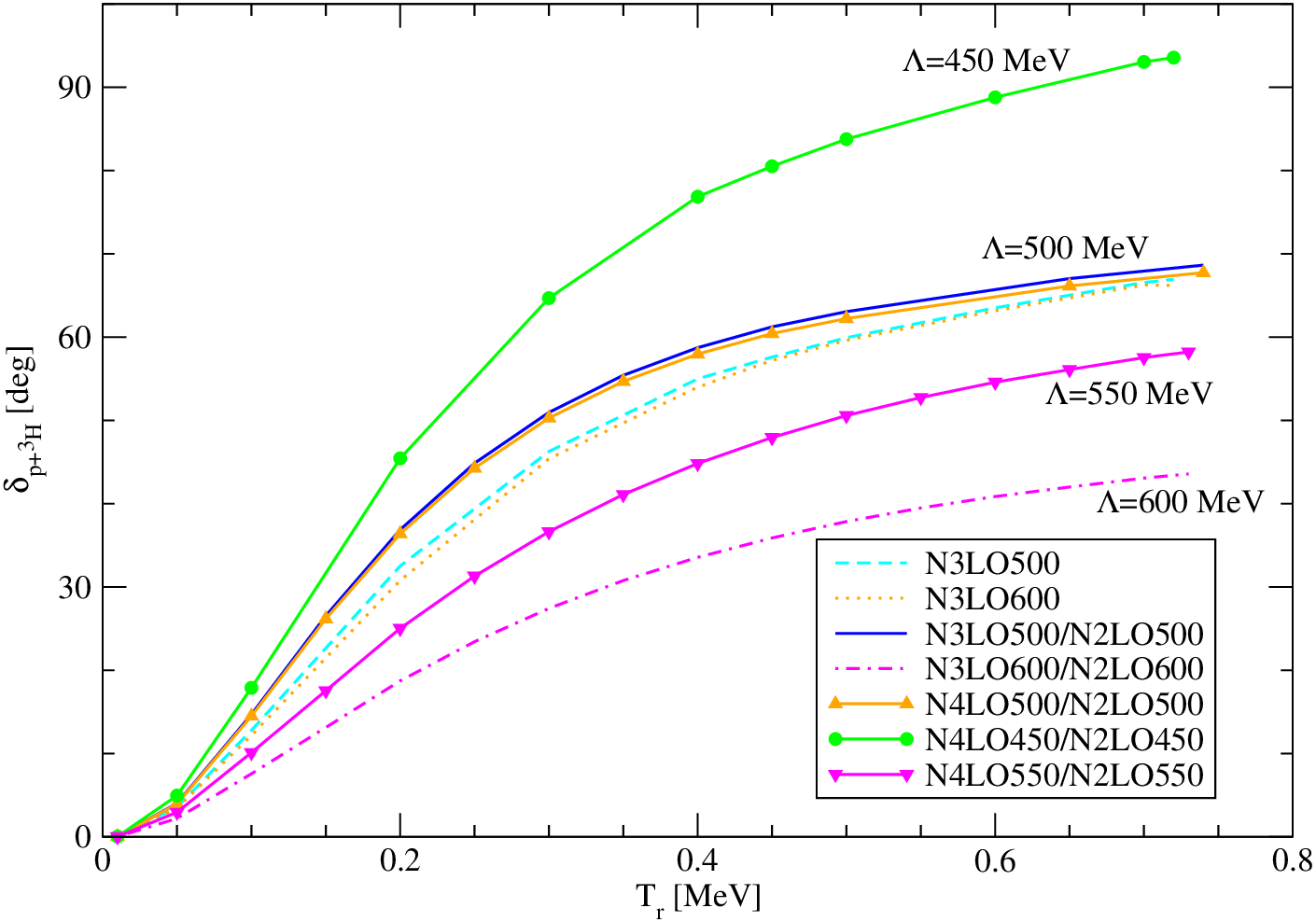}
  \end{center}
  \caption{(color online) ${}^1S_0$ $p+\tri$ phase-shift as function of the c.m. kinetic energy $T_{r}$
    calculated with several interactions.} 
   \label{fig:phases_1s0}
\end{figure*}

We now present the results for some $p+\tri$ phase-shifts calculated with
the N3LO500, N3LO600, N3LO500/N2LO500, and N3LO600/N2LO600
interactions in Fig.~\ref{fig:phases_pt}. 
We note rather large differences for the $^1S_0$ phase-shift when the
3N force is added, while for the $P$-wave phase-shifts the results with
and without the 3N force are rather close. Again, below the
threshold of the $n+\het$ channel, the inclusion of the ``closed''
component as given in Eq.~(\ref{eq:psiclosed}) could improve the
convergence, in particular for the ${}^1S_0$ case. Work in this
direction is in progress.
In any case, for the ${}^1S_0$ phase-shift, below the opening of the
$n+\het$ channel, the results obtained using the N3LO500/N2LO500 and 
N3LO600/N2LO600 differ considerably. In order to explore this result,
we have performed additional calculations using the N4LO450/N2LO450,
N4LO500/N2LO500, and N4LO550/N2LO550 interactions. The results
obtained for the ${}^1S_0$ phase-shift at low energies are reported in 
Fig.~\ref{fig:phases_1s0}.

We note that the results obtained with the N3LO500/N2LO500 and
N4LO500/N2LO500 interactions are very close (we remember that the values of
$c_1$, $c_3$, $c_4$, $c_D$, and $c_E$ in these two 3N force
interactions are different). On the other hand we observe again 
a strong dependence on the cutoff values. The interactions with the
softer cutoff corresponds to larger values of the $p+\tri$
phase-shift. Note that the differences between the phase-shifts
calculated with the various $\Lambda$ are significantly larger than
the theoretical uncertainties connected to the extrapolation procedure
discussed in the previous section, which at $E_p=0.60$ MeV
(corresponding to $T_r=0.45$ MeV) was estimated to be around $1.4$
deg. Therefore, these differences cannot be ascribed to the
uncertainties in the extrapolation of the phase-shifts.

\begin{table}
\caption{Energy of the $0^+$ resonance and its width as extracted from the
phase-shifts reported in Fig.~\ref{fig:phases_1s0}. The experimental values are
extracted from the R-matrix analysis reported in Ref.~\protect\cite{A4b}.}
\begin{center}
\begin{tabular}{l|cc}
\hline
\hline
 Interaction & $E_R$ (MeV) & $\Gamma$ (MeV)  \\
 \hline
 N3LO500         & 0.126 & 0.556 \\
 N3LO600         & 0.134 & 0.588 \\
 \hline
 N3LO500/N2LO500 & 0.118 & 0.484 \\
 N3LO600/N2LO600 & 0.130 & 0.989 \\
 \hline
 N4LO450/N2LO450 & 0.126 & 0.400 \\
 N4LO500/N2LO500 & 0.118 & 0.490 \\
 N4LO550/N2LO550 & 0.130 & 0.740 \\
 \hline
  Expt.          & 0.39 & 0.50 \\
\hline
\end{tabular}
\end{center}
\label{tab:res1}
\end{table}

From these phase-shifts it is possible to extract 
the resonance parameters as discussed previously. We
report in Table~\ref{tab:res} the values obtained using method 1.
The values of $E_R$ is somewhat independent on the
inclusion of the 3N force and the value of the cutoff, and it results to be
around $0.1$ MeV, somewhat at variance with respect to the experimental datum.
On the contrary, the width is very sensitive to the cutoff. The potentials
with cutoff $\Lambda>500$ MeV predict a too large width when compared to the
experimental value.

\begin{center}
  \begin{table}[t]
  \caption{\label{tab:res} Energies $E_R$ and widths $\Gamma$ of the resonances in the
    different waves obtained using the chiral interactions. The
    experimental values are taken from Ref.~\protect\cite{A4b}
    and obtained from an R-matrix analysis.}
  \begin{tabular}{lcccc}
\hline
\hline
    & \multicolumn{2}{c}{${}^3P_0$} &
      \multicolumn{2}{c}{${}^1P_1$} \\
 Interaction & $E_R$ (MeV) & $\Gamma$ (MeV)  &
               $E_R$ (MeV) & $\Gamma$ (MeV)  \\
  \hline
 N3LO500         & 0.89 & 0.46\phan & 1.7 & \phan 8.2 \\
 N3LO600         & 1.05 & 0.57\phan & 1.7 & \phan 8.6 \\
 N3LO500/N2LO500 & 0.90 & 0.46\phan & 1.8 & \phan 8.6 \\
 N3LO500/N2LO500 & 0.98 & 0.54\phan & 1.8 & \phan 8.7 \\
 \hline
 R-matrix           & 1.20 & 0.84 & 6.13 & 12.7\phan  \\
 \hline
 \hline
   &    \multicolumn{2}{c}{${}^3P_1$} &
        \multicolumn{2}{c}{${}^3P_2$} \\
 Interaction & $E_R$ (MeV) & $\Gamma$ (MeV)  &
               $E_R$ (MeV) & $\Gamma$ (MeV)  \\
  \hline
 N3LO500         &  1.0\phan &  4.7\phan  & 1.4 &  3.1\phan   \\
 N3LO600         &  1.0\phan &  4.8\phan  & 1.5 &  3.3\phan   \\
 N3LO500/N2LO500 &  1.3\phan &  4.7\phan  & 1.4 &  2.7\phan   \\
 N3LO600/N2LO600 &  1.3\phan &  4.4\phan  & 1.7 &  2.9\phan   \\
 \hline
 R-matrix           & 4.43 & 6.10 & 2.02 &  2.01 \\
\hline
\hline
  \end{tabular}
  \end{table}
\end{center}
  
We have also extracted the resonance parameters from the ${}^3P_0$,
${}^1P_1$, ${}^3P_1$, and ${}^3P_2$ phase shifts (some of them are
reported in Fig.~\ref{fig:phases_pt}). The results are listed
in Table~\ref{tab:res}. The experimental information is obtained using
an R-matrix method as discussed
in Ref.~\cite{A4b}, so it is not clear whether the two methods would give
consistent results. Work to clarify this issue is still in progress.
From inspection of the table, we can see that the values of $E_R$
are consistently smaller than the experimental ones.
The width of the resonance in the $0^-$ wave is
predicted to be smaller than that reported by the R-matrix
analysis. From the calculation, this resonance is found to
have approximately the same width as the $0^+$ resonance studied
earlier. The dependence on the cutoff and on the inclusion of the 3N
interaction is not critical. The width of the resonances found in the $1^-$ wave are
noticeably large. Very likely in this case the extracted values of $E_R$ and $\Gamma$ are not significant.
On the other hand, the resonance in the $2^-$ wave is
well established, and the energy and width are in reasonable agreement
with the values extracted from the R-matrix analysis.

\subsection{$p+\tri$ and $n+\het$ scattering}
\label{sec:rest3}

\begin{figure*}
  \includegraphics[width=0.8\textwidth,clip,angle=0]{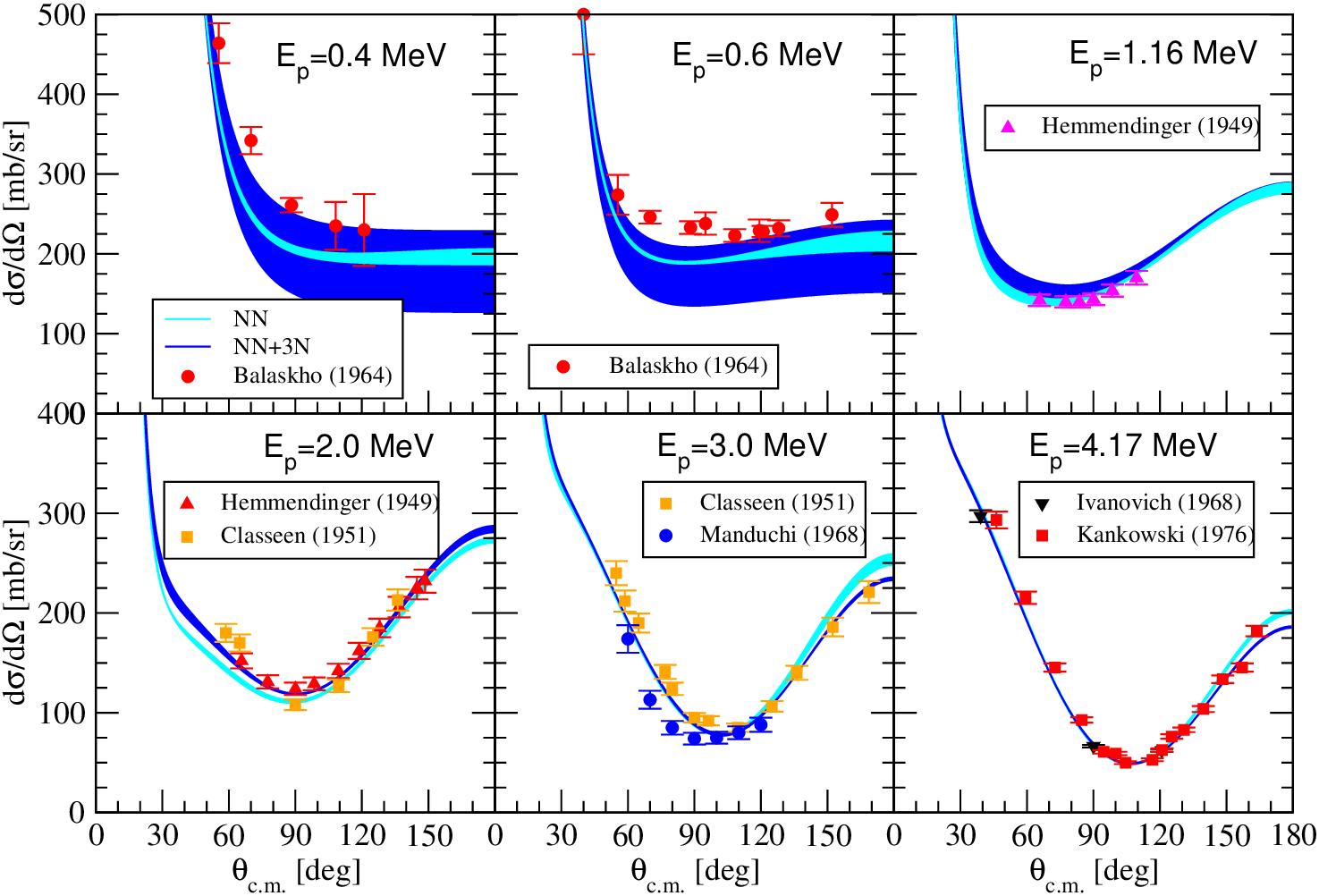}
  \caption{(color online)  Same as in Fig.~\ref{fig:xsu_ph} but for
    the $p+\tri$ differential cross section. The experimental data are from
    Refs.~\protect\cite{Hemme49,Clas51,Balas65,Mandu68,Iva68,Kanko76}.} 
   \label{fig:xsu_pt}
\end{figure*}

\begin{figure*}
  \includegraphics[width=0.8\textwidth,clip,angle=0]{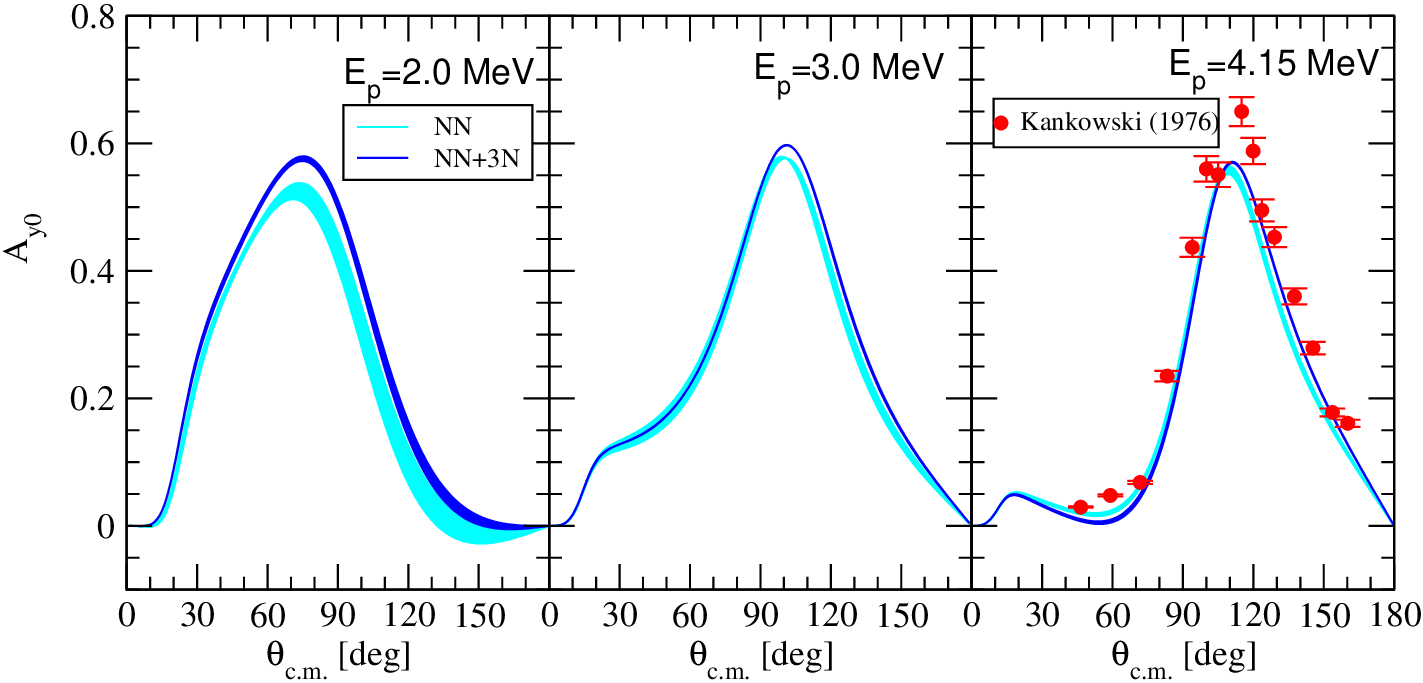}
  \caption{(color online) Same as in Fig.~\ref{fig:xsu_ph}, but for
    the $p+\tri$ proton analyzing power.
    The experimental data are from
    Refs.~\protect\cite{All93,Dan10}.}
  \label{fig:ay0_pt}
\end{figure*}

Let us now consider the results obtained for the $p+\tri$ and $n+\het$
observables. We have reported the results for the $p+\tri$ unpolarized differential
cross section in Fig.~\ref{fig:xsu_pt} at various  energies of the
incident proton. Again, the results obtained with N3LO500 and N3LO600 (N3LO500/N2LO500
and N3LO600/N2LO600) potentials are collected in the light cyan (darker blue) bands.
By inspecting the figure, at the two lowest energies
the effect of 3N force is sizable.
We also note that, at the two lowest energies, the observable becomes very cutoff
dependent when including the 3N force.
For those energies the $n+\het$ channel is  closed, and the cross
section considerably depends on the position of the first excited state of
$\heq$. In fact, the differences in the cross section originate mainly from the
${}^1S_0$ phase-shift.  Above the $n+\het$ threshold,
the width of the band is small, as observed before  for $p+\het$. In
this case, we find that the contribution of the 3N force is small. The
$p+\tri$ analyzing powers are reported in Fig.~\ref{fig:ay0_pt}, where
we show only the results obtained at energies larger than the
$n+\het$ threshold (below it this observable is tiny). For these
energies, the effect of 3N force is not very important. We find that
the height of the peaks is only slightly increased when the 3N force
are included, but this does not significantly help in reducing the
disagreement observed with the experimental data at $E_p=4.15$ MeV, as
it can be seen in Fig.~\ref{fig:ay0_pt}. 

\begin{figure*}
  \begin{center}
  \includegraphics[width=0.8\textwidth,clip,angle=0]{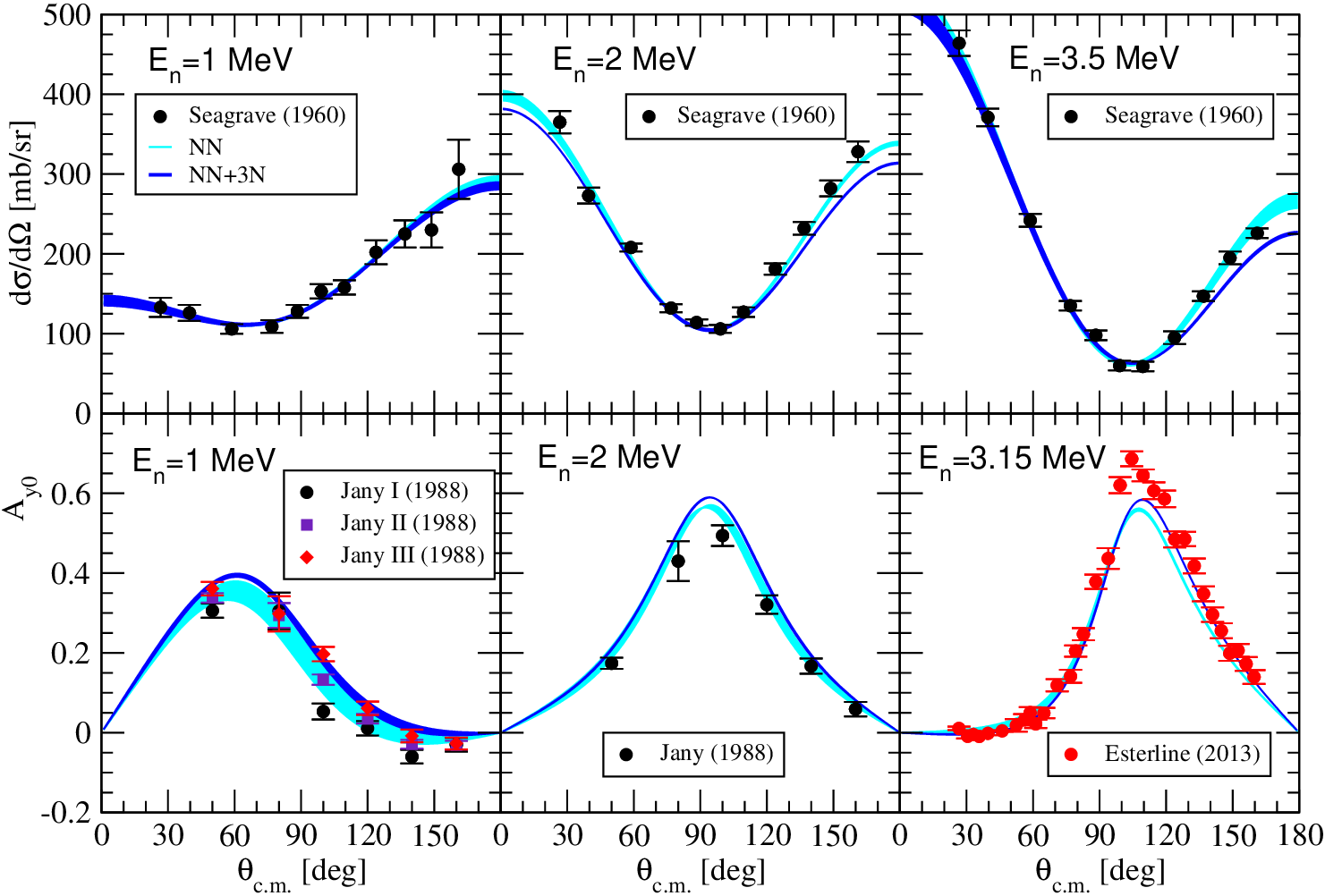}
  \end{center}
  \caption{(color online) The same as Fig.~\protect\ref{fig:xsu_ph} but for the $n+\het$
    differential cross section and neutron analyzing power.
    The experimental data are from
    Refs.~\protect\cite{Sea60,Jany88,Este13}.} 
   \label{fig:nh}
\end{figure*}

Some of the results obtained for $n+\het$ elastic scattering are reported in Fig.~\ref{fig:nh}.
The results obtained using NN interaction only or including also 
the 3N force are as usual shown by bands. As it can be seen inspecting the figure,
the widths of the bands are always small, showing that the dependence on $\Lambda$ is not
critical. Also the effects of the inclusion of the 3N forces are small.

\begin{figure*}
  \begin{center}
  \includegraphics[width=0.8\textwidth,clip,angle=0]{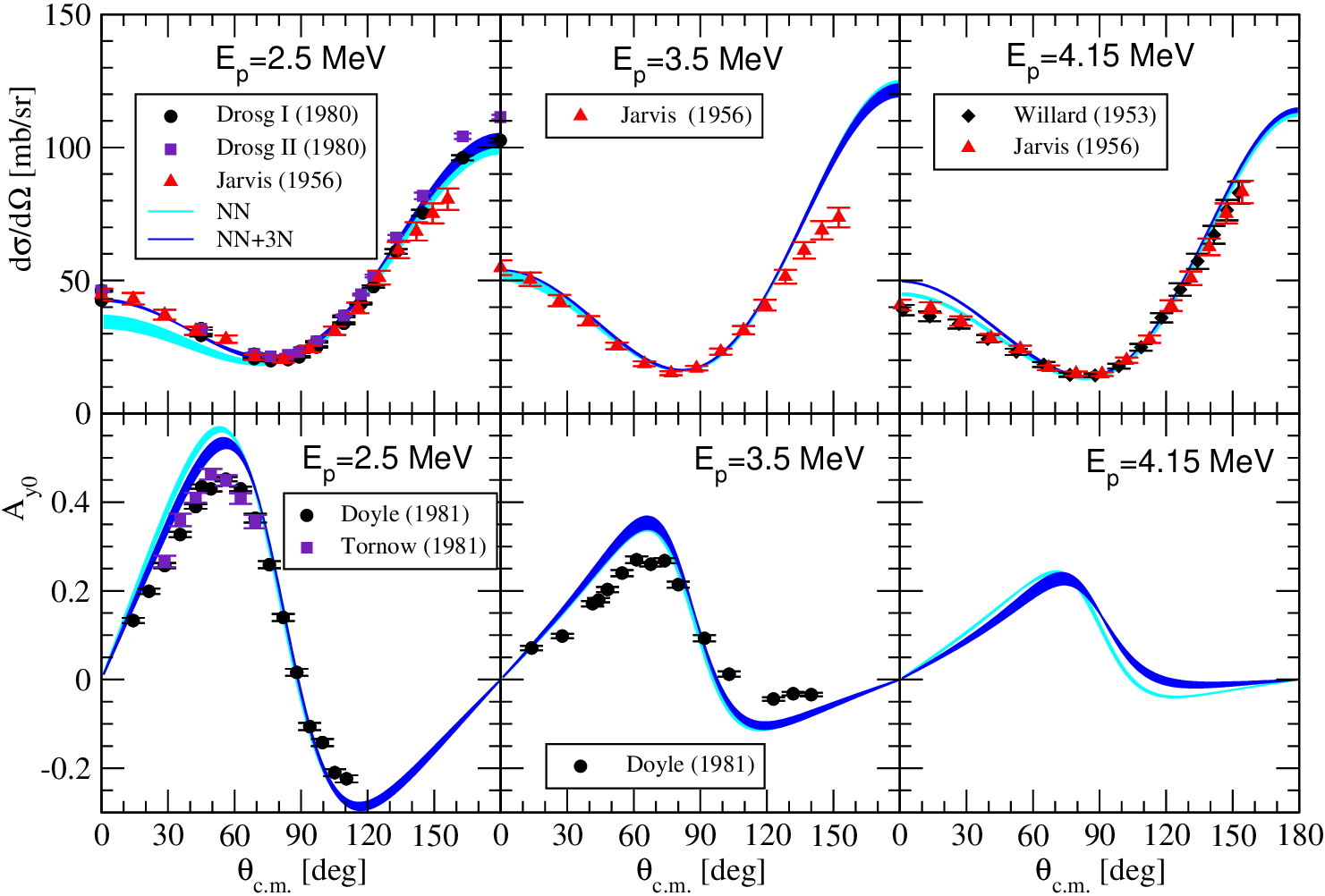}
  \end{center}
  \caption{(color online) The same as in Fig.~\protect\ref{fig:xsu_ph} but for the
    $p+\tri\rightarrow n+\het$ differential cross section and proton analyzing power.
    The experimental data are from
    Refs.~\protect\cite{Will53,Jarvis56,Drosg80,Doy81,Tornow81}.} 
   \label{fig:ptx}
\end{figure*}

The results for some $p+\tri\rightarrow n+\het$ charge exchange reaction observables
are reported in Fig.~\ref{fig:ptx}, together with the available experimental data.
We see that the contribution of the 3N force is small for these observables.
Again, the dependence on the cutoff is not critical.

\section{Conclusions and perspectives}
\label{sec:conc}

We have discussed in detail the application of the HH
method to the 4N scattering problem, limiting our study to processes
with only two clusters in the asymptotic regions (but below the
energies for which the channel $d+d$ is open). We have discussed
the issues of convergence and numerical stability, showing that they are 
under control. The convergence of the HH expansion is usually well
achieved for chiral potentials, except for the $p+\tri$ $J^\pi=0^+$
phase-shift, where large extrapolations are needed in order to take
into account the contribution of HH states of large $K$. However, we have
also discussed the procedure used to estimate the ``missing'' phase-shift,
believed to be reliable. 

In the paper we have also included the results of a first
campaign of calculations of various low energy elastic and
charge-exchange processes. In particular, we have studied the effect of
including the N2LO 3N forces, constrained to reproduce the $\tri$ binding
energy and the GTME in tritium $\beta$-decay. For $n+\tri$ elastic
scattering, the inclusion of the 3N forces is very helpful in
reproducing the scattering lengths and the total cross section, in
particular in the resonance region. For $p+\het$ the main effect of
the inclusion of 3N force is to reduce the disagreement between theory
and experiment in the observable $A_{y0}$, which is present when only
NN forces are taken into account.

For $n+\het$ elastic scattering and the charge
exchange reaction $p+\tri\rightarrow n+\het$, the inclusion of the 3N forces
is tiny, although in general it helps to obtain a slightly better
description of the data. On the other hand, for the $p+\tri$ elastic
scattering (at energies below the opening of the $n+\het$ channel)
sizable effects of the 3N force are observed mainly in the ${}^1S_0$ wave.
In particular, a rather strong dependence on the cutoff used to regularize
the chiral potential is found when the 3N force is included in the calculations. 
We have speculated that this effect might be related to a critical dependence on
the 3N force of the position and width of the resonance representing
the first excited state of $\heq$. Further studies of this resonance
are currently in progress. Moreover, it would be rather important to
have new and more  accurate measurements of $p+\tri$ elastic
scattering at these low energies, as the available experimental data
are rather old and of limited angular range. 

More refined calculations of the same processes with the new
generation of chiral potentials up to N4LO~\cite{MEN17} are currently
in progress. First of all, we would like to improve the
calculations of the $p+\tri$ phase shifts just below the 
opening of the $n+\het$ channel including explicitly in the wave
functions the ``closed'' component given in Eq.~(\ref{eq:psiclosed}).
Moreover, calculations with larger sets of HH functions will be
undertaken. From the observables calculated using the interactions
at different chiral orders, we plan also to estimate the ``theoretical
uncertainties'' due to our incomplete knowledge of the nuclear
dynamics, following the procedure proposed in Ref.~\cite{EE15}. Further
calculations performed with the local EFT interactions developed in
Refs.~\cite{Pea16,Pea18,Bea18} are planned. These latter interactions
take into account also the $\Delta$-particle degrees of
freedom. Finally, we propose also to explore the effect of the
3N force contact terms appearing at N4LO~\cite{Girla11,Girla19}. These
terms are currently studied in the $A=3$ system in order to solve the
$A_y$ puzzle found in $N+d$ scattering. It would be very interesting
to see if these terms can help in solving also the various disagreements
discussed in this paper for $A=4$ scattering.

The availability of the  $n+\tri$,  $p+\het$, $p+\tri$, and $n+\het$
scattering wave functions will allow for the study of various
radiative capture reactions, as 
$p+\tri\rightarrow \heq+\gamma$ and $n+\het\rightarrow \heq+\gamma$,
of electron scattering elastic and transition form factors, as for
the $\heq(e,e')\heq^*$ process, of reactions of astrophysical
interest, as the ``hep'' reaction $\het(p,e^-\nu_e)\heq$, and of the
process $\tri(p,e^+e^-)\heq$, recently exploited experimentally
in order to demonstrate the existence of a new kind of particle~\cite{Kras19}.

In the near future, we plan also to extend the formalism to $d+d$
scattering and to energies above the threshold for the breakup in
three or more clusters in the final state. Work in this direction
has been already undertaken.

% -----------------------------------------------------------------------------

\appendix

\section{The regularization of the function $G_L$}
\label{app:g}

In this appendix, we describe the functions $f_L(y)$ used
to regularize the irregular Coulomb functions using method 1, namely
\begin{equation}
  {\widetilde G_{L}(\eta,qy)\over qy}  =  {G_{L}(\eta,qy)\over qy}
    -{ f_L(y)\over y^{L+1}} \exp(-\beta y) \ ,\label{eq:reg1b}
\end{equation}
where, in general, 
\begin{eqnarray}
  f_L(y)  &=&  a_0 + a_1 y+ a_2 y^2 + \cdots + a_N y^N \nonumber \\
          & +& ( b_1 y+ b_2 y^2 + \cdots + b_M y^M)\log(2qy)\ .
     \label{eq:flregb}
\end{eqnarray}
An important aspect of this method is that the function
\begin{eqnarray}
 \overline G_L &=& -\biggl\{ f_L''-\biggl(2\beta+{2L\over y}\biggr)
 f_L'\nonumber \\
    &+&\Bigl(\beta^2+2{\beta L-\eta q\over y } +q^2\Bigr) f_L\biggr\}
     {e^{-\beta y}\over y^{L+1}} \ ,\label{eq:ovrlggb}
\end{eqnarray}
where $f'=df/dy$, etc. becomes a smooth function, without any oscillatory
behavior. The function $f_L$ is chosen (as discussed below) so that both 
$\widetilde G_L/qy $ and $\overline G_L$ are
regular at the origin. Let us first discuss the cases $L=0$ and $L=1$
separately, and then we give the general expression for $L>1$.

\subsection{Case $L=0$}
Let us start from the small-$y$ behavior of the irregular Coulomb function $G_0$, which
reads~\cite{abra}
\begin{eqnarray}
  {G_0(\eta,qy)\over qy} &\rightarrow & {1\over C_0(\eta)q}
   \biggl[ \biggl( 2\eta q + 2\eta^2 q^2 y + O(y^2) \biggr)
          \log(2qy)\nonumber \\
   & +&  \biggl({1\over y} + O(y) \biggr) \biggr]\ .\label{eq:g0}
\end{eqnarray}
The quantities $C_L(\eta)$ are defined as~\cite{abra}
\begin{equation}
  C_0(\eta)=\sqrt{ 2\pi\eta\over e^{2\pi\eta}-1}\ , \qquad
  C_L(\eta)= { \sqrt{L^2+\eta^2}\over L (2 L+1) }C_{L-1}(\eta)\ . \label{eq:cfactor}
\end{equation}
Note that $C_L(0)=1/(2L+1)!$. Let us look for a function $f_0(y)$
expressed as
\begin{equation}
  f_0(y) = a_0 + a_1 y
           + ( b_1 y +  b_2 y^2 + b_3 y^3)\log(2qy)\ .
   \label{eq:f0reg1}
\end{equation}
For $y\rightarrow 0$
\begin{eqnarray}
  {f_0(y)e^{-\beta y} \over y} &\!\rightarrow\!& {a_0\over y} +
  {a_1\!-\!a_0\beta} +O(y) \nonumber \\    
      &\!\!+\!\!& \left[ b_1 + (b_2 \!-\!\beta b_1)y+O(y^2)\right]\log(2qy)\ .
   \label{eq:f0reg2}
\end{eqnarray}
In order to have $\widetilde G_0/qy= G_0/qy - f_0(y)e^{-\beta y}/y$ regular at the
origin (together with its first derivative), we have to make vanish the coefficients
of the terms $1/y$, $\log(2qy)$, and  $y\log(2qy)$, namely
\begin{equation}
  a_0= {1\over C_0(\eta) q}\ , \quad 
  b_1= 2 \eta q a_0\ , \quad
  b_2=2\eta q (\eta q+\beta ) a_0\ .\label{eq:f0reg3}
\end{equation}
The other two coefficients $a_1$ and $b_3$ are determined so that $\overline
G_0$ is regular at the origin. From Eq.~(\ref{eq:ovrlggb}), and using the
expressions for the coefficients $a_0$, $b_1$, and $b_2$ given above, we find
for $y\rightarrow 0$
\begin{eqnarray}
  \overline G_0 &=& \biggl\{ {3 b_2 - 2\beta (a_1+b_1)+(\beta^2+q^2) a_0-2\eta q a_1
       \over y} +O(y^0) \nonumber \\
    && + \biggl[ 6b_3 -(4\beta+2\eta q) b_2 +(\beta^2+q^2)
    b_1\nonumber \\
   && \qquad\qquad +     O(y)\biggr]\log(2qy)\biggr\} 
    e^{-\beta y}\ . \label{eq:f0reg4}
\end{eqnarray}
Therefore, we choose
\begin{eqnarray}
  a_1&=& {\beta^2+q^2  +2 \eta q (\beta+3\eta q)\over 2(\beta+\eta q)} a_0\ , \label{eq:f0reg5}\\
  b_3&=& {1\over 3} (2\beta+\eta q) b_2 -{\beta^2+q^2\over 6} b_1 \ .\label{eq:f0reg6}
\end{eqnarray}
In this way $\overline G_0$ is regular at the origin.

\subsection{Case $L=1$}
As before, we start from the small-$y$ behavior of $G_1$, which
reads~\cite{abra}
\begin{eqnarray}
  {G_1(\eta,qy)\over qy} &\!\!\rightarrow\!\! & {1\over 3 C_1(\eta) q^2}
   \biggl[ {2\over 3} q^2 \eta (1+\eta^2)\nonumber \\
     && \times \Bigl(qy+ {\eta\over 2} (qy)^2 + O(y^3)\Bigr)
          \log(2qy)\nonumber \\
    &&  + \Bigl({1\over y^2} -{\eta q\over y} + {1+2\eta^2 \over
            2} q^2 + O(y) \Bigr) \biggr]\ .\label{eq:g1}
\end{eqnarray}
Let us look for a function $f_1(y)$
expressed as
\begin{equation}
  f_1(y) = a_0 + a_1 y+ a_2 y^2 
           + ( b_3 y^3 +  b_4 y^4)\log(2qy)\ .
   \label{eq:flreg1}
\end{equation}
For $y\rightarrow 0$
\begin{eqnarray}
  {f_1(y)e^{-\beta y} \over y^2} &\!\rightarrow\!& {a_0\over y^2} +
  {a_1-a_0\beta\over y} +(a_2- a_1\beta +a_0 {\beta^2\over 2}) +O(y)
  \nonumber \\
     &      \!+\!& \left[ b_3 y +  (b_4\! -\!b_3\beta)y^2+O(y^3)\right]\log(2qy)\ .
   \label{eq:flreg2}
\end{eqnarray}
In order to have $\widetilde G_1/qy= G_1/qy - f_1(y)e^{-\beta y}/y^2$ regular at the
origin (together with its first derivative), we have to make vanish the coefficients
of the terms $1/y^2$, $1/y$, and  $y\log(2qy)$, namely
\begin{equation}
  a_0= {1\over 3C_1(\eta) q^2}\ , \quad 
  a_1= (\beta - \eta q)a_0\ , \quad
  b_3={2\over 3} q^3\eta (1+\eta^2) a_0\ .\label{eq:flreg3}
\end{equation}
The other two coefficients $a_2$ and $b_4$ are determined so that $\overline
G_1$ is regular at the origin. From Eq.~(\ref{eq:ovrlggb}), and using the
expressions for the coefficients $a_0$, $a_1$, and $b_3$ given above, we find
for $y\rightarrow 0$
\begin{eqnarray}
  \overline G_1 &=& \biggl\{ {2 a_2 - \Bigl(\beta^2+q^2+2\eta q(\beta-\eta
      q)\Bigr) a_0 \over y^2}  \nonumber \\
    && + { \Bigl( 2 q^3\eta (1+\eta^2) + (\beta^2+q^2)(\beta-\eta q)\Bigr)
      a_0 \over y}\nonumber \\
     && \qquad - { 2 (\beta+\eta q) a_2 \over y} + O(y^0)\nonumber \\
    && + \biggl[ 4b_4 -(4\beta+2\eta q)b_3 +
    O(y)\biggr]\log(2qy)\biggr\}\nonumber \\
    && \qquad \times e^{-\beta y}\ . \label{eq:flreg4}
\end{eqnarray}
Therefore, we choose
\begin{equation}
  a_2= \left[{\beta^2+q^2 \over 2} -\eta q (\beta-\eta q)\right]a_0\ , \qquad
  b_4= {2\beta +\eta q\over 2} b_3\ .\label{eq:flreg5}
\end{equation}
With this choice also the coefficient of the term $1/y$ automatically
vanishes. In this way $\overline G_1$ is regular at the origin. 

\subsection{Cases $L\ge2$}
Now for $y\rightarrow 0$ the logarithmic term in the expression of
$G_L/qy$ does not give problem (it is multiplied by a factor
$y^{L+1}$). Therefore, we can retain all coefficients $b=0$
in Eq.~(\ref{eq:flregb}). The coefficients $a$'s are then fixed using the same
procedure as described above. Now, we define $f_L(y)$ as
\begin{equation}
  f_L(y) = a_0 + a_1 y+ a_2 y^2 +  \cdots + a_{L+2} y^{L+2}\ , \qquad L\ge 2\ ,
   \label{eq:flregc}
\end{equation}
and we introduce
\begin{equation}
   \tilde f_L(y)= f_L(y)e^{-\beta y}\ , \qquad
   \tilde f_L(y)=\sum_{k=0}^{\infty} \tilde a_k y^k \ ,
   \label{eq:flregc2}
\end{equation}
where
\begin{equation}
    \tilde a_k = \sum_{k'=0}^{L+2} { a_{k'}
      (-\beta)^{k-k'} \over (k-k')! } \ ,\qquad k=0,\ldots,\infty\ .
    \label{eq:flregc3} 
\end{equation} 
The coefficients $\tilde a_k$, $k=0,\ldots,L+2$ can be fixed by requiring that
$\widetilde G_L/qy $ and $\overline G_L$ be regular at the origin, 
with the result that 
\begin{eqnarray}
  &k=0       &\ \ \tilde a_0 = {1\over (2L+1)C_L(\eta)q^{L+1} }  \ ,\nonumber \\
  &k=1       &\ \  \tilde a_{1} = - {\eta q \over L} \tilde a_0 \ , \label{eq:flregc4}\\
  &k=2,\ldots,L+2 &\ \      \tilde a_{k} = - {2\eta q \tilde a_{k-1} - q^2 \tilde a_{k-2}
    \over k(k-2L-1) }\ .\nonumber
\end{eqnarray}
The parameters $a_k$, $k=0,\ldots,L+2$, can be readily obtained from $\tilde
a_k$ by recurrence, using Eq.~(\ref{eq:flregc3}). In fact
\begin{eqnarray}
  a_0&\!=\! & \tilde a_0 = {1\over (2L+1)C_L(\eta)q^{L+1} }  \ ,\label{eq:flregc5} \\
  a_k&\!=\! &  \tilde a_k\! -\! \sum_{k'=0}^{k-1} { a_{k'}
      (-\beta)^{k-k'} \over (k-k')! } \ ,\  k=1,\ldots,L+2\ .
    \label{eq:flregc6}
\end{eqnarray}
Since we need to fix $\tilde a_k$, $k=0,\ldots,L+2$, we need to have at least $L+3$
parameters $a_0,\ldots,a_{L+2}$ in the expansion for $f_{L\ge 2}$ given in
Eq.~(\ref{eq:flregc}), i.e. we can set $a_{k\ge L+3}=0$.

%\section{Convergence}
%\label{app:c}

\begin{acknowledgments}
The Authors would like to acknowledge the National Supercomputing Consortium
CINECA where part of the calculations presented in this paper were performed.
%Moreover, the Authors would like to thank also  the assistance and help of the staff of the computer center of 
%INFN-Pisa, 
\end{acknowledgments}

\end{document}